\documentclass{article}
\usepackage[utf8]{inputenc}
\usepackage[a4paper, margin=0.8in]{geometry}
\usepackage[style=nature,doi=false,isbn=false,eprint=false,maxbibnames=99]{biblatex}
\AtEveryBibitem{\clearlist{language}}
\AtEveryBibitem{\clearlist{urldate}}

\DeclareUnicodeCharacter{2212}{-}

\usepackage{siunitx}
\usepackage{tabularx}[]
\usepackage{array}
\usepackage{amsmath}
\usepackage{amssymb}
\usepackage{booktabs}
\usepackage{makecell}
\usepackage{enumitem}
\usepackage{soul} 
\usepackage{xr} 
\usepackage[symbol]{footmisc} 

\usepackage{graphicx}
\usepackage{caption}
\usepackage{subcaption}
\usepackage{float}
\usepackage{hyperref}
\usepackage[protrusion=true]{microtype}
\usepackage[capitalise,noabbrev]{cleveref}
\usepackage{fancyhdr}
\usepackage{authblk}

\title{Dynamics and Control of Oscillatory Bioreactors}
\author[1]{Pavan K. Inguva$^{\ddag,}$}
\author[1]{Krystian Ganko$^{\ddag,}$}
\author[1]{Alexis B. Dubs}
\author[1]{Richard D. Braatz*$^{,}$}
\affil[1]{Massachusetts Institute of Technology, 77 Massachusetts Avenue, Cambridge, MA 02139, USA}

\date{}

\begin{document}

\maketitle
\footnotetext[3]{Authors contributed equally to this work.}
\footnotetext[1]{Corresponding author. Email: braatz@mit.edu}
\renewcommand{\thefootnote}{\arabic{footnote}} 

\begin{abstract}
\noindent
Bioreactors are widely used in many industries to generate a range of products using various host cells e.g., yeast, insect, and mammalian cells. Depending on the process, product, and host cell, some bioreactors exhibit sustained periodic behavior in key process variables such as metabolite concentrations, biomass, and product titer. Such dynamical behavior can arise from different mechanisms, including predator-prey dynamics, substrate inhibition, and cell sub-population synchrony. Oscillatory dynamical behavior is undesirable as it can impact downstream processes, especially in a continuous operation, and can make process operations and product quality control more challenging. This article provides an overview of oscillatory dynamics. The mechanisms that give rise to the oscillations and process control strategies for suppressing the oscillations are discussed, while providing insights that go beyond past studies. Alternative process configurations are proposed for bypassing the mechanisms that generate oscillations.
\end{abstract}

\section{Introduction}

Bioreactors are widely used in the manufacturing of a variety of products ranging from bulk chemicals such as bioethanol \cite{aditiya_second_2016} and lactic acid \cite{tejayadi_lactic_1995}, fine chemicals such as pigments and polyunsaturated fatty acids \cite{sharmila_production_2020}, and complex biological products such as monoclonal antibodies \cite{warnock_bioreactor_2006} and viral particles \cite{frensing_continuous_2013}. In addition to the diversity of products manufactured in bioreactors, depending on process and product requirements, different host cell types (e.g., mammalian, insect, yeast, and plant cells) \cite{demain_production_2009} and bioreactor configurations \cite{sharma_advances_2022,kiani_deh_kiani_different_2022} can be used for production. Most combinations of a product, host cell type, and bioreactor configuration, when considered together with the intrinsically complex intracellular kinetics present in biological systems result in a system with highly nonlinear dynamics.

Complex dynamical systems are able to demonstrate a range of behaviors such as a multiplicity of steady states, oscillations, and period doubling \cite{epstein_introduction_1998} which need to be addressed when optimizing bioreactors. The study of dynamical systems is extensive and provides a mathematical framework for understanding and predicting such complex behaviors. Nonlinear dynamics can be characterized by bifurcation analysis (e.g., see \cite{garhyan_bifurcation_2004,zhang_bifurcation_2001,epstein_introduction_1998} and citations therein), which is facilitated by modern software packages \cite{veltz_bifurcationkitjl_2020}. Understanding the basis for nonlinear dynamics in bioreactors, as well as strategies to mitigate the impact of and/or eliminate such behavior, is important during process development and for subsequent production-scale operations. 

One particular complex dynamical behavior of interest in bioreactors is steady-state oscillations in key process variables such as product titers and biomass concentration. Oscillations have been experimentally observed in multiple types of bioreactor operations and have been attributed to multiple causes such as predator-prey dynamics \cite{canova_mechanistic_2023,frensing_continuous_2013} and complex substrate and/or product inhibition \cite{alvarez-ramirez_existence_2009,mclellan_incidence_1999,daugulis_experimental_1997}. The mathematical and physical basis for oscillations in some bioreactors have been explored theoretically (e.g., see \cite{frensing_continuous_2013,alvarez-ramirez_existence_2009,garhyan_bifurcation_2004,garhyan_exploration_2003,zhang_bifurcation_2001} and citations therein) and can provide insights on how to mitigate oscillations by selecting stable operating conditions. Oscillatory behavior in bioreactors is problematic for multiple reasons,
\begin{enumerate}
    \item Process productivity such as quantified by production rate and/or substrate utilization can vary significantly during operation, leading to material loss and inefficient processes.
    \item The biopharmaceutical industry has high product quality requirements, and unstable bioreactor conditions arising of oscillatory behavior can adversely impact product quality.
    \item The design of downstream processes to handle process variables that oscillate significantly over time is challenging as key process variables such as product titers are important design parameters for sizing subsequent unit operations, especially for continuous manufacturing operations.  
\end{enumerate}

While identifying open-loop strategies to mitigate oscillatory behavior can be explored for a given process configuration, it may sometimes be insufficient to ensure robust process operation as process disturbances and upsets may drive the system to an oscillatory state (e.g., see \cite{garhyan_bifurcation_2004}). As such, closed-loop control which employs feedback to control a process can be a more robust strategy. This study explores how simple output feedback can be used to stabilize and optimize bioreactors. A series of four case studies, each with a different physical basis for oscillatory behavior, are discussed that demonstrate how output feedback control can suppress and sometimes even eliminate oscillatory behavior. Alternative process configurations and bioreactor types which bypass the source of the oscillatory behavior are also explored. 

\section{Case Study 1: Continuous Stirred Tank Bioreactor}

\begin{figure}[htb]
    \centering
    \begin{subfigure}{0.48\textwidth}
        \centering
        \includegraphics[height=4cm]{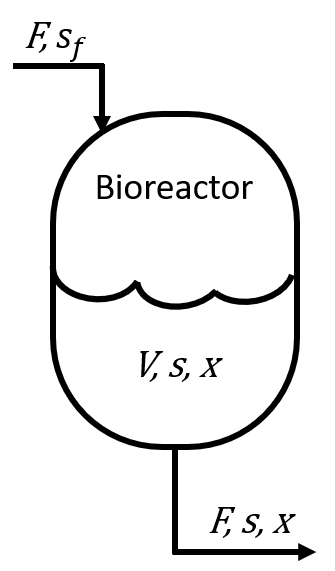} 
        \vspace{-0.2cm}
        \caption{Continuous stirred tank bioreactor}
        \label{fig:bioreactor}
    \end{subfigure}
    \hfill
    \centering
    \begin{subfigure}{0.48\textwidth}
        \centering
        \includegraphics[height=4cm]{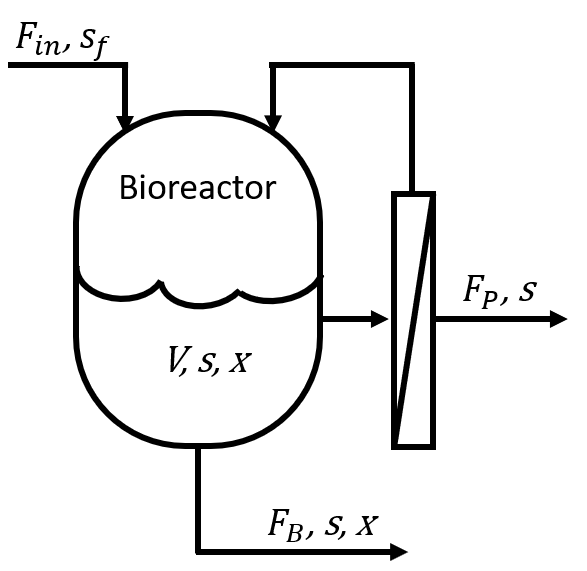} 
        \vspace{-0.2cm}
        \caption{Continuous stirred tank bioreactor with perfusion}
        \label{fig:bioreactorperf}
    \end{subfigure}
    \caption{Schematic of continuous bioreactor configurations considered in Case Study 1.}
    \label{fig:bioreactor_schematic}
\end{figure}

This case study considers the open- and closed-loop dynamics of a continuous, stirred tank bioreactor (Fig.~\ref{fig:bioreactor}). Sections \ref{alvarez:description}, \ref{alvarez:openloop}, and \ref{alvarez:closed} summarize the model, analysis, and feedback control results of \cite{alvarez-ramirez_existence_2009}, while providing some additional insights, whereas Section \ref{alvarez:config} describes the results of modeling an alternative bioreactor configuration that removes the oscillations without using feedback control.

\subsection{Model Description}
\label{alvarez:description}
A continuous two-state bioreactor model describing one substrate (e.g., glucose) and the biomass useful for understanding bioreactor dynamics while being amenable to analytical methods is 
\begin{align}
    \frac{\textrm{d}s}{\textrm{d}t} &= D(s_\textrm{f} - s) - \frac{\mu(s)}{v(s)}x, \label{eq:dsdt} \\
    \frac{\textrm{d}x}{\textrm{d}t} &= (\mu(s) - D)x,
    \label{eq:dxdt}
\end{align}
where $s$ and $x$ are the substrate (e.g., glucose) and biomass concentrations, $s_\textrm{f}$ is the feed substrate concentration, and $D$ is the dilution rate which is the ratio of the volumetric flow rate and the bioreactor working volume $\left(\frac{F}{V}\right)$, $\mu$ is the growth rate which can be a function of the substrate concentration, and $v$ is the yield function which describes the amount of substrate consumed to produce new cells and which can also be a function of the substrate concentration \cite{alvarez-ramirez_existence_2009}. The functional forms for $\mu$ and $v$ are 
\begin{equation}
     \mu(s) = \frac{\mu_{\max} s}{K_\textrm{s} + s}
     \label{eq:mu}
\end{equation}
\begin{equation}
     v(s) = \left(v_{\min}+a s\right)^\rho, \quad a \geq 0.
     \label{eq:v}
\end{equation}
where $\mu(s)$ follows the widely used Monod kinetics form, and $v(s)$ models a type of cell growth in which a minimum amount of substrate is required to produce a new unit weight of cells but, as the amount of available substrate increases, the amount of substrate used in the production of each new unit weight of cells increases. Both $\mu$ and $v$ are monotonically increasing functions of the substrate concentration $s$, $\mu(s)$ is bounded from above by $\mu_{\textrm{max}}$, and $v(s)$ is bounded from below by $v_{\textrm{min}}$.
A list of the parameter and initial condition values for the model \eqref{eq:dsdt}--\eqref{eq:dxdt} used in this work can be found in Table \ref{tbl:alvarez_params}.

\begin{table}[htbp]
\centering
\caption{Description of variables and parameters in the -continuous two-state bioreactor model \eqref{eq:dsdt}--\eqref{eq:dxdt}.}\renewcommand{\arraystretch}{1.4}
    \begin{tabular}{ccccc}
    \toprule
     \makecell[c]{Variable/ \\ Parameter}& Description & Nominal Value & Initial Condition & Units\\
     \hline 
     $s$ & Substrate concentration & -- & 200 & $\frac{\text{g}}{\text{L}}$ \\
     $x$ & Biomass concentration & -- & 1 & $\frac{\text{g}}{\text{mL}}$ \\
     $D$ & Dilution rate & 0.25 & -- & $\frac{\text{1}}{\text{hr}}$ \\
     $s_\textrm{f}$ & Concentration of substrate in the feed & 200 & -- & $\frac{\text{g}}{\text{L}}$ \\
     $\mu_\textrm{max}$ & Monod maximum value of growth rate & 0.3 & -- & $\frac{\text{g}}{\text{g}}$ \\
     $K_\textrm{s}$ & Monod half velocity constant & 1.75 & -- & $\frac{\text{g}}{\text{L}}$ \\
     $v_\textrm{min}$ & Minimum yield & 0.1 & -- & $\frac{\text{g}}{\text{g}}$ \\
     $a$ & Yield function parameter & 0.03 & -- & $\frac{\textrm{L}}{\textrm{g}}$ \\
     $\rho$ & Yield function parameter & 0.75 & -- & -- \\
     \bottomrule
    \end{tabular}
    \renewcommand{\arraystretch}{1}
    \label{tbl:alvarez_params}
\end{table}

\subsection{Open-loop Analysis}
\label{alvarez:openloop}
A detailed mathematical analysis of the model equations can be found in \cite{alvarez-ramirez_existence_2009} and key results are presented here to illustrate specific concepts that are relevant for understanding oscillatory systems. Stable oscillations in a dynamical system can occur when the states remain bounded over time and the equilibrium point (the system state at which the system is not changing and the time derivatives of the system state are equal to zero) of the system is unstable \cite{del_vecchio_biomolecular_2015}.  

\begin{figure}[htb]
    \centering
    \begin{subfigure}{0.45\textwidth}
        \centering
        \includegraphics[width=\textwidth]{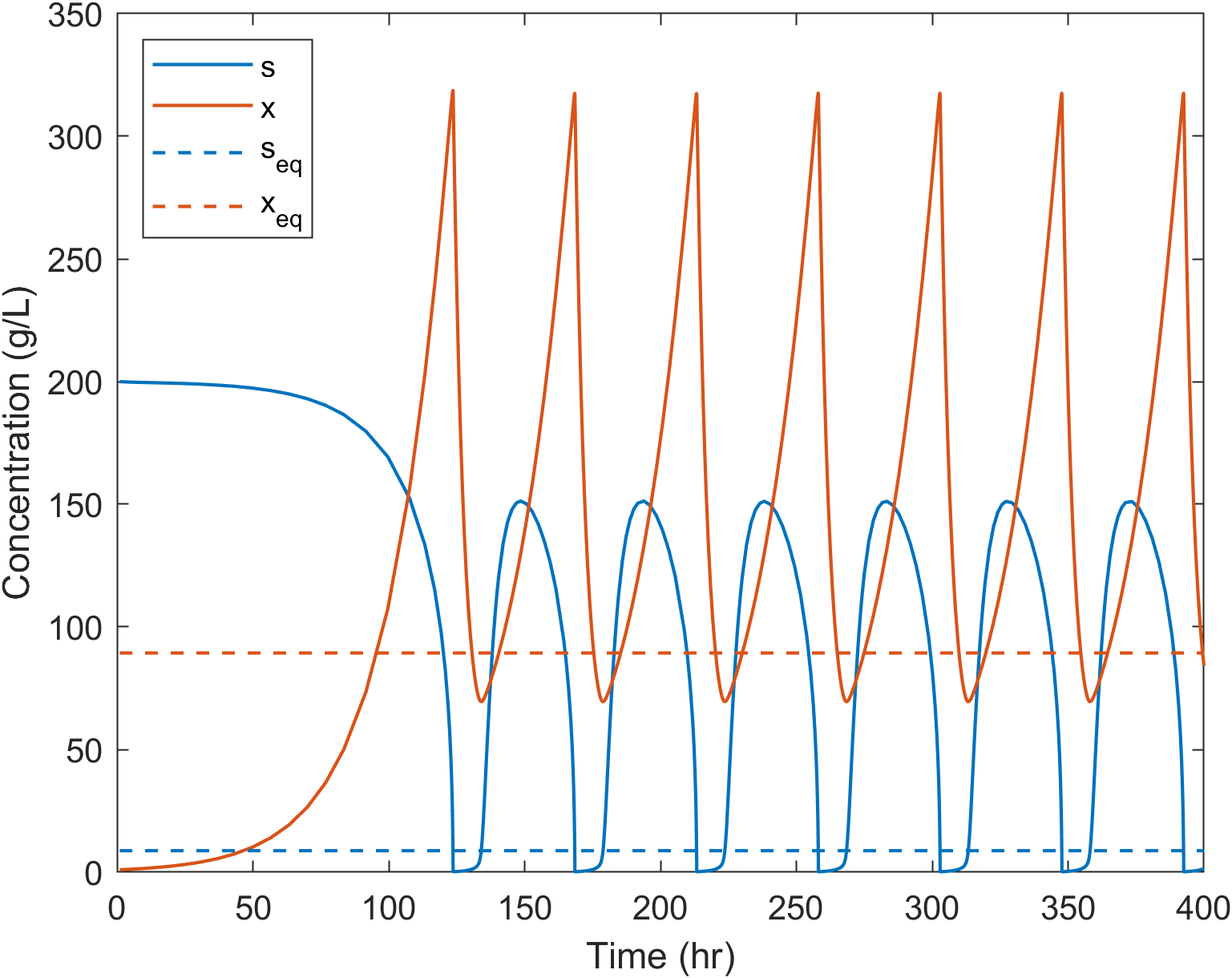} 
        \vspace{-0.4cm}
        \caption{Concentration time series}
        \label{fig:1}
    \end{subfigure}
    \hfill
    \centering
    \begin{subfigure}{0.45\textwidth}
        \centering
        \includegraphics[width=\textwidth]{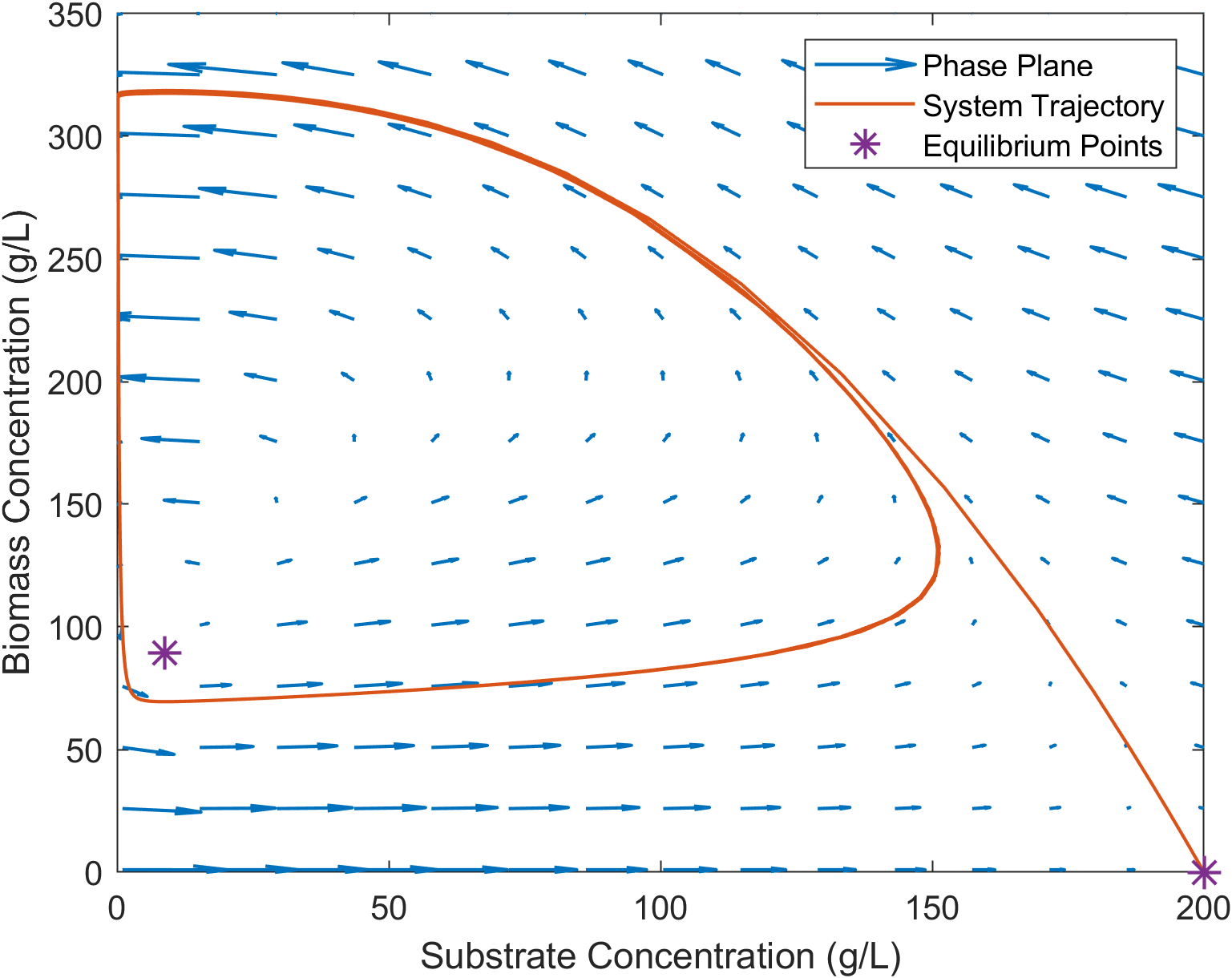} 
        \vspace{-0.4cm}
        \caption{Phase portrait}
        \label{fig:2}
    \end{subfigure}
    \vskip\baselineskip
    \begin{subfigure}{0.45\textwidth}
        \centering
        \includegraphics[width=\textwidth]{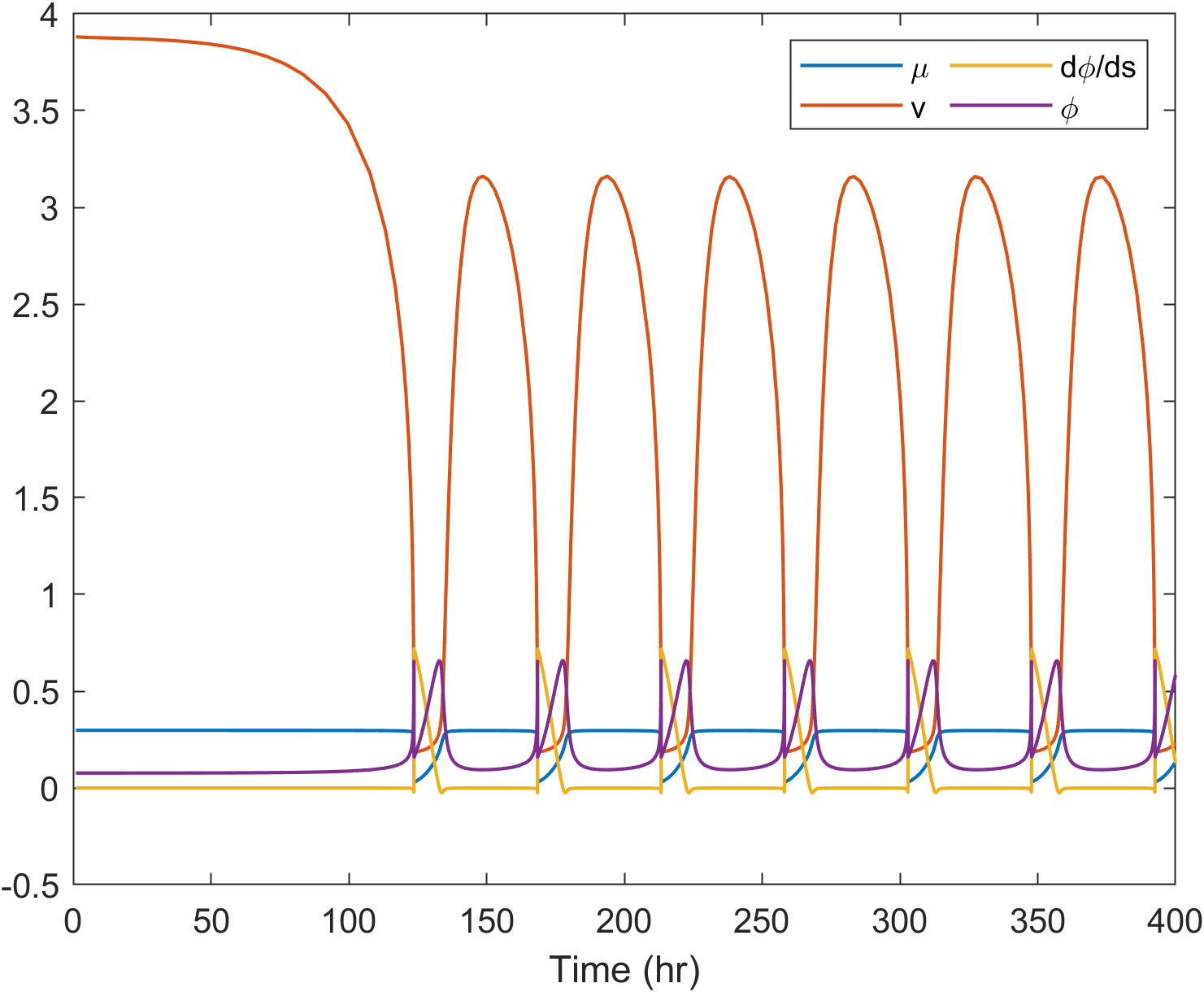} 
        \vspace{-0.4cm}
        \caption{Time series of $\phi$}
        \label{fig:3}
    \end{subfigure}
    \hfill
    \centering
    \begin{subfigure}{0.45\textwidth}
        \centering
        \includegraphics[width=\textwidth]{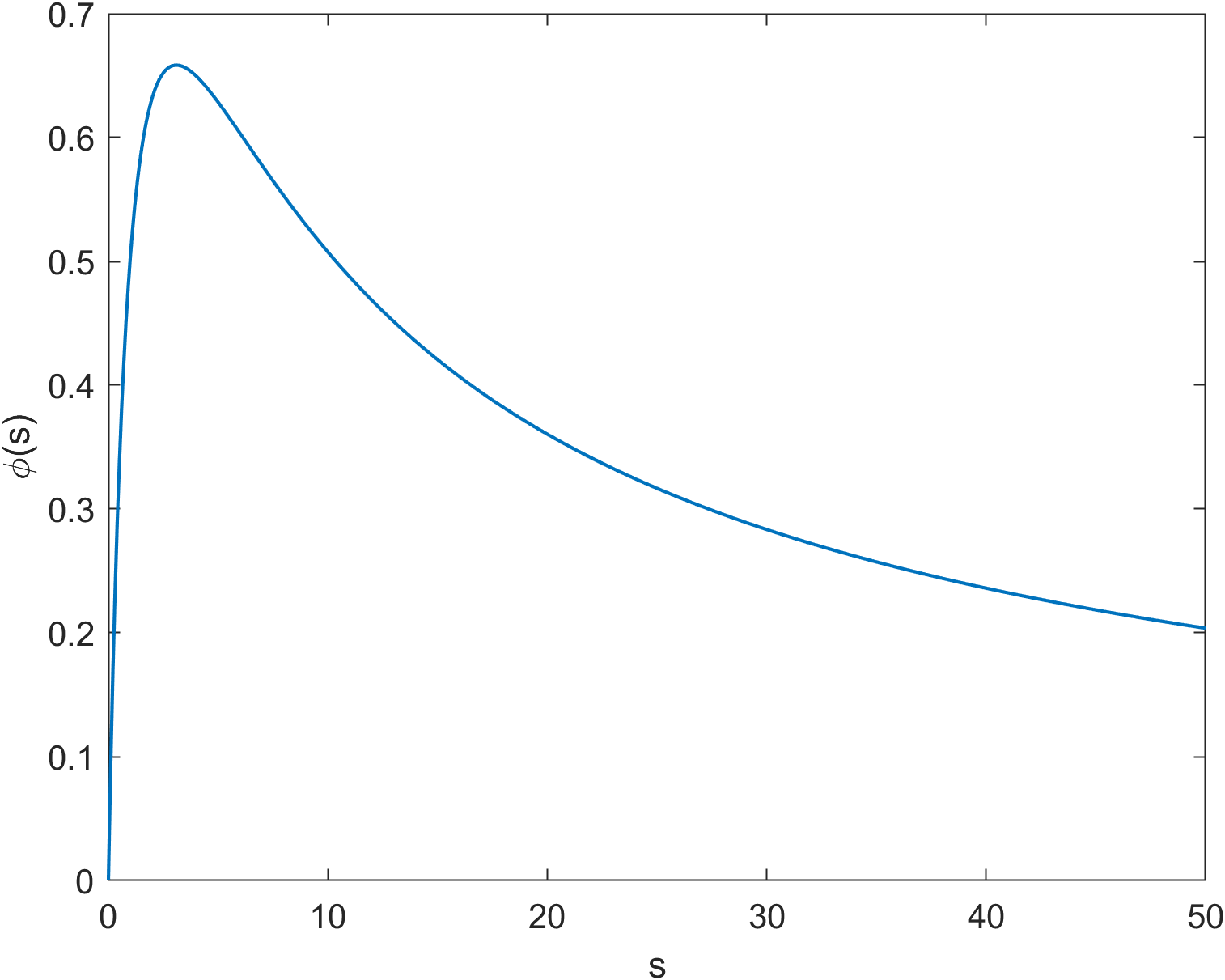} 
        \vspace{-0.4cm}
        \caption{$\phi(s)$ with \eqref{eq:mu} for $\mu$ and \eqref{eq:v} for $v$}
        \label{fig:phi}
    \end{subfigure}
    \caption{Open-loop dynamics of the bioreactor model \eqref{eq:dsdt}--\eqref{eq:dxdt}. Large oscillations occur. The equilibrium points are ($s_{\textrm{eq}},x_{\textrm{eq}}) = (8.75, 89.35)$ and $(200, 0)$, the latter being the washout point.}
    \label{fig:ogmodel}
\end{figure}

The system \eqref{eq:dsdt}--\eqref{eq:dxdt} has two equilibrium points. One is the trivial ``washout point", which occurs when the growth rate is too low to maintain a population of cells in the bioreactor. The washout point has equal inlet and outlet substrate concentrations, $s_{\textrm{wo}} = s_{\textrm{f}}$, and a biomass concentration equal to zero, $x_{\textrm{wo}} = 0$. The nontrivial equilibrium point can be described by 
\begin{align}
    \mu(s_\textrm{eq}) &= D, \\
    x_\textrm{eq} &= (s_\textrm{f}-s_\textrm{eq})v(s_\textrm{eq}),
    \label{eq:eq}
\end{align}
with the first equation being that the growth rate is equal to the dilution rate. 
The stability condition of an equilibrium point can be derived by considering the eigenvalues of the Jacobian at that point. For the system \eqref{eq:dsdt}--\eqref{eq:dxdt}, the equilibrium point becomes unstable when
\begin{equation}
    -D-\left.\frac{\textrm{d} \phi}{\textrm{d} s}\right|_{s_\textrm{eq}} x > 0,
    \label{eq:osc}
\end{equation}
where $\phi(s) = \frac{\mu(s)}{v(s)}$ is the rate of substrate consumption per unit weight of cells. This inequality can only hold for a negative $\left.\frac{\textrm{d} \phi}{\textrm{d} s}\right|_{s_\textrm{eq}}$ which can occur for the above $\mu(s)$ and $v(s)$, and for a constant $v$ with some more complicated expressions for $\mu(s)$  (e.g., for Monod kinetics extended to account for substrate inhibition) \cite{alvarez-ramirez_existence_2009}. This condition \eqref{eq:osc} can only hold for an $\phi(s)$ that has a maximum (e.g., see Fig.~\ref{fig:phi}).
The stable oscillations in the biomass and substrate concentrations can have a high magnitude for some values of the model parameters (Figs.~\ref{fig:1}b). 

Qualitatively, the origins of the oscillatory behavior can be understood by examining Fig.~\ref{fig:ogmodel}. Initially the bioreactor starts at high substrate and low biomass concentrations. As the cells grow, the biomass concentration initially increases exponentially while the substrate concentration decreases as substrate is consumed. As the substrate concentration approaches 0, $\phi$ increases, as more substrate is needed to maintain the same rate of cell growth. This coupled with a high biomass concentration causes the substrate concentration to further drop. When the substrate concentration drops below $s^*$ (where $\phi(s^*) = \phi_\textrm{max}$), the biomass concentration sharply declines. As the substrate concentration continues to drop, $\phi$ sharply decreases, and it takes less substrate to maintain the same rate of cell growth. (This occurs after an spike in $\phi$, which can be explained by $v$ initially decreasing faster than $\mu$.) After some delay with the continuous addition of fresh media, the lower biomass concentration and lower $\phi$ allows the substrate concentration to recover, leading to an increase in both substrate and biomass concentrations. The cycle then repeats.

\subsection{Output feedback control}
\label{alvarez:closed}
As demonstrated in \cite{alvarez-ramirez_existence_2009}, process control can be used to eliminate the oscillations in the system. One of the simplest forms of feedback control is proportional (P) control. 
A single-loop proportional control strategy that uses the substrate concentration $s$ in the bioreactor as the controlled variable and the substrate concentration of the feed, $s_\textrm{f}$, as the manipulated variable is
\begin{equation}
    s_\textrm{f} = K_c(s_\textrm{sp} - s) + s_\textrm{nom}
\end{equation}
where $s_\textrm{nom}$ is the nominal substrate concentration (aka the controller bias), $s_\textrm{sp}$ is the setpoint for the substrate concentration $s$, and $K_{c}$ is the controller gain. The open-loop equilibrium substrate concentration $s_\textrm{eq}$ can be shown to be independent of $s_{f}$ and the substrate concentration $s$ of the controlled system tends towards $s_\textrm{eq}$. Hence, the setpoint and equilibrium substrate concentrations are set to be the same ($s_\textrm{sp}=s_\textrm{eq}$). In practice, the model may be modified to account for additional physics such as biomass density-dependent inhibition in which case this result would not hold. The nominal substrate concentration $s_\textrm{nom}$ is set to 200 g/L, which is the value used for the substrate concentration of the feed, $s_\textrm{f}$, in the open-loop simulations. 

\begin{figure}[h]
    \centering
    \includegraphics[scale = 0.6]{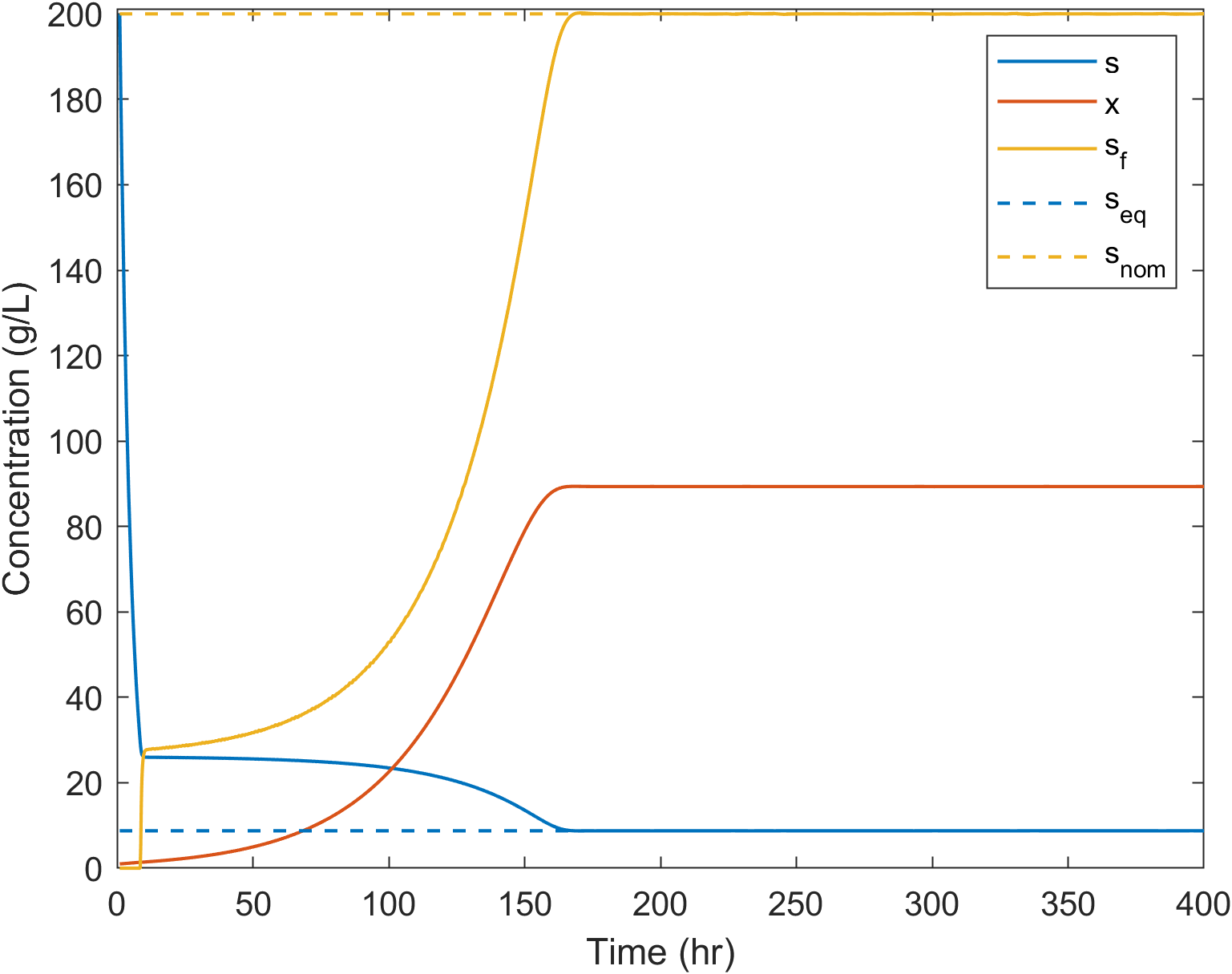}
    \vspace{-0.2cm}
    \caption{Closed-loop dynamics of the continuous bioreactor \eqref{eq:dsdt}--\eqref{eq:dxdt} for $s_\textrm{nom} = 200$ and $K_c = 10$. The system approaches a stable equilibrium, demonstrating the use of feedback control to eliminate oscillation.}
    \label{fig:control}
\end{figure}

The criterion for the minimum value of the controller gain $K_{c}$ that attenuates oscillations can be developed by considering the eigenvalues of the Jacobian of the closed-loop system \cite{alvarez-ramirez_existence_2009}. Assuming the model parameter regime gives rise to self-oscillations in the open-loop system, the closed-loop system is stable if and only if
\begin{equation}
    K_c>\frac{-\left.\frac{\textrm{d} \phi}{\textrm{d} s}\right|_{s_\textrm{eq}} x_\textrm{eq}}{D}-1.
    \label{eq:kccrit}
\end{equation}
When this condition is satisfied, the steady-state values of the substrate and biomass concentrations approach the values of the previously unstable equilibrium (Fig.~\ref{fig:control}).

\begin{figure}[htb]
    \centering
    \begin{subfigure}{0.45\textwidth}
        \centering
        \includegraphics[width=\textwidth]{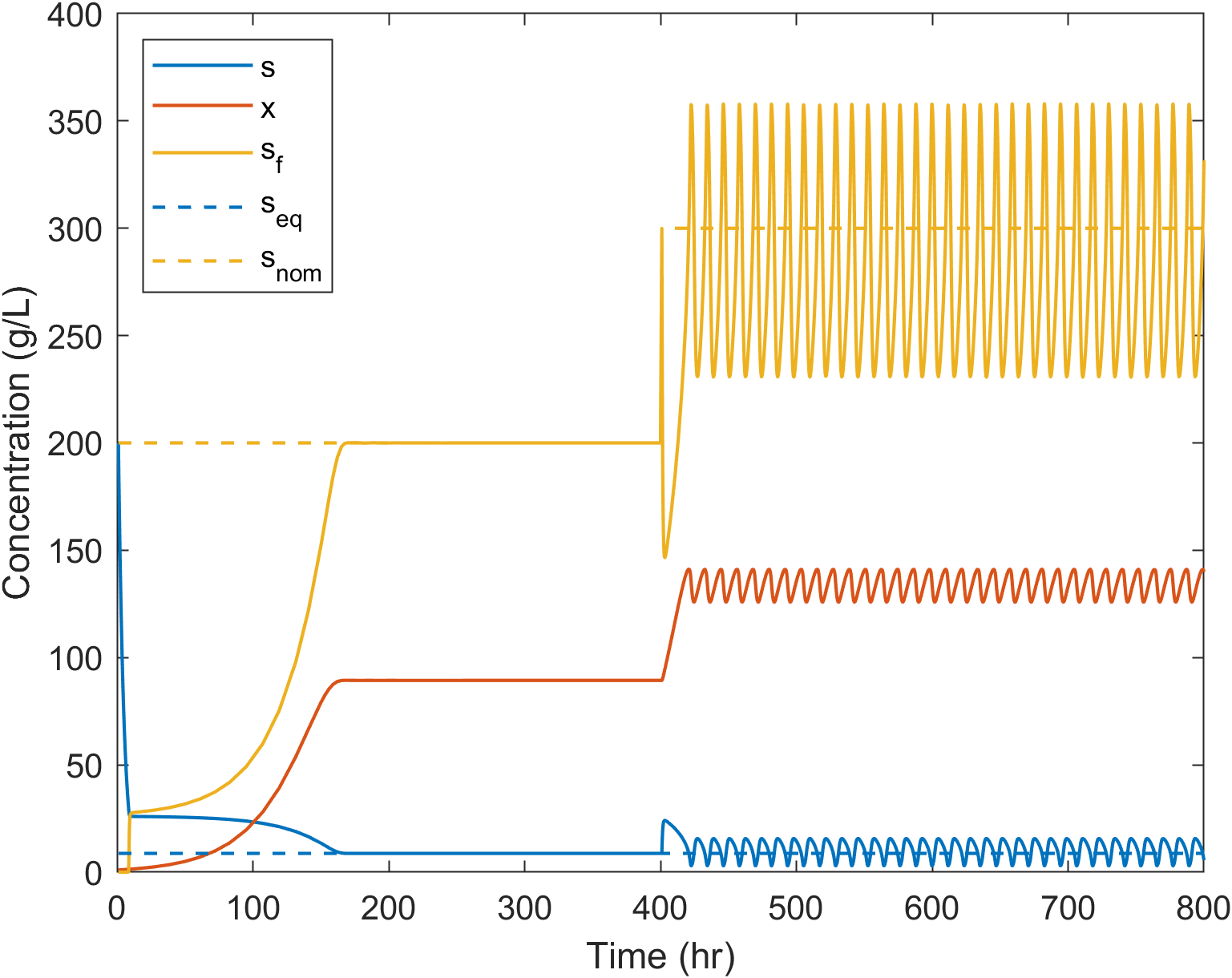} 
        \vspace{-0.4cm}
        \caption{$K_c = 10$}
        \label{fig:sfkc10}
    \end{subfigure}
    \hfill
    \centering
    \begin{subfigure}{0.45\textwidth}
        \centering
        \includegraphics[width=\textwidth]{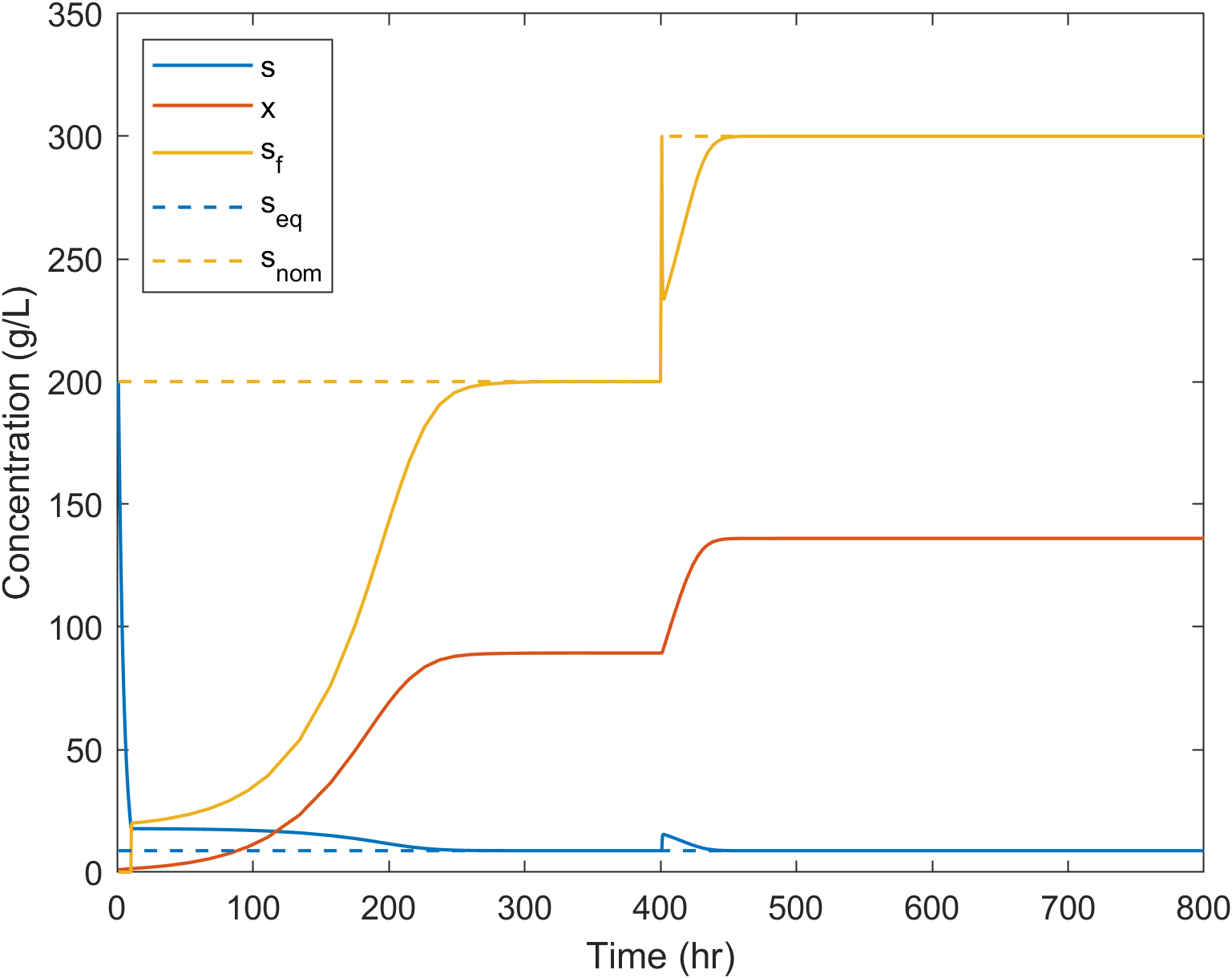} 
        \vspace{-0.4cm}
        \caption{$K_c = 20$}
        \label{fig:sfkc20}
    \end{subfigure}
    \caption{Closed-loop dynamics of the continuous bioreactor \eqref{eq:dsdt}--\eqref{eq:dxdt} with controller bias $s_\textrm{nom}$ changing from 200 to 300 at $t = 400$ hr. Increasing controller bias $s_\textrm{nom}$ can lead to oscillations even in a closed-loop system. These oscillations can be eliminated by increasing $K_c$.}
\end{figure}

The equilibrium biomass concentration  $x_\textrm{eq}$ can be shown to be dependent on the substrate concentration of the feed, $s_\textrm{f}$, and hence the controller bias $s_\textrm{nom}$.
If the process economics favors operating at a higher cell density, the equilibrium biomass concentration $x_\textrm{eq}$ can be increased by increasing $s_\textrm{nom}$.
However, the stability criterion \eqref{eq:kccrit} for the closed-loop system shows that the value of the controller gain $K_\textrm{c}$ necessary to attenuate oscillations is dependent on equilibrium biomass concentration $x_\textrm{eq}$.
Therefore, it is possible that increasing $s_\textrm{nom}$ can push the system from a stable state to an unstable state, as shown in Fig.~\ref{fig:sfkc10}.
The stability criterion \eqref{eq:kccrit} at the higher $s_\textrm{nom}$ can be fulfilled by increasing the controller gain $K_\textrm{c}$.
However, this increase in $K_\textrm{c}$ can lead to a slower approach to the system's steady state (Fig.~\ref{fig:sfkc20}).

\subsection{Alternative Process Configuration}
\label{alvarez:config}
\begin{figure}[htb]
    \centering
    \begin{subfigure}{0.45\textwidth}
        \centering
        \includegraphics[width=\textwidth]{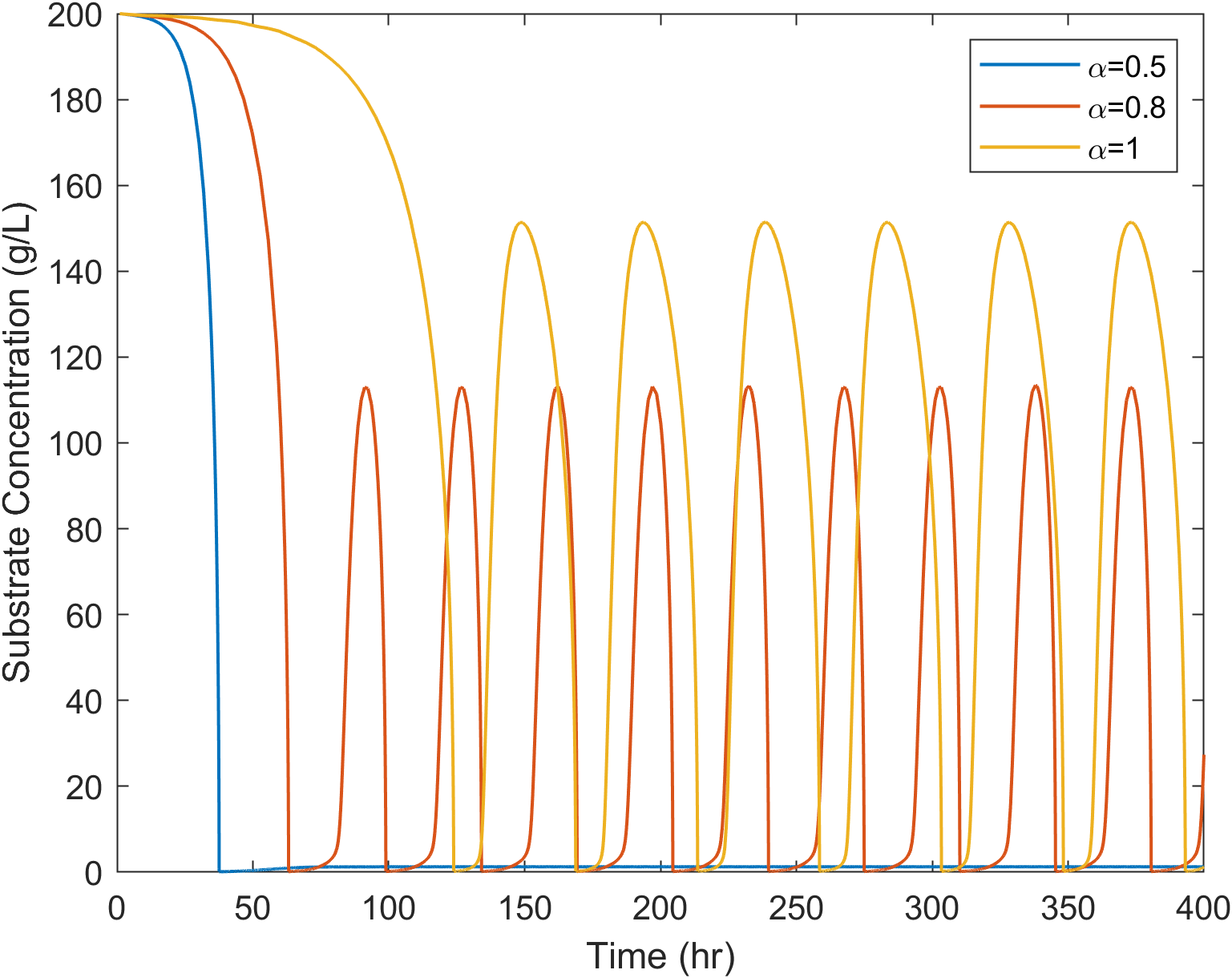} 
        \vspace{-0.4cm}
        \caption{Substrate concentration time series}
        \label{fig:perfs}
    \end{subfigure}
    \hfill
    \centering
    \begin{subfigure}{0.45\textwidth}
        \centering
        \includegraphics[width=\textwidth]{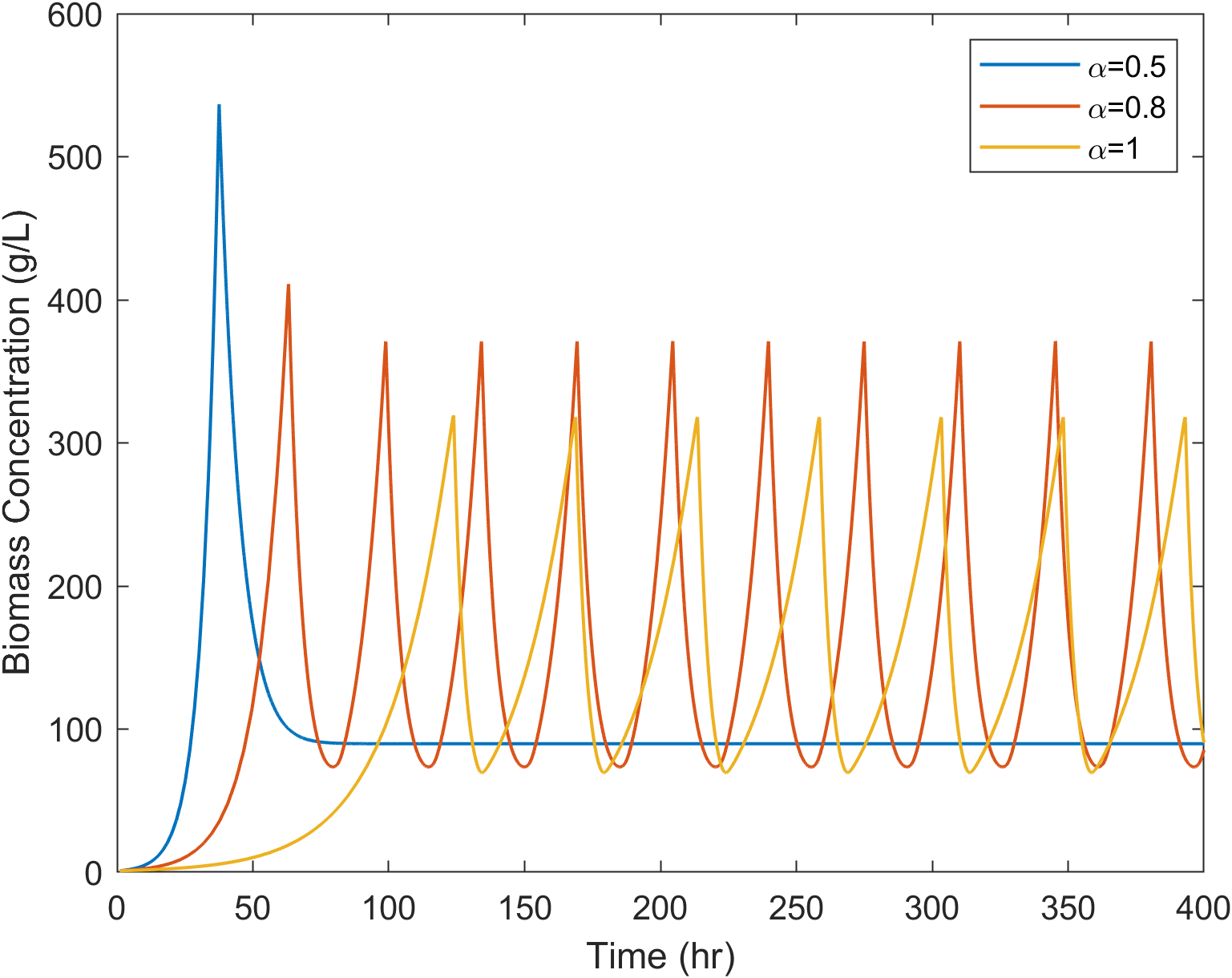} 
        \vspace{-0.4cm}
        \caption{Biomass concentration time series}
        \label{fig:perfx}
    \end{subfigure}
    \caption{Open-loop dynamics of the perfusion bioreactor \eqref{eq:dsdtperf}--\eqref{eq:dxdtperf}. There exists a value of $\alpha$ below which the system exhibits stable behavior and above which the system exhibits oscillatory behavior. The value $\alpha = 1$ recaptures the behavior of the system without perfusion and is included as a reference.}
\end{figure}

This section shows that the oscillations can be eliminated by modifying the bioreactor to operate in perfusion mode rather than by using feedback control.
Perfusion bioreactors include a cell separation device, which leads to two outlet streams, one with cells (the cell bleed $F_\textrm{B}$) and one without (the perfusate $F_\textrm{P}$), as shown in Fig.~\ref{fig:bioreactorperf} \cite{chotteau_perfusion_2015}.
The model equations describing a perfusion bioreactor are
\begin{align}
    \frac{\textrm{d}s}{\textrm{d}t} &= D(s_\textrm{f} - s) - \frac{\mu(s)}{v(s)}x, \label{eq:dsdtperf} \\
    \frac{\textrm{d}x}{\textrm{d}t} &= (\mu(s) - \alpha D)x,
    \label{eq:dxdtperf}
\end{align}
where $\alpha = \frac{F_\textrm{B}}{F_\textrm{in}}$ is the ratio of the cell bleed to the inlet stream.
This model assumes that the volume of the bioreactor remains constant ($F_\textrm{in} = F_\textrm{B} + F_\textrm{P}$), which is reasonable given that the constant volume control for a bioreactor would be fast-acting compared to the dynamics of \eqref{eq:dsdtperf} and \eqref{eq:dxdtperf}.
This model is very similar to that of a non-perfusion bioreactor, only differing by replacing the dilution rate $D$ by $\alpha D$ in 
 \eqref{eq:dxdtperf}. A sufficiently large value of $\alpha$ removes the oscillations in the bioreactor (Figs.\ \ref{fig:perfs}b).

\begin{figure}[h]
    \centering
    \includegraphics[scale = 0.6]{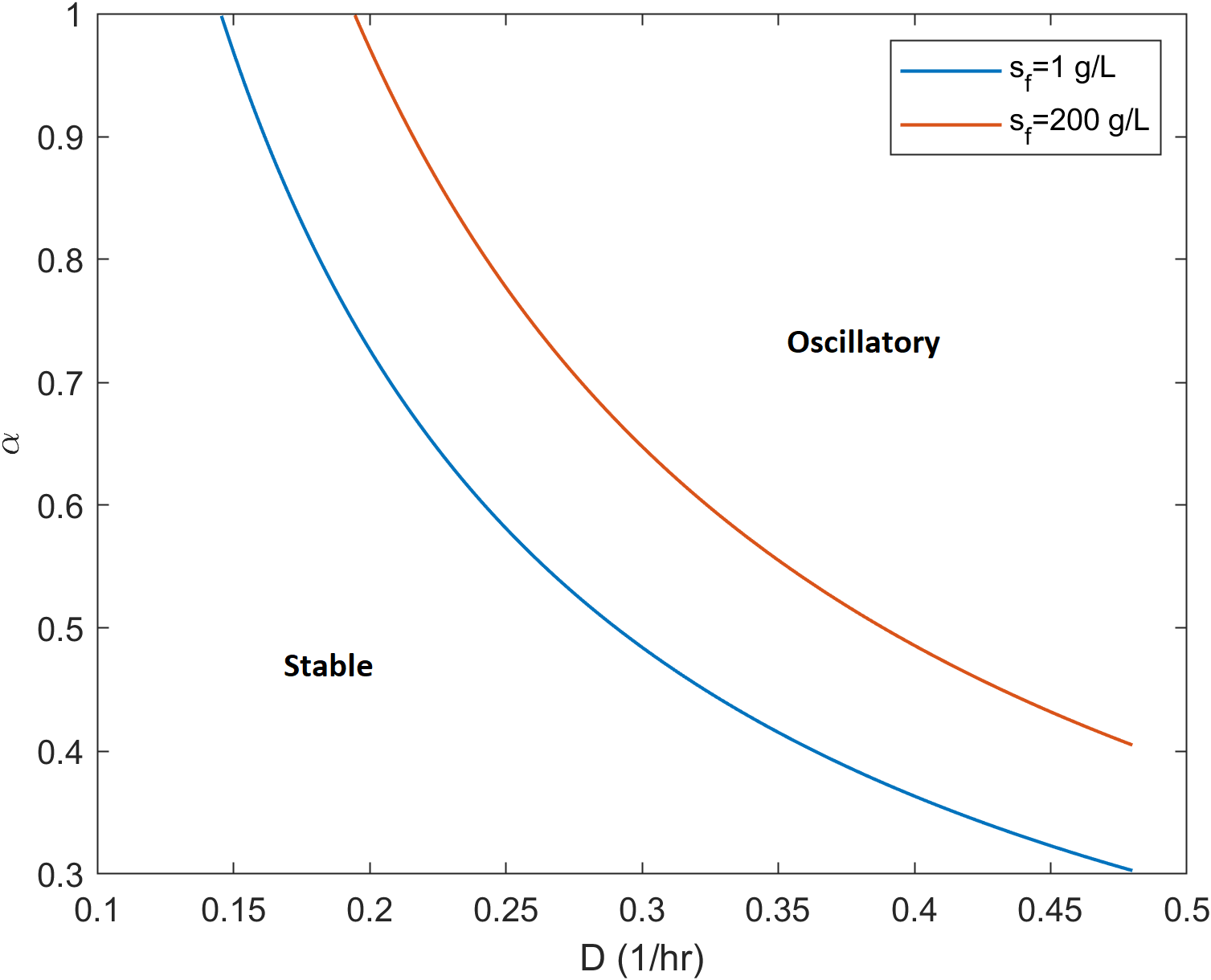}
    \vspace{-0.2cm}
    \caption{Regions of stability and oscillation for the perfusion bioreactor \eqref{eq:dsdtperf}--\eqref{eq:dxdtperf} for two values of the feed substrate concentration $s_\textrm{f}$. For a wide range of values of the dilution rate $D$ and feed substrate concentration $s_\textrm{f}$, there exists a value of the cell bleed ratio $\alpha^*$ for which the system is stable for $\alpha < \alpha^*$. }
    \label{fig:stabcrit}
\end{figure}

The same procedure used to analyze the stability of the bioreactor without perfusion can be used for this bioreactor with perfusion, which arrives at the same condition for instability \eqref{eq:osc}. For a non-perfusion bioreactor with stable oscillations,  $\phi(s_\textrm{eq})$ is in the region with a negative slope
(Fig.\ \ref{fig:phi}).
By setting the left-hand sides of\eqref{eq:dsdtperf} and \eqref{eq:dxdtperf} to zero and plugging the expression for $\mu(s)$ into \eqref{eq:mu}, we can analytically solve for the equilibrium concentration of substrate to give the expressions\medskip\\
\noindent
For the bioreactor without perfusion:
\begin{equation}
    s_{\textrm{eq,without}} = \frac{D K_\textrm{s}}{\mu_{\textrm{max}} - D}
\end{equation}
For the perfusion bioreactor:
\begin{equation}
    s_{\textrm{eq,perfusion}} = \frac{\alpha D K_\textrm{s}}{\mu_\textrm{max} - \alpha D}
    \label{eq:seqperf}
\end{equation}
The fact that $\alpha$ is less than one implies that
\begin{equation}
\frac{s_\textrm{eq,perfusion}}{s_\textrm{eq,without}} \le 1,
\end{equation}
given all other system parameters remain the same. For a large enough $\alpha$, the lower value of the equilibrium substrate concentration $s_\textrm{eq}$ for the perfusion bioreactor pushes $\phi(s_\textrm{eq})$ into the region with a positive slope (Fig.\ \ref{fig:phi}). \eqref{eq:osc} shows that a large enough value of $\frac{\textrm{d} \phi(s_\textrm{eq})}{\textrm{d}s}$ will lead to a stable system. Therefore, we can solve for the boundary between the unstable and stable systems using \eqref{eq:osc}, which occurs for

\begin{align}
    \frac{\textrm{d} \phi(s_\textrm{eq})}{\textrm{d}s}&=\frac{\mu_\textrm{max}(v_\textrm{min}+a s_\textrm{eq})^{-\rho-1}[K_\textrm{s}(v_\textrm{min}-a(\rho-1) s_\textrm{eq})-a \rho s_\textrm{eq}^2]}{(K_\textrm{s}+s_\textrm{eq})^2} ,\\
    x_\textrm{eq} &=\frac{(v_\textrm{min}+a s_\textrm{eq})^\rho}{\alpha}(s_\textrm{f}-s_\textrm{eq}), \\
    s_\textrm{eq} &= \frac{\alpha D K_\textrm{s}}{\mu_\textrm{max} - \alpha D}.
\end{align}
The values of $\alpha$ and dilution rate $D$ that yield stable steady state and  oscillatory operations are separated by a smooth curve (Fig.\ \ref{fig:stabcrit}). The boundary between steady-state and oscillatory behavior shifts to higher values of $\alpha$ and $D$ for higher feed substrate concentration  $s_\textrm{f}$.

\section{Case Study 2: Continuous Ethanol Fermenter}

The fermentation of a carbohydrate source to produce ethanol is of interest as bioethanol is used as a sustainable chemical feedstock and fuel source \cite{aditiya_second_2016}. The large production rates have motivated the development of continuous fermentation processes to improve volumetric productivity; however, continuous operation can have complex nonlinear dynamics such as steady-state multiplicity and oscillations in process variables \cite{paz_astudillo_importance_2011}. To better understand the complex dynamics in continuous ethanol fermenters, a range of process models have been proposed and analyzed (see \cite{paz_astudillo_importance_2011} and citations therein). 

This case study considers the open- and closed-loop dynamics of an experimentally validated continuous ethanol fermenter model developed by \cite{mclellan_incidence_1999,daugulis_experimental_1997}. The first two subsections outline the model and open-loop dynamics. The last subsection describes the implementation of an output feedback controller that is able to eliminate oscillations while maintaining a productivity comparable to or higher than the open-loop system. 

\subsection{Model Description}

This case study considers the model developed in \cite{mclellan_incidence_1999,daugulis_experimental_1997} for ethanol production in a continuous stirred tank bioreactor with \textit{Zymomonas mobilis}. The model has five states and tracks three measurable states of the bioreactor: Biomass, $X$, substrate, $S$, and product, $P$. The two other states, $W$ and $Z$, describe the time history of the rate of change of ethanol in the bioreactor. The model equations are
\begin{align}
    \frac{\textrm{d}X}{\textrm{d}t} &= (\mu(S,P,Z) - D) X,\label{eq:first-ethanol} \\
    \frac{\textrm{d}S}{\textrm{d}t} &= \frac{1}{Y_\textrm{PS}}Q_{P}X + D(S_\textrm{f} - S), \\
    \frac{\textrm{d}P}{\textrm{d}t} &= Q_\textrm{d}{P}X - DP,  \\
    \frac{\textrm{d}Z}{\textrm{d}t} &= \beta (W-Z), \\
    \frac{\textrm{d}W}{\textrm{d}t} &= \beta \!\left(\! \frac{\textrm{d}P}{\textrm{d}t} - W \!\right)\!,
    \label{eq:ethanol}
\end{align}
with the auxiliary variable definitions
\begin{align}
    &\mu(S,P,Z) = \frac{1}{2} (1 - \tanh\!{(\lambda Z - \delta)} ) \frac{\mu_{\max}S\! \left( 1 - \left(\frac{P}{P_{\text{ma}}} \right)^{\!a} \right) ( 1 - \tilde{P}^{\,b} )}{K_S + S + \frac{S\max\{0, S - S_i\}}{K_i - S_i}},  \\
    &Q_p = Q_\textrm{P,max} \dfrac{S}{K_{\text{mp}} + S}\! \left( 1 - \left( \frac{P}{P_{\text{me}}}\right)^{\!\!\alpha}\, \right),
\end{align}
where 
\begin{equation}
    \tilde{P} = \begin{cases}
    0 & \text{for} \ P \leq P_{\text{ob}}, \\
\frac{P - P_\text{ob}}{P_\text{mb} - P_\text{ob}} & \text{for} \ P_\text{ob} < P < P_\text{mb}, \\
    1 & \text{for} \ P \geq P_{\text{mb}}.
    \end{cases} \label{eq:last-ethanol}
\end{equation}
A description of the model variables and parameters can be found in Table~\ref{tbl:ethanol_params}. Additional details on the model equations and parameters are available in the original publications \cite{veeramallu_structured_1990,lee_fermentation_1983}.

\begin{table}[htbp]
\centering
\caption{Description of the variables and parameters in the ethanol bioreactor model \eqref{eq:first-ethanol}--\eqref{eq:last-ethanol}.}\renewcommand{\arraystretch}{1.4}
    \begin{tabular}{ccccc}
    \toprule
     \makecell[c]{Variable/ \\ Parameter}& Description & Nominal Value & Initial Condition & Units\\
     \hline 
     $X$ & Biomass concentration & -- & 10 & $\frac{\text{g}}{\text{L}}$ \\
     $S$ & Substrate concentration & -- & 100 & $\frac{\text{g}}{\text{L}}$ \\
     $P$ & Product concentration & -- & 0 & $\frac{\text{g}}{\text{L}}$ \\
     $W$ & \makecell[c]{First-order weighted average of the \\ ethanol concentration change rate} & -- & 0 & $\frac{\text{g}}{\text{L hr}}$ \\
     $Z$ & \makecell[c]{Second-order weighted average of the \\ ethanol concentration change rate} & -- & 0 & $\frac{\text{g}}{\text{L hr}}$ \\
     $D$ & Dilution rate & 0.06 & -- & $\frac{1}{\text{hr}}$ \\
     $S_\textrm{f}$ &  Feed substrate concentration & 200 & -- & $\frac{\text{g}}{\text{L}}$ \\
     $Y_\textrm{PS}$ & Product yield on substrate & 0.495 & -- & $\frac{\text{g}}{\text{g}}$ \\
     $\beta$ & Fitted parameter & 0.0366 & -- & $\frac{1}{\text{hr}}$ \\
     $\lambda$ & Fitted parameter & 21.05 & -- & $\frac{\text{L hr}}{\text{g}}$ \\
     $\delta$ & Fitted parameter & 0.8241 & -- & -- \\
     $\alpha$ & Fitted parameter & 8.77 & -- & -- \\
     $a$ & Fitted parameter & 0.3142 & -- & --  \\
     $b$ & Fitted parameter & 1.415 & -- & -- \\
     $\mu_{\max}$ & Maximum specific growth rate & 0.41 & -- & $\frac{1}{\text{hr}}$ \\
     $Q_{P,\max}$ & Maximum specific ethanol production rate & 2.613 & -- & $\frac{\text{g}}{\text{g hr}}$ \\
     $K_\textrm{S}$ & Substrate inhibition parameter & 0.5 & -- & $\frac{\text{g}}{\text{L}}$ \\
     $K_i$ & Substrate inhibition parameter & 200 & -- & $\frac{\text{g}}{\text{L}}$ \\
     $K_{\text{mp}}$ & Substrate saturation parameter & 0.5 & -- & $\frac{\text{g}}{\text{L}}$ \\
     $S_i$ & \makecell[c]{Lower threshold substrate \\ concentration for substrate inhibition} & 80 & -- &  $\frac{\text{g}}{\text{L}}$ \\
     $P_{\text{ma}}$ & \makecell[c]{Threshold product concentration \\ where growth is still possible} & 217 & -- & $\frac{\text{g}}{\text{L}}$ \\ 
     $P_{\text{ob}}$ & \makecell[c]{Lower threshold product concentration for \\ the product inhibition term with exponent $b$} & 50 & -- & $\frac{\text{g}}{\text{L}}$ \\ 
     $P_{\text{mb}}$ & \makecell[c]{Lower maximum product concentration for \\ the product inhibition term with exponent $b$} & 108 & -- & $\frac{\text{g}}{\text{L}}$ \\
     $P_{\text{me}}$ & \makecell[c]{Upper threshold product concentration \\ for ethanol production} & 127 & -- &  $\frac{\text{g}}{\text{L}}$ \\
     \bottomrule
    \end{tabular}
    \renewcommand{\arraystretch}{1}
    \label{tbl:ethanol_params}
\end{table}

\subsection{Open-loop Dynamics}
\begin{figure}[htbp]
    \centering
    \includegraphics[width= 0.9\linewidth]{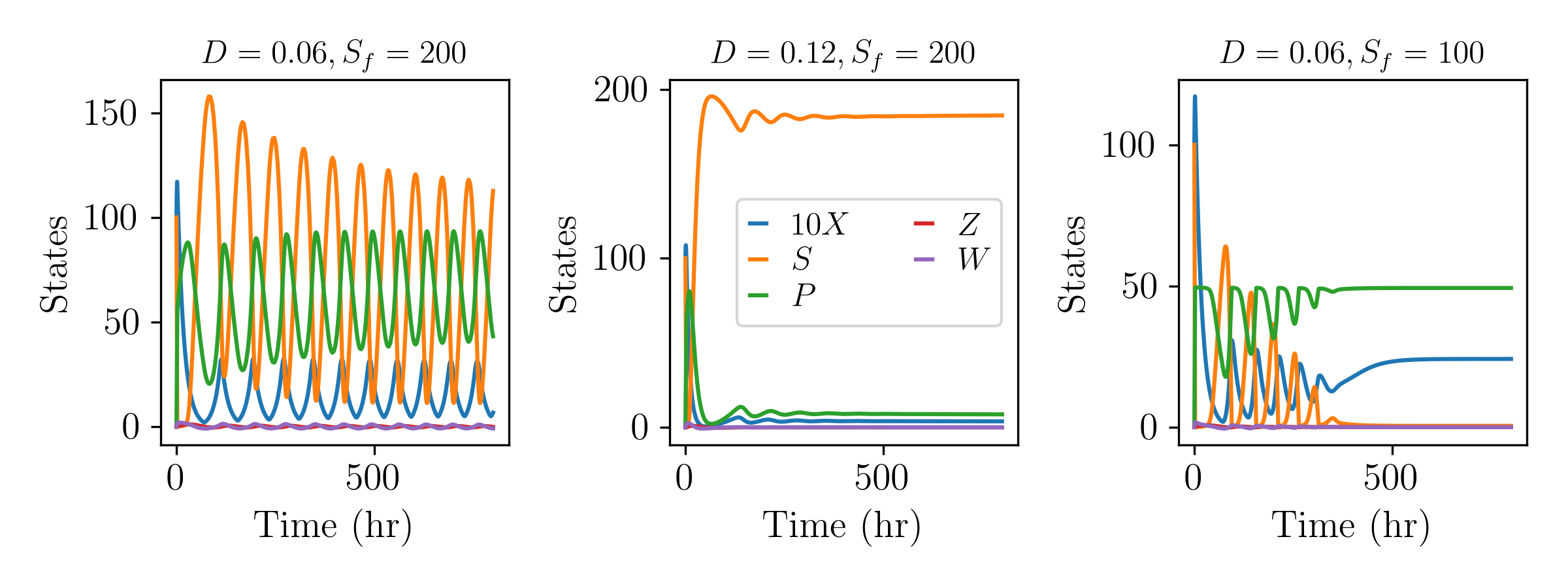}
    \caption{Open-loop dynamics of the continuous ethanol bioreactor \eqref{eq:first-ethanol}--\eqref{eq:last-ethanol} for several values of dilution rate $D$ and feed substrate concentration  $S_\textrm{f}$. Operating the bioreactor at the nominal values of $D$ and $S_\textrm{f}$ results in oscillations. Increasing $D$ and/or decreasing $S_\textrm{f}$ suppresses the oscillations, though at the expense of lower productivity.}
\label{fig:ethanol_openloop}
\end{figure}

Oscillations occur for some 
values for the dilution rate $D$ and feed substrate concentration $S_\textrm{f}$ (Fig.~\ref{fig:ethanol_openloop}), including for the  nominal conditions in Table~\ref{tbl:ethanol_params}. The oscillations arise from the states $W$ and $Z$ which depend on the rate of change of the ethanol product concentration, to account for the delayed product inhibition believed to be the cause of the oscillations \cite{mclellan_incidence_1999,daugulis_experimental_1997}. Qualitatively, when the ethanol concentration is increasing, there is a delayed effect which inhibits biomass growth at a later time. Consequently, as biomass growth slows and the ethanol concentration decreases, the reverse effect occurs and the cells eventually produce more ethanol---overall resulting in the oscillatory behavior.  

The oscillations are suppressed by increasing the dilution rate and/or decreasing the substrate feed concentration (Fig.~\ref{fig:ethanol_openloop}). At higher dilution rates, the bioreactor approaches the washout point, where the substrate concentration $S$ is equal to the feed substrate concentration $S_{\text{f}}$ while all other states are zero. 
When the feed substrate concentration is decreased, the rate of product concentration changes are less severe, which results in less-stressed biomass,  attenuating the cause of the oscillatory behavior \cite{mclellan_incidence_1999}. However, eliminating the oscillations by increasing the dilution rate $D$ and/or reducing the feed substrate concentration $S_\textrm{f}$ is undesirable as the production rate and/or the process efficiency, which can be quantified by the substrate conversion, are adversely impacted.

\subsection{Output Feedback Control}

A proportional feedback control law of the form,
\begin{equation}
    u(t) = K_{c}x + u_{0},
    \label{eq:feedback_controllaw}
\end{equation}
is introduced where $u$ is the manipulated variable which corresponds to either the dilution rate $D$ or the substrate feed concentration $S_\textrm{f}$, $K_{c}$ is the controller gain, $x$ is the measured variable which is one of the system states, and $u_0$ is the controller bias. Of the five system states in \eqref{eq:first-ethanol}--\eqref{eq:last-ethanol}, only the substrate, biomass, and product concentrations ($S$, $X$, and $P$ respectively) are readily measurable with inline sensors and therefore are considered for use as the measured variable. The values of the manipulated variables are bounded with the ranges $D\in [0,0.06]$ 1/hr and $S_\textrm{f} \in [0, 200]$ g/L.

\begin{figure}[htb]
    \centering
    \includegraphics[width=0.8\textwidth]{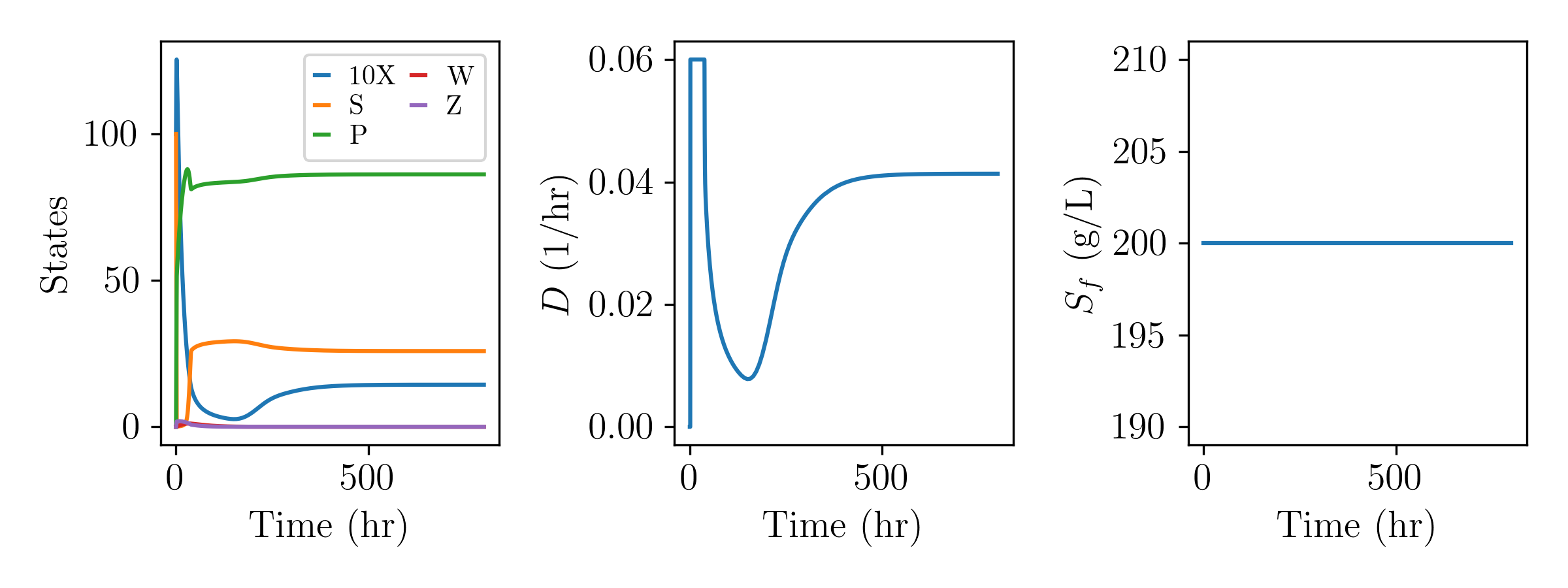} 
    \vspace{-0.2cm}
    \caption{Closed-loop dynamics for the ethanol bioreactor \eqref{eq:first-ethanol}--\eqref{eq:last-ethanol} with a single-loop proportional feedback control law using the substrate concentration $S$ and the dilution rate $D$ as the measured and manipulated variables. The controller gain and bias are $K_c= -0.01$ and $u_0 =0.3$.}
    \label{fig:ethanol_feedback}
\end{figure}

Closed-loop simulations for a wide range of values of the controller gain $K_{c}$ and bias $u_{0}$ for the different combinations of measured and manipulated variables are given in Section S1 of the Supplementary Information. To understand the controller performance, it is useful to first consider the ways in which poor controller performance can manifest in the closed-loop system: 
\begin{enumerate}
    \item Comparable performance to the open-loop system, where just reducing the feed substrate concentration $S_\textrm{f}$ can help suppress oscillations. 
    \item Sustained closed-loop oscillations.
    \item Slow closed-loop response, that is, the stabilization of the bioreactor takes a prohibitively long time.
    \item Controller-induced instability for high values of the controller gain $K_c$. 
    \item Driving the system to an unproductive state e.g., low/no dilution rate or low feed substrate concentration.
\end{enumerate}
Using the dilution rate $D$ as the manipulated variable enabled stabilization of the system for all choices of measured variables, whereas using the feed substrate concentration $S_\textrm{f}$ as the manipulated variable in most cases resulted in poor closed-loop performance in one or more of the modes described above (see Section S1 in the Supplementary Information). Of the different combinations of measured and manipulated variables, using the dilution rate $D$ with the substrate concentration $S$ as the measured variable is the best performing, with the fastest stabilization of the system for a large range of control parameters, yielding a non-oscillatory system that is able to maintain a high product concentration for dilution rate $D \sim 0.04$ 1/hr (see Fig.~\ref{fig:ethanol_feedback}). In comparison, the other measured and manipulated combinations resulted in poor closed-loop performance (manifesting in one or more of the modalities described above) or was able to suppress oscillations but had the system dynamics very sensitive to the values of the control parameters, indicating poor robustness of the closed-loop system.\footnote{Given that sensors always have some bias in practice.}

\section{Case Study 3: Two-Stage Continuous Influenza Viral Bioreactor}
\label{sec:frensing}

This case study considers the open- and closed-loop dynamics of a two-stage continuous, stirred tank bioreactor used for the production of influenza viral particles. The first three subsections consider the experimentally validated model developed by \cite{frensing_continuous_2013} whereas the last subsection provides simulation results for an alternative bioreactor configuration that eliminates the source of oscillations without using feedback control.

\subsection{Model Description}

The setup consists of two stages: a first-stage cell bioreactor which expands a cell culture of uninfected target cells which are fed into the second-stage viral bioreactor where viral replication takes place. The species balances in the viral bioreactor are \cite{canova_mechanistic_2023,frensing_continuous_2013}
\begin{align}
    \frac{\textrm{d}T}{\textrm{d}t} &= \mu T - k_{1}(V_\textrm{s} + V_\textrm{d})T + D(T_\textrm{in} - T), \label{eq:frensing-1} \\
    \frac{\textrm{d} I_\textrm{d}}{\textrm{d}t} &= k_{1}V_\textrm{d}T - (k_{1}V_\textrm{s} - \mu)I_\textrm{d} - DI_\textrm{d},  \\
    \frac{\textrm{d} I_\textrm{s}}{\textrm{d}t} &= k_{1}V_\textrm{s}T - (k_{1}V_\textrm{d} + k_{2})I_\textrm{s} -DI_\textrm{s},  \\
\frac{\textrm{d}I_\textrm{c}}{\textrm{d}t} &= k_{1}(V_\textrm{s}I_\textrm{d} + V_\textrm{d}I_\textrm{s}) -k_{2}I_\textrm{c} -DI_\textrm{c}, \\
\frac{\textrm{d}V_\textrm{s}}{\textrm{d}t} &= k_{3}I_\textrm{s} - (k_{1}(T + I_\textrm{d} + I_\textrm{s} + I_\textrm{c}) + k_{4} + D)V_\textrm{s},\\
   \frac{\textrm{d}V_{d}}{\textrm{d}t} &= k_{33}I_\textrm{c} + fk_{3}I_\textrm{s} -(k_{1}(T+I_\textrm{d}+I_\textrm{s}+I_\textrm{c}) + k_{4} + D)V_\textrm{d},
    \label{eq:frensing}
\end{align}
which account for the presence of defective interfering particles (DIPs). Descriptions of the variables and parameters in the model can be found in Table~\ref{tbl:frensing_params}.

\begin{table}[htbp]
\centering
\caption{Description of variables and parameters in the continuous influenza bioreactor model \eqref{eq:frensing-1}--\eqref{eq:frensing}.}\renewcommand{\arraystretch}{1.4}
    \begin{tabular}{ccccc}
    \toprule
     \makecell[c]{Variable/ \\ Parameter}& Description & Nominal Value & Initial Condition & Units\\
     \hline 
     $T$ & Concentration of uninfected target cells & -- & $5$$\times$$10^{6}$ & $\frac{\text{cells}}{\text{mL}}$ \\
     $I_\textrm{d}$ & Concentration of cells infected with DIPs & -- & 0 & $\frac{\text{cells}}{\text{mL}}$ \\
     $I_\textrm{s}$ & Concentration of cells infected with STVs & -- & 0 & $\frac{\text{cells}}{\text{mL}}$ \\
     $I_\textrm{c}$ & Concentration of co-infected cells & -- & 0 & $\frac{\text{cells}}{\text{mL}}$ \\
     $V_\textrm{s}$ & Concentration of STVs & -- & $1.25$$\times$$10^5$ & $\frac{\text{virions}}{\text{mL}}$ \\
     $V_\textrm{d}$ & Concentration of DIPs & -- & 0 & $\frac{\text{virions}}{\text{mL}}$ \\
     $T_\textrm{in}$ & Cell concentration of the feed & $3$$\times$$10^6$ & -- & $\frac{\text{cells}}{\text{mL}}$ \\
     $D$ & Dilution rate of virus reactor & 0.0396 & -- & $\frac{1}{\text{hr}}$ \\
     $\mu$ & Cell growth rate constant & 0.027 & -- & $\frac{1}{\text{hr}}$ \\
     $k_{1}$ & STV and DIP infection rate constant & $2.12$$\times$$10^{-9}$ & -- & $\frac{\text{mL}}{\text{virion hr}}$  \\
     $k_{2}$ & Virus-induced cell apoptosis rate constant & $7.13$$\times$$10^{-3}$ & -- & $\frac{1}{\text{hr}}$ \\
     $k_{3}$ & STV production rate constant & 168 & -- & $\frac{\text{virions}}{\text{cell hr}}$ \\
     $k_{33}$ & DIP production rate constant & 168 & -- & $\frac{\text{virions}}{\text{cell hr}}$ \\
     $k_{4}$ & Virus degradation rate constant & 0.035 & -- & $\frac{1}{\text{hr}}$ \\
     $f$ & Fraction of STV-infected cells that produce DIPs & $10^{-3}$ & --  & --\\
     \bottomrule
    \end{tabular}
    \renewcommand{\arraystretch}{1}
    \label{tbl:frensing_params}
\end{table} 

\subsection{Open-loop Dynamics}

The viral particles that infect target cells in viral bioreactors follow predator-prey dynamics, which are well-known to often have oscillatory behavior \cite{may_limit_1972}. Even the simplified model obtained by removing the dynamics induced by DIPs in \eqref{eq:frensing-1}--\eqref{eq:frensing} has a Hopf bifurcation \cite{guckenheimer_nonlinear_1983} which indicates oscillation under some conditions, albeit at nonphysical values of the parameters and inputs \cite{frensing_continuous_2013}. 

\begin{figure}[htb]
    \centering
    \begin{subfigure}{0.3\textwidth}
        \centering
        \includegraphics[width=\textwidth]{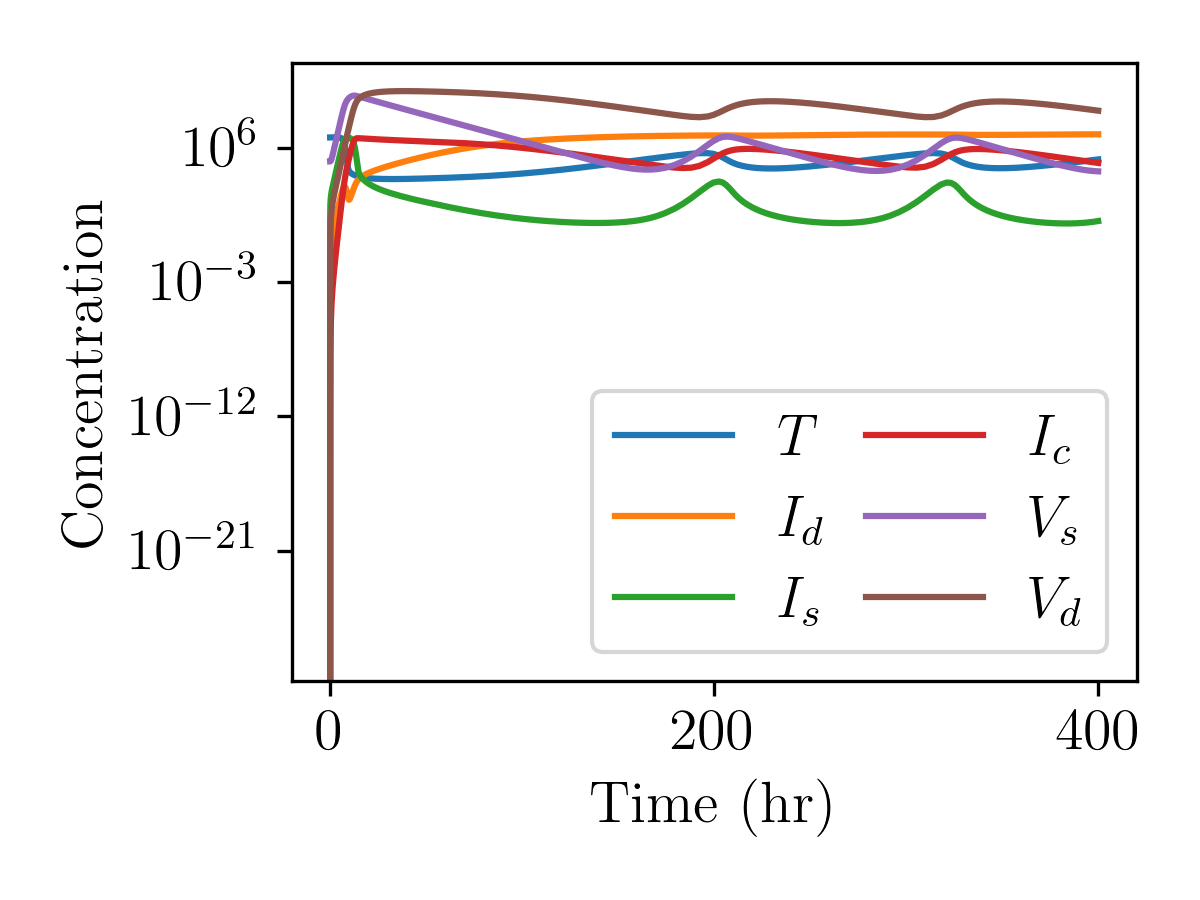} 
        \vspace{-0.6cm}
        \caption{$D=0.0396$}
    \end{subfigure}
    \hfill
    \centering
    \begin{subfigure}{0.3\textwidth}
        \centering
        \includegraphics[width=\textwidth]{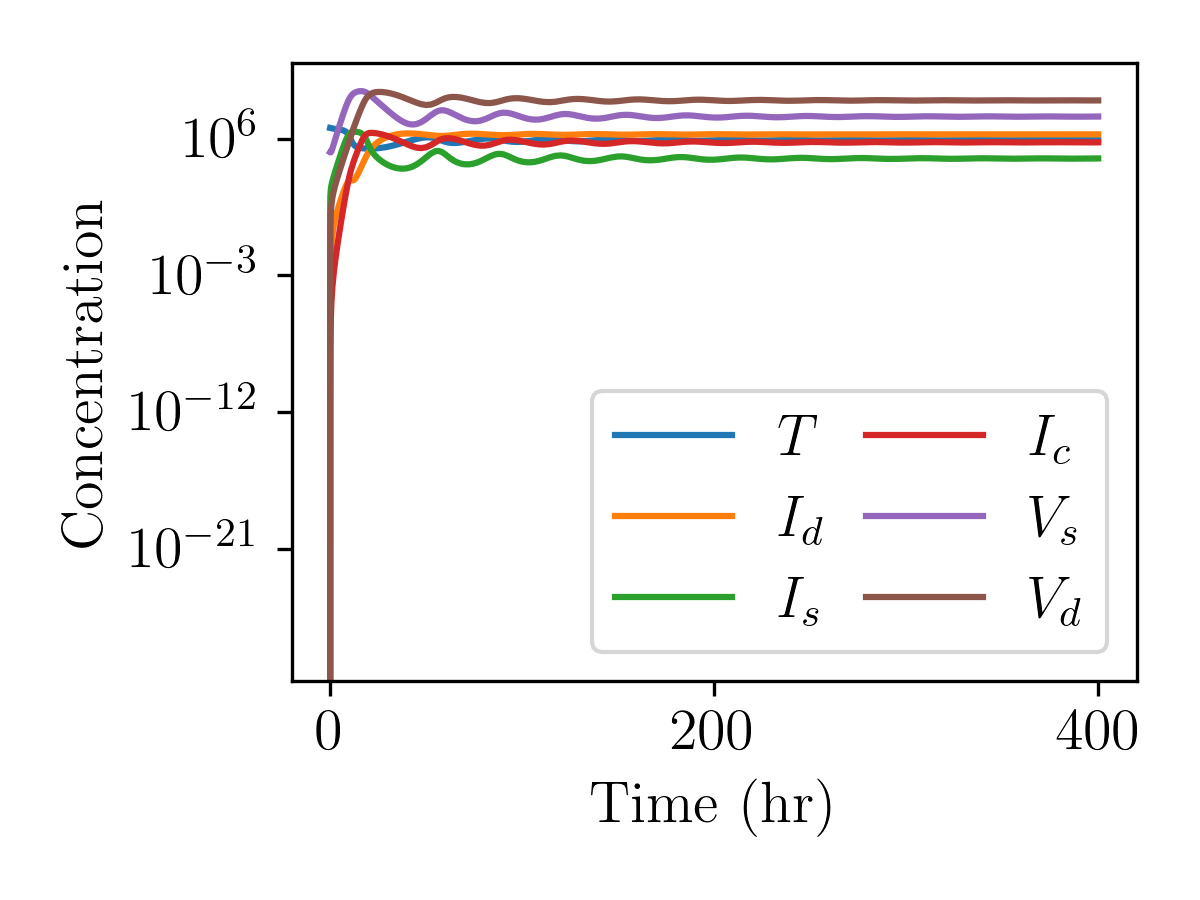} 
        \vspace{-0.6cm}
        \caption{$D=0.25$}
    \end{subfigure}
    \hfill
    \centering
    \begin{subfigure}{0.3\textwidth}
        \centering
        \includegraphics[width=\textwidth]{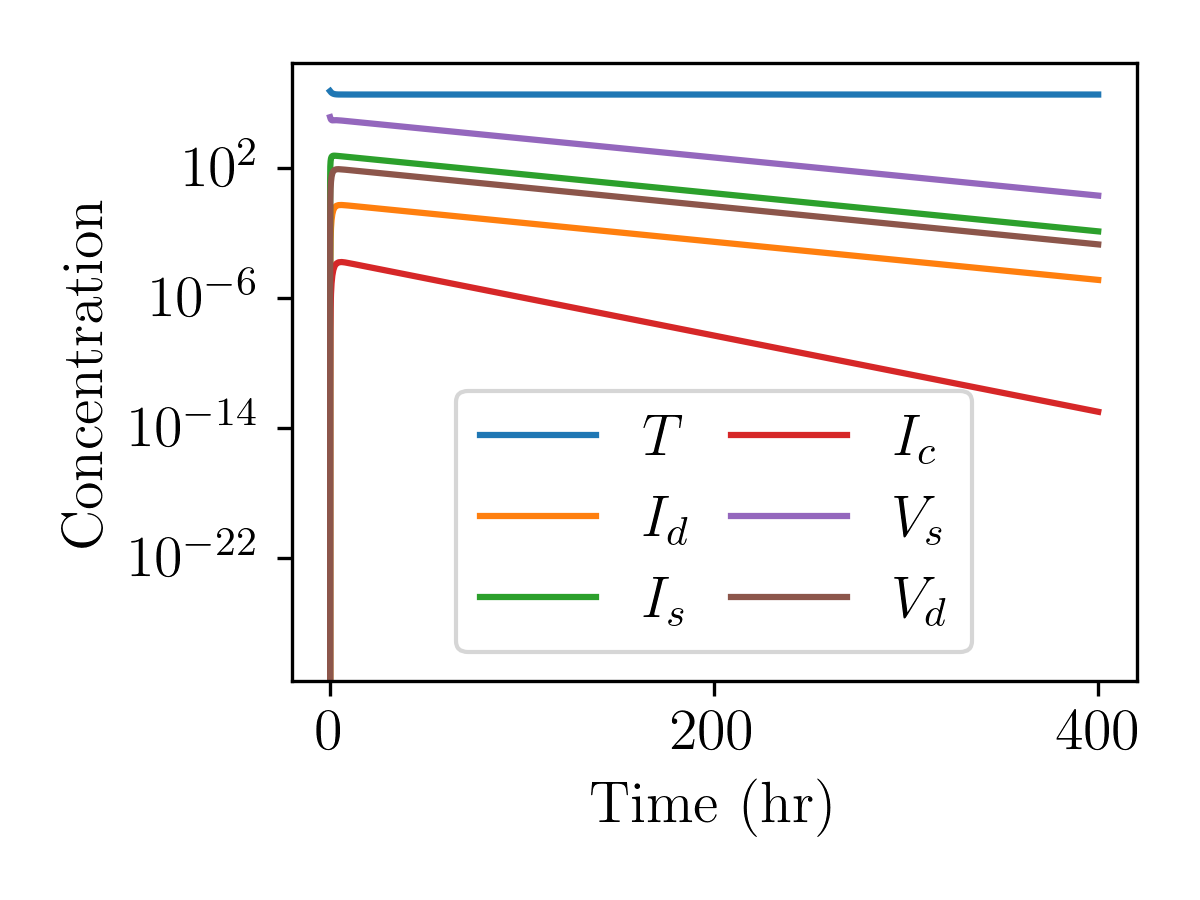}
        \vspace{-0.6cm}
        \caption{$D=1.05$}
    \end{subfigure}
    \caption{Open-loop dynamics of the continuous influenza viral bioreactor \eqref{eq:frensing-1}--\eqref{eq:frensing} with the nominal parameters and input values for three values for the dilution rate $D$. As $D$ increases, the oscillations are suppressed; when $D$ becomes too large e.g., $\gtrsim 1.05$ 1/hr, washout occurs.}
    \label{fig:viral_openloop}
\end{figure}

For influenza viral production, DIPs are significant and contribute to oscillatory behavior which has been observed experimentally, consistent with the dynamics of the bioreactor model \cite{frensing_continuous_2013} as can be seen in Fig.~\ref{fig:viral_openloop}. DIPs interfere with standard virus particle (STV) replication as cells co-infected with DIPs and STVs, $I_\textrm{c}$, produce mostly DIPs instead of STVs \cite{frensing_defective_2015}. 
The source of the oscillatory behavior can be understood through coupled interactions between the species. As coinfected cells mostly produce DIPs instead of STVs, DIPs accumulate within the system while STV concentrations drop. Consequently, as DIP production is impacted (since there are fewer STV present) and with dilution, the total viral particle concentration decreases and the bioreactor is able to operate at a lower multiplicity of infection (MOI) which enables STV concentration to increase again. Operating at a higher dilution rates (in the absence of a control system) can help stabilize the system, as the bioreactor is operating at a lower MOI (Fig.~\ref{fig:viral_openloop}b).

\subsection{Output Feedback Control}

\begin{figure}[htb]
    \centering
    \begin{subfigure}{0.8\textwidth}
        \centering
        \includegraphics[width=\textwidth]{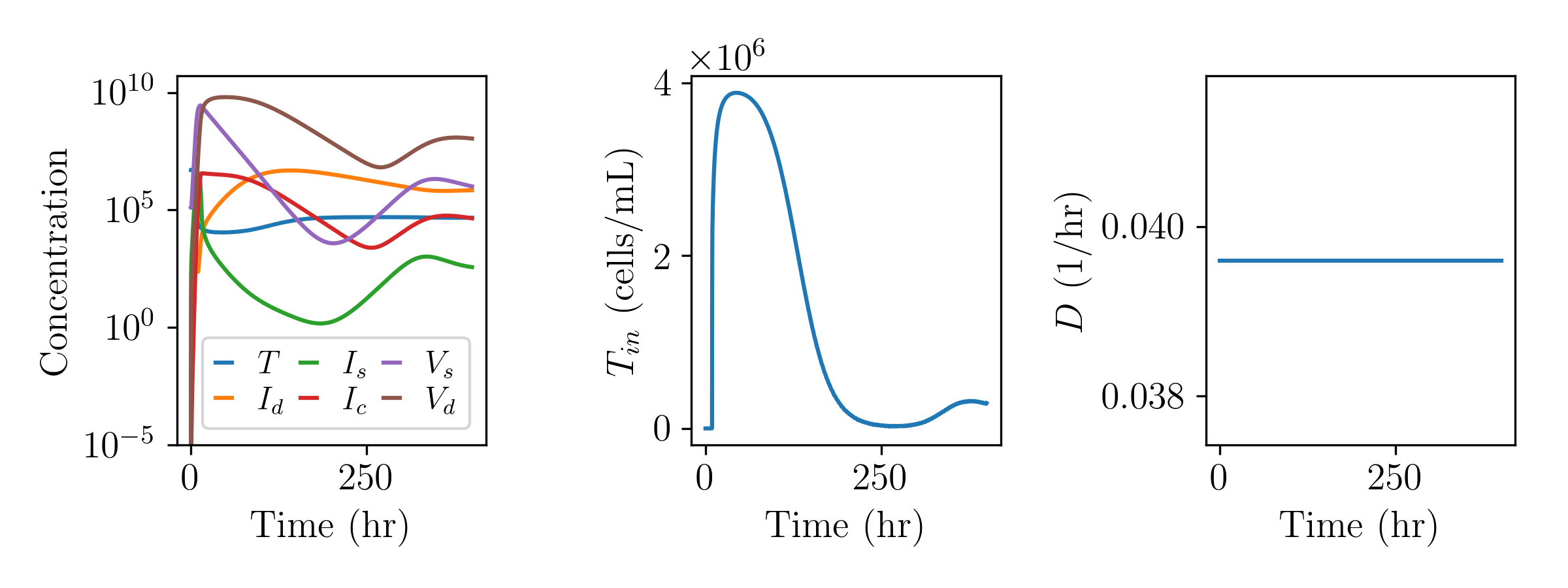} 
        \vspace{-0.8cm} 
        \caption{Manipulated variable $T_\textrm{in}$ and control parameters $K_c= -100$ and $u_0 = 5$$\times$$10^6$}
    \end{subfigure}
    \hfill
    \centering
    \begin{subfigure}{0.8\textwidth}
        \centering
        \includegraphics[width=\textwidth]{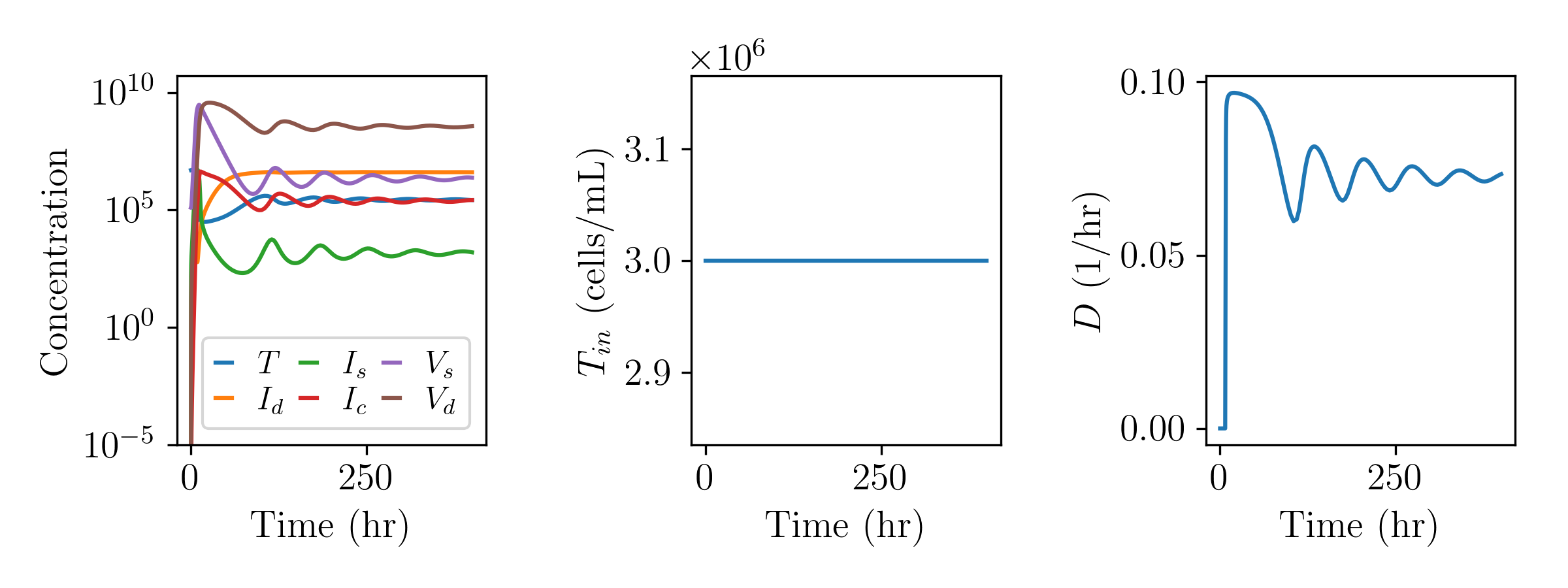} 
        \vspace{-0.8cm}
        \caption{Manipulated variable $D$ and control parameters $K_c= -1$$\times$$ 10^{-7}$ and $u_0 = 0.1$}
    \end{subfigure}
    \caption{Closed-loop dynamics for the continuous influenza viral bioreactor \eqref{eq:frensing-1}--\eqref{eq:frensing} with the target cell concentration $T$ as the measured variable for two choices of the manipulated variable. Both the feed cell concentration $T_\textrm{in}$ and dilution rate $D$ can be used as the manipulated variable to suppress the oscillations.}
    \label{fig:viral_feedback}
\end{figure}

Consider application of the single-loop proportional control law \eqref{eq:feedback_controllaw} and the dilution rate $D$ or the fee cell concentration $T_\textrm{in}$ as options for the manipulated variable. Of the six system states, the uninfected target cell concentration $T$ is the most easily measurable state with inline measurements such as optical sensors \cite{pais_holographic_2020}, which motivates its choice as the measured variable. The controller was implemented so as to bound the manipulated variables by $T_\textrm{in} \in [0, 1$$\times$$10^{7}]$ cells/mL and $D \in [0, 0.1]$ 1/hr. For a given combination of measured and manipulated variable, the other potential manipulated variable is set to the nominal value given in Table~\ref{tbl:frensing_params}. The closed-loop dynamics can be stabilized using either the dilution rate $D$ and feed cell concentration $T_\textrm{in}$ as the manipulated variable (Fig.~\ref{fig:viral_feedback}). 

Assuming that DIPs do not adversely impact product quality (e.g., for use in vaccines), the performance of the system and controller can be quantified by considering the total amount of virus particles produced over the entire run, given by
\begin{equation}
    J = \int_0^{t_\textrm{f}} \! D(V_\textrm{s} + V_\textrm{d}) \textrm{d}t,
    \label{eq:viral_obj}
\end{equation}
where $t_\textrm{f}=400$ hr is the final time of the simulation. For the ranges of controller and system parameter values, controlling the target cell concentration $T$ using the feed cell concentration $T_\textrm{in}$ is able to yield a higher productivity, especially at higher values of the controller bias $u_0$ which results in a higher operating cell density and consequently a higher productivity. Control of the bioreactor using the dilution rate $D$ as the manipulated variable introduces another layer of complexity when considering continuous operation. Manipulating the flowrate into and out of the bioreactor would result in fluctuating flowrates for downstream operations, potentially necessitating surge tanks for smooth continuous operation.

\begin{figure}[htb]
    \centering
    \begin{subfigure}{0.45\textwidth}
        \centering
        \includegraphics[width=\textwidth]{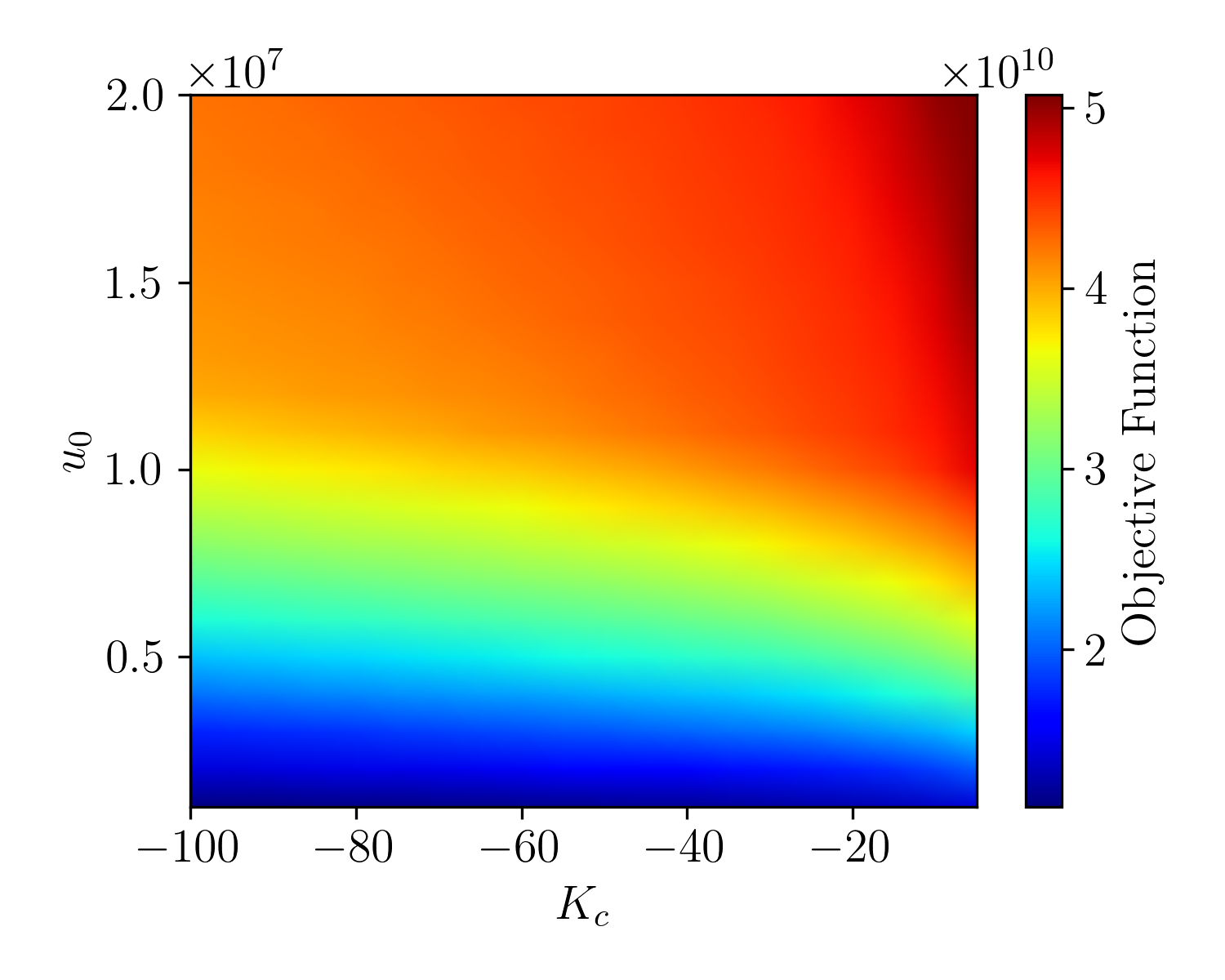} 
        \vspace{-0.8cm}
        \caption{Manipulated variable $T_\textrm{in}$}
    \end{subfigure}
    \hfill
    \centering
    \begin{subfigure}{0.45\textwidth}
        \centering
        \includegraphics[width=\textwidth]{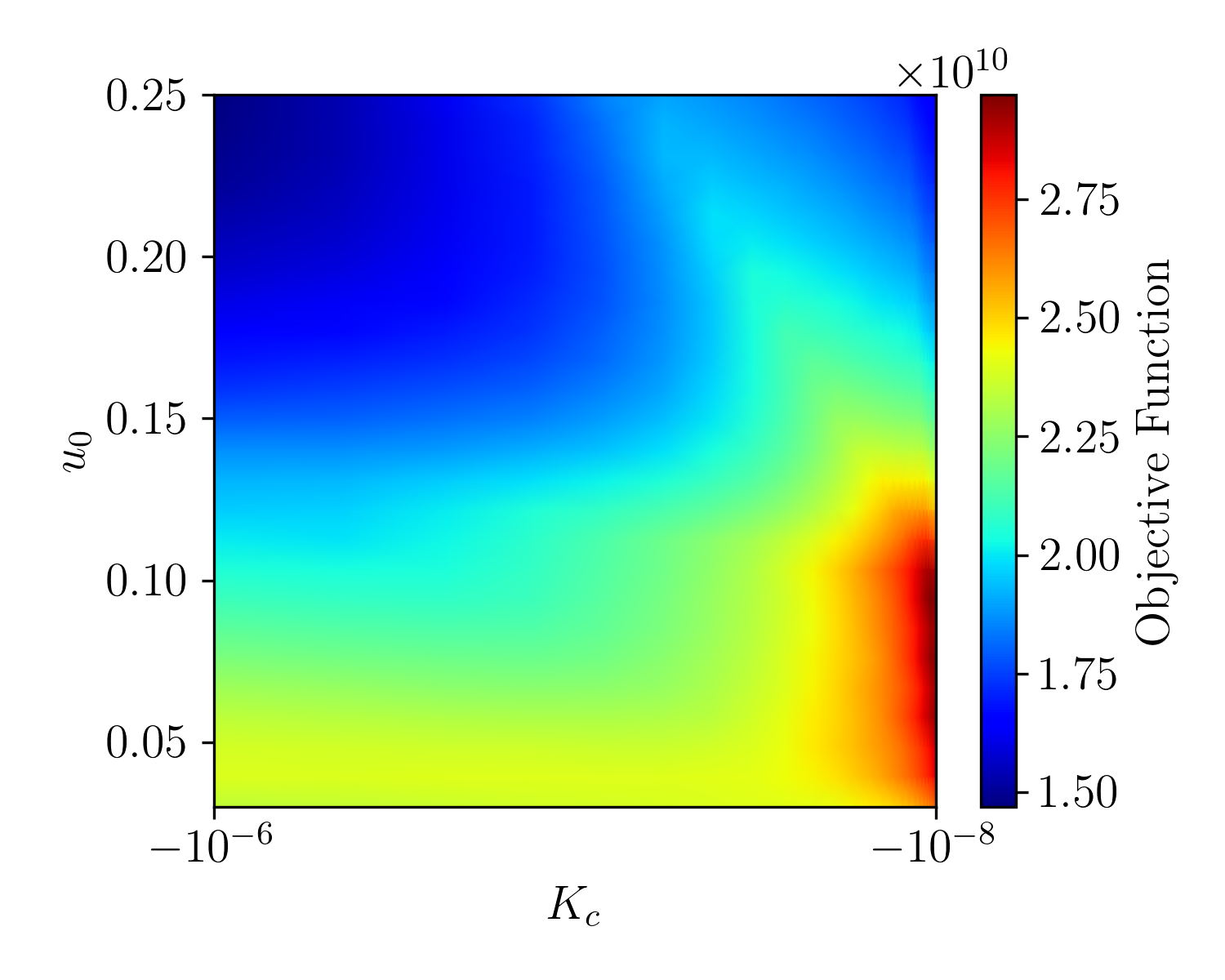} 
        \vspace{-0.8cm}
        \caption{Manipulated variable $D$}
    \end{subfigure}
    \caption{Objective function \eqref{eq:viral_obj} values for the continuous influenza viral bioreactor \eqref{eq:frensing-1}--\eqref{eq:frensing} for two choices of manipulated variable and for a wide range of controller parameters ($K_c$ and $u_0$). The measured variable is the target cell concentration.}
    \label{fig:viral_feedback_opt}
\end{figure}

The selection of the controller parameters involve a tradeoff between productivity \eqref{eq:viral_obj} and suppressing oscillatory behavior. As seen in Fig.~\ref{fig:viral_feedback_opt}, there are controller parameter values that maximize productivity: less negative controller gain $K_{c}$ for both control-loop pairings, and larger controller bias $u_{0}$ for the $(T,T_{\textrm{in}})$ pairing and lower controller bias for the $(T,D)$ pairing. However, less negative controller gains result in slow/no suppression of oscillatory behavior (see Figs.\ S7--S8 in Section S2 of the Supplementary Information for exemplar simulations).

\subsection{Alternative Process Designs}

Alternative process configurations can be used to avoid oscillatory behavior in a viral bioreactor. Instead of using two-stage stirred-tank reactors as considered above, a tubular flow bioreactor can be used for the viral bioreactor stage which has been demonstrated to eliminate oscillations by avoiding the accumulation of DIPs \cite{tapia_continuous_2019}. A model for a tubular flow bioreactor can be constructed assuming the same mechanistic expressions for the kinetic phenomena as used in \eqref{eq:frensing-1}--\eqref{eq:frensing} by introducing the residence time $\tau$ as an intrinsic variable \cite{inguva_efficient_2022}, converting the model into a system of partial differential equations (PDEs),
\begin{align}
    \frac{\partial T}{\partial t} + \frac{\partial T}{\partial \tau} &= \mu T - k_{1}(V_\textrm{s} + V_\textrm{d})T , \label{eq:frensing_pfr-1}\\
    \frac{\partial I_\textrm{d}}{\partial t} + \frac{\partial I_\textrm{d}}{\partial \tau} &= k_{1}V_\textrm{d}T - (k_{1}V_\textrm{s} - \mu)I_\textrm{d},  \\
    \frac{\partial I_\textrm{s}}{\partial t} + \frac{\partial I_\textrm{s}}{\partial \tau} &= k_{1}V_\textrm{s}T - (k_{1}V_\textrm{d} + k_{2})I_\textrm{s} ,  \\
    \frac{\partial I_\textrm{c}}{\partial t} + \frac{\partial I_\textrm{c}}{\partial \tau} &= k_{1}(V_\textrm{s}I_\textrm{d} + V_\textrm{d}I_\textrm{s}) -k_{2}I_\textrm{c}, \\
    \frac{\partial V_\textrm{s}}{\partial t} + \frac{\partial V_\textrm{s}}{\partial \tau} &= k_{3}I_\textrm{s} - (k_{1}(T + I_\textrm{d} + I_\textrm{s} + I_\textrm{c}) + k_{4} )V_\textrm{s},  \\
    \frac{\partial V_\textrm{d}}{\partial t} + \frac{\partial V_\textrm{d}}{\partial \tau} &= k_{33}I_\textrm{c} + fk_{3}I_\textrm{s} -(k_{1}(T+I_\textrm{d}+I_\textrm{s}+I_\textrm{c}) + k_{4})V_\textrm{d}.
    \label{eq:frensing_pfr}
\end{align} 
This model was simulated using the FiPy PDE solver  \cite{guyer_fipy_2009,inguva_introducing_2021}. Dirichlet boundary conditions corresponding to the initial conditions  were applied at $\tau = 0$ and zero-gradient outflow boundary conditions were applied at a point further downstream of the simulation domain. An initial condition of zero concentration for all species in the bioreactor was applied. 

\begin{figure}[htbp]
    \centering
    \includegraphics[width= 0.5\linewidth]{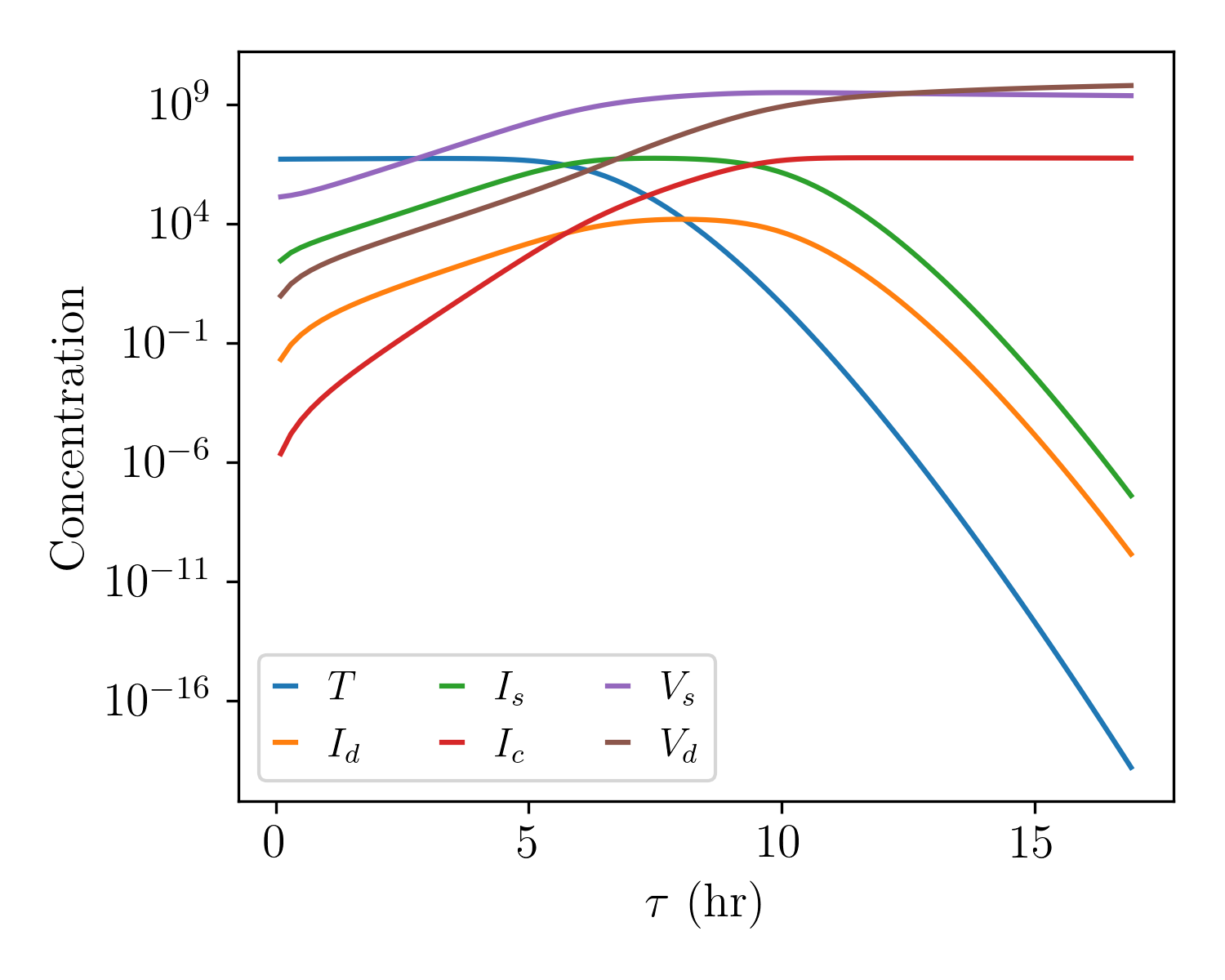} \vspace{-0.6cm}

    \caption{Spatial variation of the states in the continuous tubular flow viral bioreactor model \eqref{eq:frensing_pfr-1}--\eqref{eq:frensing_pfr} for nominal parameter and initial conditions at time $t=17$ hr.}
    \label{fig:frensing_pfr}
\end{figure}

As seen in experiments \cite{tapia_continuous_2019}, the simulations indicate that oscillatory behavior is avoided when using a tubular flow bioreactor configuration, which prevents the accumulation of DIPs at lower residence times ($\lesssim 10$ hr, see Fig.~\ref{fig:frensing_pfr}). 
As DIPs cannot accumulate at any specific location in the tubular bioreactor and target cells are only fed at the inlet, the oscillatory dynamics inherent in the stirred tank bioreactor configuration are eliminated since periodic shifts to high and low MOI conditions are not possible. Another advantage of the tubular flow bioreactor is that it enables manufacturers to obtain a high-purity STV product stream simply by specifying the residence time (i.e., length of the bioreactor for a given flowrate). 

\section{Case Study 4: Continuous Yeast Cell Bioreactor}

\textit{S. cerevisiae}, aka budding yeast, is widely used in academia and industry. In the brewing and baking sectors, the desired products come from yeast catabolism of sugars. In the biomanufacturing sector, recombinant proteins synthesized by yeast cells are the product. In the use of yeast as a model organism for the study of eukaryotic cell biology, the asexually-reproducing yeast itself is the desired product \cite{botstein_yeast_1997}. Both induced and undesired cell density and metabolite oscillations have been observed in continuous yeast culture since the 1970s \cite{kaspar_von_meyenburg_energetics_1969,kaspar_von_meyenburg_stable_1973,strassle_predictive_1989,parulekar_induction_1986}.

This case study considers the open- and closed-loop dynamics of a continuous, stirred tank bioreactor culturing budding yeast cells. Subsections \ref{yeast:description} and \ref{yeast:openloop} summarize the model and provide a comprehensive explanation for the existence of oscillations in continuous production of yeast cells. Subsection \ref{yeast:closed} describes our feedback control stabilization of oscillations and controller selection given several performance objectives. The physical intuition used in selecting the best performing controllers is provided.

\subsection{Model Description}
\label{yeast:description}
The dynamics of the mechanistic model \eqref{eq:yeast-PBM} have been experimentally validated for operations in which sustained oscillations occur in substrate concentrations and biomass for yeast cells in continuous bioreactors \cite{kurtz_control_1998, zhang_bifurcation_2001}. Consider a continuous bioreactor in which budding yeast is cultured with equal inlet and outlet flow rates and a single substrate $S$. Yeast cells consume the substrate and progress through different growth states as quantified by their wet cell mass $m$: non-fissioning daughter cell ($m<m_\textrm{t}^*$), non-fissioning mother cell with growing bud ($m_\textrm{t}^* < m < m_\textrm{t}^* + m_\textrm{a}$), and mother cell with scissioning daughter cell bud ($m>m_\textrm{t}^*$). The partitioning of mass between a budding daughter cell and a mother cell is asymmetric. An investigation of this system suggests that oscillations are associated with interactions between the transition state mass $m_\textrm{t}^*$ and critical fissioning mass $m_\textrm{d}^*$ boundary movement as substrate concentrations change and yeast cell sub-populations synchronize \cite{henson_dynamic_2003}.

The population balance model for continuous culturing of yeast cells is
\begin{align}
    \frac{\partial N(m,t)}{\partial t} &+ \frac{\partial [ k(S^\prime) N(m,t) ]}{\partial m} = \Psi(m,t,N),
    \label{eq:yeast-PBM}\\
    \Psi(t,m,N) &= 2\int_0^{m_{\textrm{max}}}\Gamma(m^\prime,S^\prime)p(m,m^\prime,m_\textrm{t}^*(S^\prime))N(m^\prime,t)dm^\prime - [D+\Gamma(m,S^\prime)]N(m,t),\\
    N(m,0) &= N_0(m) = \frac{N_{00}}{\sigma_0\sqrt{2\pi}} \exp\!\left(\frac{-(m-\mu_0)^2}{2\sigma_0^2}\right),  \\
    N(0,t) &= 0,\\ 
    \frac{\textrm{d}m_0}{\textrm{d}t} &= -Dm_0+\int_0^{m_{\textrm{max}}}\Gamma(m,S^\prime)N(m,t)\textrm{d}m, 
    \label{eq:yeast-zeroth-moment-balance} \\
    m_0(0) &= N_{00}, \\
    \frac{\textrm{d}S}{\textrm{d}t} &= D(S_\textrm{f}-S)-\frac{k(S^\prime)}{Y}m_0(t), 
    \label{eq:yeast-substrate-balance}\\
    S(0) &= S_0,  \\
    \frac{\textrm{d}S^\prime}{\textrm{d}t} &= \alpha (S-S^\prime), 
    \label{eq:yeast-substrate-filtered}\\
    S^\prime(0) &= S^\prime_0,  \\
    k(S') &= \frac{\mu_\textrm{m} S^\prime}{K_\textrm{m} + S^\prime},  
    \label{eq:yeast-growth-rate-function}\\
    \Gamma(m, S') &= 
    \begin{cases} 
    0 & \text{for } m < m_\textrm{t}^*(S^\prime)+m_\textrm{a} \\
    \gamma \operatorname{exp}(-\epsilon(m-m_\textrm{d}^*(S^\prime))^2) & \text{for } m \in [m_\textrm{t}^*(S^\prime)+m_\textrm{a}, m_\textrm{d}^*(S^\prime)] \\
    \gamma & \text{for } m \in (m_\textrm{d}^*(S^\prime),m_\textrm{max}]
    \end{cases}  \\
    p(m,m',m_\textrm{t}^*) &= 
    \begin{cases}
    A \operatorname{exp}(-\beta (m-m_\textrm{t}^*)^2)+A\operatorname{exp}(-\beta(m-m^\prime+m_\textrm{t}^*))^2 & \text{for } (m < m^\prime) \vee (m^\prime > m_\textrm{t}^*+m_\textrm{a}) \\
    0 & \textrm{otherwise}
    \end{cases}  \\
    m_\textrm{t}^*(S^\prime) &= 
    \begin{cases} 
    m_\textrm{t0}+K_\textrm{t}(S_\textrm{l}-S_\textrm{h}) &\text{for }  S^\prime < S_\textrm{l} \\
    m_\textrm{t0}+K_\textrm{t}(S^\prime-S_\textrm{h}) &\text{for }  S^\prime\in[S_\textrm{l},S_\textrm{h}] \\
    m_\textrm{t0} & \text{for } S^\prime>S_\textrm{h}
    \end{cases}  \\
    m_\textrm{d}^*(S^\prime) &= 
    \begin{cases} 
    m_\textrm{d0}+K_\textrm{d}(S_\textrm{l}-S_\textrm{h}) &\text{for }  S^\prime < S_\textrm{l} \\
    m_\textrm{d0}+K_\textrm{d}(S^\prime-S_\textrm{h}) & \text{for } S^\prime\in[S_\textrm{l},S_\textrm{h}] \\
    m_\textrm{d0} & \text{for } S^\prime>S_\textrm{h}
    \end{cases}
\end{align}
This hyperbolic partial differential equation (PDE) \eqref{eq:yeast-PBM} has source terms for the birth, death, and reactor outflow of yeast cells, typical of segregated cell, unstructured kinetic models \cite{fredrickson_statistics_1967,tsuchiya_dynamics_1966}. The yeast cell number distribution $N(m,t)$ is coupled to a substrate mass balance \eqref{eq:yeast-substrate-balance} and a filtered substrate $S^\prime$ response \eqref{eq:yeast-substrate-filtered}. The filtered substrate $S^\prime$ models a commonly observed delayed change in cell metabolism due to changes in extracellular substrate concentrations $S$. Movement in cell mass boundaries $m_\textrm{t}^*$ and $m_\textrm{d}^*$ is modeled as a linear dependence on filtered substrate. Full specification of the model requires functions for the single cell growth rate $k(S^\prime)$, the fission rate $\Gamma(m,S^\prime)$, the partition probability distribution $p(m,m^\prime,S^\prime)$, an initial yeast seed number distribution in mass $N_0(m,t)$, a boundary condition enforcing non-zero cell mass $N(0,t)$, and initial substrate concentrations $S(0)$ and $S^\prime(0)$. 
\begin{table}[htbp]
\centering
\caption{Description of variables and parameters in \eqref{eq:yeast-PBM}. Adapted from \cite{zhang_bifurcation_2001}.}\renewcommand{\arraystretch}{1.4}
    \begin{tabular}{ccccc}
    \toprule
     \makecell[c]{Variable/ \\ Parameter}& Description & Nominal Value & Initial Condition & Units\\
     \hline 
     $N_{00}$ & Yeast inoculum cell density & -- & $1$$\times$$10^{4}$ & $\frac{1}{\text{L}}$ \\
     $\mu_0$ & Yeast inoculum mean cell mass & -- & $3$$\times$$10^{-11}$ & $\text{g}$ \\
     $\sigma_0$ & Yeast inoculum cell mass standard deviation & -- & $1$$\times$$10^{-11}$ & $\text{g}$ \\
     $S$ & Substrate concentration & -- & $25$ & $\frac{\text{g}}{\text{L}}$ \\
     $S^\prime$ & Filtered substrate concentration & -- & $25$ & $\frac{\text{g}}{\text{L}}$ \\
     $D$ & Dilution rate of reactor & $0.4$ & -- & $\frac{\text{1}}{\text{hr}}$ \\
     $S_\textrm{f}$ & Feed concentration of substrate & $25$ & -- & $\frac{\text{g}}{\text{L}}$ \\
     $Y$ & Yeast mass to substrate yield coefficient constant & $0.4$ & -- & $\frac{\text{g}}{\text{g}}$ \\
     $\alpha$ & Yeast metabolism adjustment rate constant & $20$ & -- & $\frac{1}{\text{hr}}$ \\
     $\mu_\textrm{m}$ & Single cell growth rate constant & $5$$\times$$10^{-10}$ & -- & $\frac{\text{g}}{\text{hr}}$ \\
     $K_\textrm{m}$ & Saturated growth Monod constant & $2$$\times$$10^{-11}$ & -- & $\frac{\text{g}}{\text{L}}$  \\
     $\gamma$ & Maximum cell fission rate constant & $200$ & -- & $\frac{1}{\text{hr}}$ \\
     $\epsilon$ & Partition probability inverse variance constant & $5$$\times$$10^{22}$ & -- & $\frac{1}{\text{g}^2}$ \\
     $m_\textrm{a}$ & Fissioning bud mass constant & $1$$\times$$10^{-11}$ & -- & $\text{g}$ \\
     $A$ & Partition probability normalization constant & $\sqrt{25/\pi}$~$\times$$10^{11}$ & -- & $\frac{1}{\text{g}}$ \\
     $\beta$ & Fission rate inverse variance constant & $100$$\times$$10^{22}$ & --  & $\frac{1}{\text{g}^2}$\\
     $m_\textrm{t0}$ & Minimal transition mass constant & $6$$\times$$10^{-11}$ & --  & $\text{g}$ \\
     $K_\textrm{t}$ & Transition mass sensitivity to substrate constant & $0.01$$\times$$10^{-11}$ & -- & $\frac{\text{g}}{\text{g L}}$ \\
     $S_\textrm{l}$ & Low substrate response limit constant & $0.1$ & --  & $\frac{\text{g}}{\text{L}}$\\
     $S_\textrm{h}$ & High substrate response limit constant & $2$ & --  & $\frac{\text{g}}{\text{L}}$ \\
     $m_\textrm{d0}$ & Minimal critical mass constant & $11$$\times$$10^{-11}$ & -- & $\text{g}$ \\
     $K_\textrm{d}$ & Critical mass sensitivity to substrate constant & $2$$\times$$10^{-11}$ & --  & $\frac{\text{g}}{\text{g L}}$ \\
     \bottomrule
    \end{tabular}
    \renewcommand{\arraystretch}{1}
    \label{tbl:yeastPBM_params}
\end{table} 
The viable cell density (VCD) is equal to the zeroth-order moment $m_0(t)$ of the yeast population, and a differential equation for $m_0(t)$ as in \eqref{eq:yeast-zeroth-moment-balance} can be derived by integrating \eqref{eq:yeast-PBM} over the mass $m$. A common practice for simplifying the analysis of a population balance model is to analyze the dynamics of the moments of the distribution \cite{witkowski_modelling_1987}, and the next subsections compare the dynamics of the cell number distribution with its zeroth-order moment. Descriptions of model parameters and initial conditions are given in Table~\ref{tbl:yeastPBM_params}. More details may be found in \cite{zhang_bifurcation_2001}.

\subsection{Open-loop Dynamics}
\label{yeast:openloop}
\begin{figure}[htbp]
    \centering
    \includegraphics[width= 0.90\linewidth]{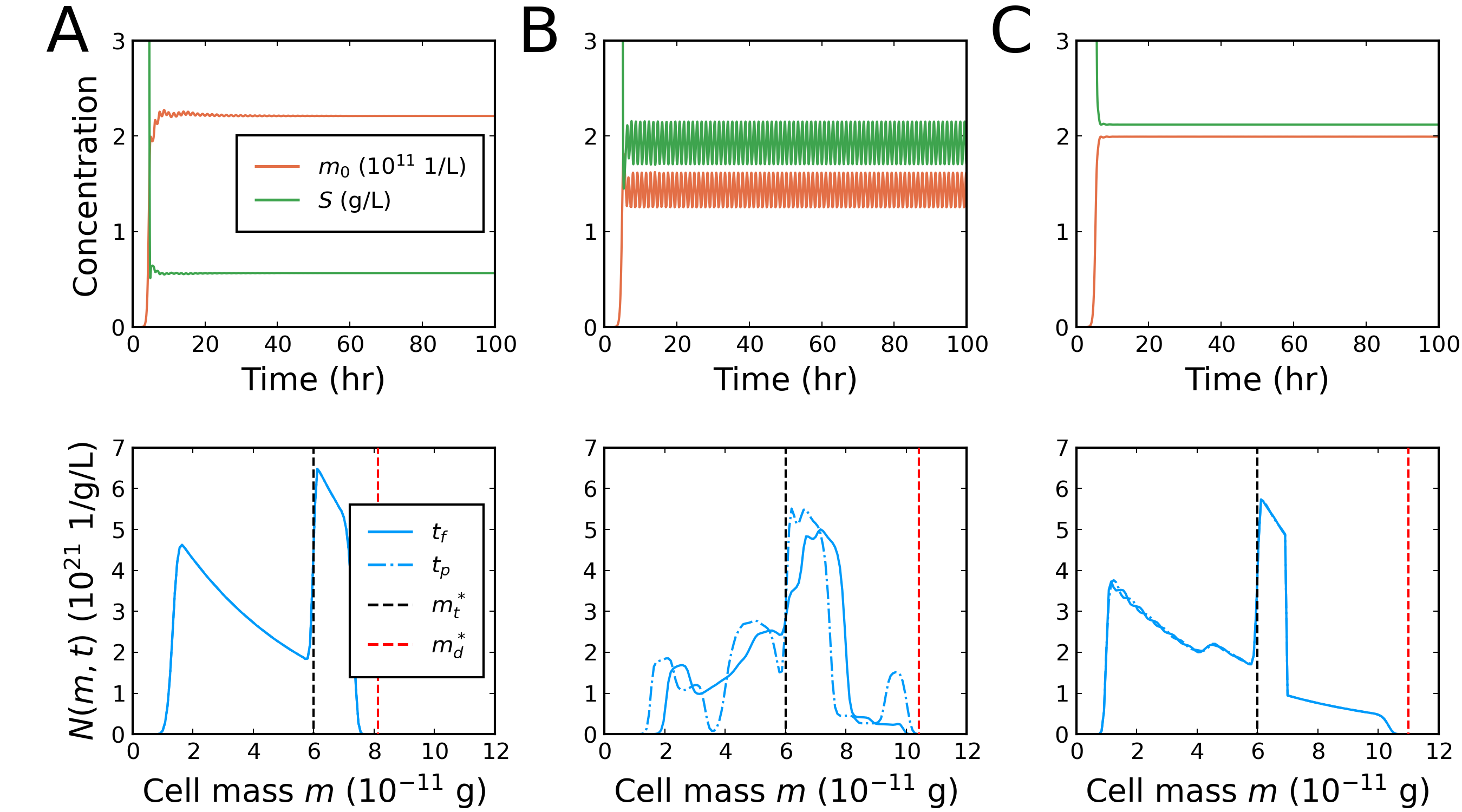}
    \caption{Open-loop simulations of the population balance model \eqref{eq:yeast-PBM} for dilution rates (A) $D=0.25$ 1/hr, (B) $D=0.55$ 1/hr, and (C) $D=0.85$ 1/hr using values from Table \ref{tbl:yeastPBM_params}. Cell mass distributions for the final time $t_\textrm{f}$ and five hours earlier ($t_\textrm{p}$) show varying degrees of synchronization of the cell populations, which relate to the appearance and disappearance of sustained oscillations. The cell mass distributions nearly perfectly overlap for $t_\textrm{f}$ and $t_\textrm{p}$ for $D=0.25$ and $0.85$ 1/hr, consistent with the stabilization of the zeroth-order moments for A and C. The intrinsic oscillations persist for the intermediate dilution rate (B, $D=0.55$ 1/hr). Fluctuations in the cell number distribution $N(m,t)$ in B occur during stable oscillations in the zeroth-order moment.
    }
    \label{fig:yeast-openloop}
\end{figure}

Here, both sets of equations are simulated for a variety of dilution rates and substrate feeds by a combination of ODE integraters in the Julia \texttt{DifferentialEquations.jl} package and hyperbolic PDE solvers in the Python \texttt{PyClaw} package  \cite{rackauckas_differentialequationsjlperformant_2017,mandli2016clawpack}. More specifically, a fifth-order accurate, adaptive time-stepping, stiff-aware ODE integrator \texttt{Rodas5} in Julia \cite{di_marzo_rodas54_1993} was explicitly coupled to \texttt{SharpClaw} \cite{KetParLev13}, which is a fifth-order accurate, one-dimensional advection high-resolution finite volume method (HRFVM) weighted-essentially non-oscillatory (WENO) algorithm \cite{hermanto_high-order_2009,gunawan_high_2004}. Sustained oscillations of yeast cell populations are observed in open-loop simulations for some operating conditions (Fig.\ \ref{fig:yeast-openloop}).


The creation of distinct mother-daughter cell sub-populations through asymmetric cell division (ACD) \cite{hartwell_unequal_1977} is a key factor in oscillations in continuous budding yeast cell culture. Being smaller, daughter cells have a longer doubling time than mother cells have, which are regenerated once a bud scissions. With \textit{cell-cycle synchrony}, which is the alignment of growth phases of cells to produce sharply peaked and distinctly propagated sub-populations in cell number distributions, daughter cells are produced in waves which periodically increase substrate consumption as the daughter cells grow to become mother cells. {The distinction of cell synchronization as a dynamic feature of a cell culture, i.e., the non-stationarity of cell number distributions which may allow distinct sub-populations to propagate, has not been highlighted in previous experimental literature. This may owe to the lack of high-resolution time-series measurements of yeast cell populations capable of demonstrating such features. On the other hand, asynchronization would be distinguished as the stationarity, or constant shape, of the corresponding cell number distributions. Stationarity under this interpretation implies stabilized and non-oscillatory cell culture behavior as shown in Fig.\  \ref{fig:yeast-openloop}.}

The earliest works on cell population synchronicity showed that the oscillation period of gas byproduct evolution was in-phase with early-stage budding cell density fluctuations, but out-of-phase with late-stage budding cell density fluctuations \cite{kaspar_von_meyenburg_stable_1973}. These observations suggest the importance of ACD and cell cycle dynamics in continuous yeast culture oscillations. 
However, synchronization in asymmetrically fissioning cell cultures alone provides an insufficient description of sustained oscillations in yeast cultures. Hirsch and Engelberg 
\cite{hirsch_determination_1965} proved that synchronized cell cultures will lose their synchrony at long culture times if the generation time\footnote{The generation time, aka doubling time, is analogous to the net growth rate $\mu$ for non-segregated cell populations and the fission rate function $\Gamma$ for segregated cell populations. The argument is that repeated passes through a fission function with finite variance will magnify the cell distribution variance and completely disperse cell sub-populations with time.} does not depend on the extracellular environment \cite{hirsch_determination_1965}. Str{\"a}ssle et al.\ \cite{strassle_predictive_1988,strassle_predictive_1989} later argued that representative models of oscillatory yeast cultures should be coupled to substrate dynamics through substrate-varying fission functions that represent repeated metabolic shifts at low substrate levels. Some later studies experimentally minimized cell metabolism disturbances in continuous yeast cultures to more strongly correlate oscillations with switches from fermentative to oxidative metabolism at low glucose feed levels \cite{parulekar_induction_1986}. Furthermore, M{\"u}nch et al.\ \cite{munch_new_1992} demonstrated that yeast cell oscillations may be induced with timed pulses of substrate that force these metabolic shifts. Importantly, these works all show that repeated metabolic shifts preserve synchronicity in the yeast cell cycle, allowing for sustained oscillations to be observed.

Altogether, yeast bioreactor oscillations involve a complex interaction between (1) continuous bioreactor operation to replenish feed and wash out accumulating mother cells, (2) low substrate levels inducing metabolic shifts, and (3) synchrony in asymmetrically budding daughter-mother cell sub-populations. Numerical solution of cell-segregated, unstructured mechanistic models of the kind in \eqref{eq:yeast-PBM} effectively lumps these three factors to produce reasonable approximations of the physical system\footnote{The three factors are, respectively: (1) continuous reactor configuration species balances, (2) substrate-dependent fission rate $\Gamma(m,S^\prime)$ that varies with movement of transition and critical mass boundaries $m_\textrm{t}^*,m_\textrm{d}^*$, and (3) a bimodal, asymmetric partition function $p(m,m^\prime)$.} \cite{zhu_model_2000}. As Fig.\ \ref{fig:yeast-openloop} illustrates, increasing the dilution rate from $D=0.25$ to $D=0.55$ 1/hr causes greater numbers of distinct cell sub-populations to form and propagate as waves. Further increasing the dilution rate to $D=0.85$ 1/hr disperses out any sub-populations previously being propagated, in turn eliminating oscillations in the process. Additional qualification of yeast system dynamics, done customarily using bifurcation analysis and continuation, is discussed at length in other works (\cite{henson_dynamic_2003} and references therein). A simpler approach is to use simulations to perform single-parameter variation studies such as presented in Section S3 of the Supplementary Information.





\subsection{Output Feedback Control}
\label{yeast:closed}

Several works have argued that oscillations in continuous yeast bioreactors can be both produced and attenuated through either the synchronization or dispersion of yeast reactor sub-populations \cite{strassle_predictive_1989,munch_new_1992,henson_dynamic_2003}. However, direct cell distribution manipulation through yeast seed injection is impractical \cite{henson_dynamic_2003}, and, although promising, robust on-line collection of cell number distributions through non-invasive, optical means remains challenging from a sterility and calibration standpoint \cite{pais_holographic_2020}. Instead, we consider the attenuation of oscillations through feedback control of traditionally more accessible bioreactor variables.

In modern high cell density (HCD) continuous yeast cell bioreactors, cell density and cell viability are measured in real time. Often, on-line measurements  include the bioreactor outlet substrate concentration $S$ and the VCD $m_0$. The dilution rate $D$ and substrate feed concentration $S_\textrm{f}$ regularly serve as manipulated variables, e.g., through peristaltic pump or valve actuation and through ratio-controlled dilution of a high concentration feed stock, respectively. Although model-based controller performance has previously been explored for this system, e.g., model predictive control \cite{henson_dynamic_2003} and input-output linearization \cite{zhang_bifurcation_2001}, oscillation attenuation was achieved in this work through simpler, model-independent single-input, single-output (SISO) feedback control laws of the form \eqref{eq:feedback_controllaw}.

Here our in-silico evaluation of controller performance compares (1) closed-loop response time and stability, (2) effects on non-controlled outputs, and (3) implementation feasibility given realistic process and controller limitations. Four candidate single-input, single-output (SISO) loops $(D,m_0)$, $(D,S)$, $(S_\textrm{f},m_0)$, and $(S_\textrm{f},S)$ are possible. Only results for the best performing loop, $(D,S)$, are presented here (the reader is referred to Section S4 of the Supplementary Information for a more comprehensive analysis of the three other controller candidates). Controller actuator limits are $D\in [0,1]$ 1/hr and $S_\textrm{f} \in [0, 200]$ g/L,  which are motivated by limitations observed in experimental continuous yeast bioreactors \cite{gomar-alba_response_2015,westman_current_2015}.\footnote{Detrimental yeast cell osmotic stress has been shown to occur at glucose substrate concentrations of $1.1$ M, which is about $200$ g/L. Recent advances in HCD yeast cell cultures have allowed for higher cell culture specific growth rates, and therefore higher reactor dilution rates---on the order of $1$ 1/hr in some cases.} 

Assuming the yeast culture is well-maintained (i.e., no adverse conditions or strong disturbances to yeast metabolism are recorded during operation), two relevant performance objectives are 
\begin{align}
    G_1(t_\textrm{f}) &= \int_0^{t_\textrm{f}} \!\left[D(t)\left(\alpha\frac{m_0(t)}{m_{0,\textrm{nom}}} - \beta \frac{S_\textrm{f}(t)}{S_{\textrm{f},\textrm{nom}}}\right) \right] \!\textrm{d}t,
    \label{eq:obj-func-yeast-econ} \\
    G_2(t_\textrm{f}) &= -\! \int_0^{t_\textrm{f}}\! \left[ \gamma_1\! \left(\frac{1}{m_{0,\textrm{nom}}}\frac{\textrm{d}m_0}{\textrm{d}t}\right)^{\!\!2}  + \gamma_2 \! \left(\frac{1}{S_{\textrm{nom}}}\frac{\textrm{d}S}{\textrm{d}t}\right)^{\!\!2} \right] \! \textrm{d}t
    \label{eq:obj-func-yeast-controller}
\end{align} 
where $t_\textrm{f}=50$ hr is a sufficiently long to capture the key dynamics of the process.
Both the economic objective $G_1(t_\textrm{f})$ and control performance objective $G_2(t_\textrm{f})$ may be maximized subject to model, controller, and physicality constraints (e.g., maintaining a positive VCD).\footnote{$G_2(t_\textrm{f})$ is a scalar measure of the extent of oscillations in $m_0$ and $S$.} The sensitivity parameters $\alpha$ (price of yeast product) and $\beta$ (cost of substrate feed) in the economic objective encode the tradeoff between biomass production and substrate expenses. Meanwhile, the sensitivity parameters $\gamma_1$ and $\gamma_2$ in the control performance objective weigh the operational benefit for candidate controllers to quickly stabilize undesirable oscillations in yeast VCD and substrate consumption.

\begin{figure}[htbp]
    \centering
    \includegraphics[width= 0.75\linewidth]{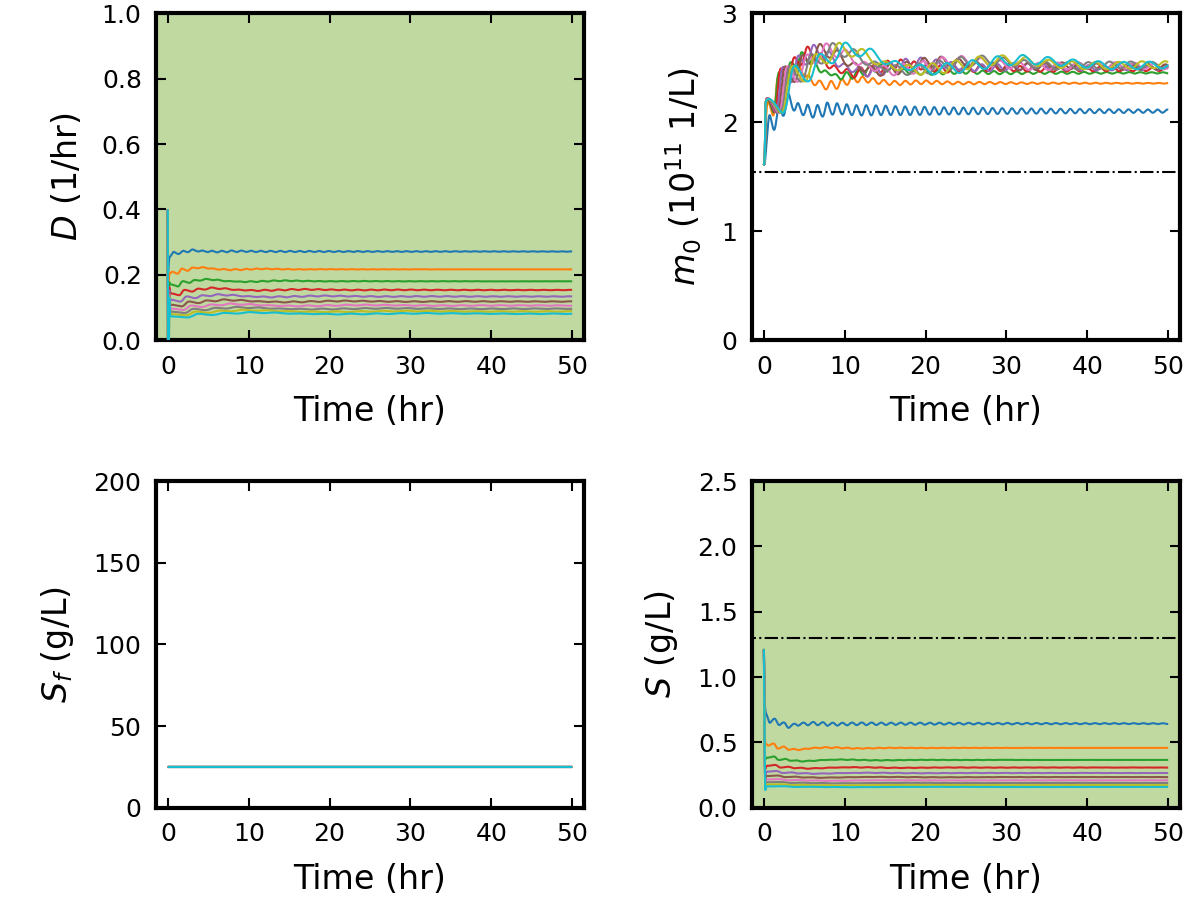}
    \caption{Closed-loop dynamics for the ($D$,$S$) control pairing for 10 equally spaced controller gains spanning $K_c \in [0.2,2]$ L/g-hr for the oscillatory nominal operating conditions defined in Table \ref{tbl:yeastPBM_params}. Dashdot lines represent the half-width of oscillations in the corresponding open-loop oscillating state variables. The best performing controller in terms of the unweighted sum of performance objectives \eqref{eq:obj-func-yeast-econ} and \eqref{eq:obj-func-yeast-controller} is marked by the orange line with controller gain $K_c = 0.4$ L/g-hr and bias $u_0=0.4$ 1/hr. Plot shading informs the control loop pairing.}
    \label{fig:yeast-closedloop-controller-eval}
\end{figure}

An initial value for the proportional gain $K_c$ in the SISO controller was set by the Internal Model Control design rule \cite{rivera_internal_1986,garcia_internal_1982},
\begin{align}
\label{eq:IMC-tuning-rule}
    K_c = \frac{2\tau_\textrm{p}+\theta}{K_\textrm{p}(2\tau_\textrm{c} + \theta)},
\end{align}
where $\tau_\textrm{p}$ (1/hr) is the process time constant,\footnote{The time-scale $\tau_\textrm{p}$ for an oscillatory process yields very naturally from the period of oscillations in the desired output.} $\theta$ (1/hr) is the process time delay, $K_\textrm{p}$ (L/g-hr) is the process gain for the ($D$,$S$) control loop, and $\tau_\textrm{c}$ is the closed-loop time constant which was set based on relatively mild assumptions\footnote{Assuming closed-loop response on the time scale of the process ($\tau_\textrm{c}=\tau_\textrm{p}$) and no time delay ($\theta=0$), the controller gain $K_c$ equals the inverse of the process gain $K_\textrm{p}$, which may be approximated as the ratio of the step response in process output substrate concentration $S$ to a step change in process input dilution rate $D$, e.g., $K_c = K_\textrm{p}^{-1} \approx \Delta D / \Delta S$ for the ($D$,$S$) control loop.} and open-loop step responses to obtain the process gain $K_\textrm{p}$ (see Section S4 in Supplementary Information). 
The objectives \eqref{eq:obj-func-yeast-econ} and \eqref{eq:obj-func-yeast-controller} are computed in closed-loop simulations for a range of proportional gains that included the above estimate of $K_c$ and controller biases $u_0$ contained in the actuator limits. A subset of these closed-loop simulations is shown in Fig.\ \ref{fig:yeast-closedloop-controller-eval}. Each control loop pairing had at least one candidate controller that attenuated oscillations within the time frame that the controllers are applied (see Figs.\ S15--S18 in Supplementary Information). 

\begin{figure}[htbp]
    \centering
    \includegraphics[width= 0.95\linewidth]{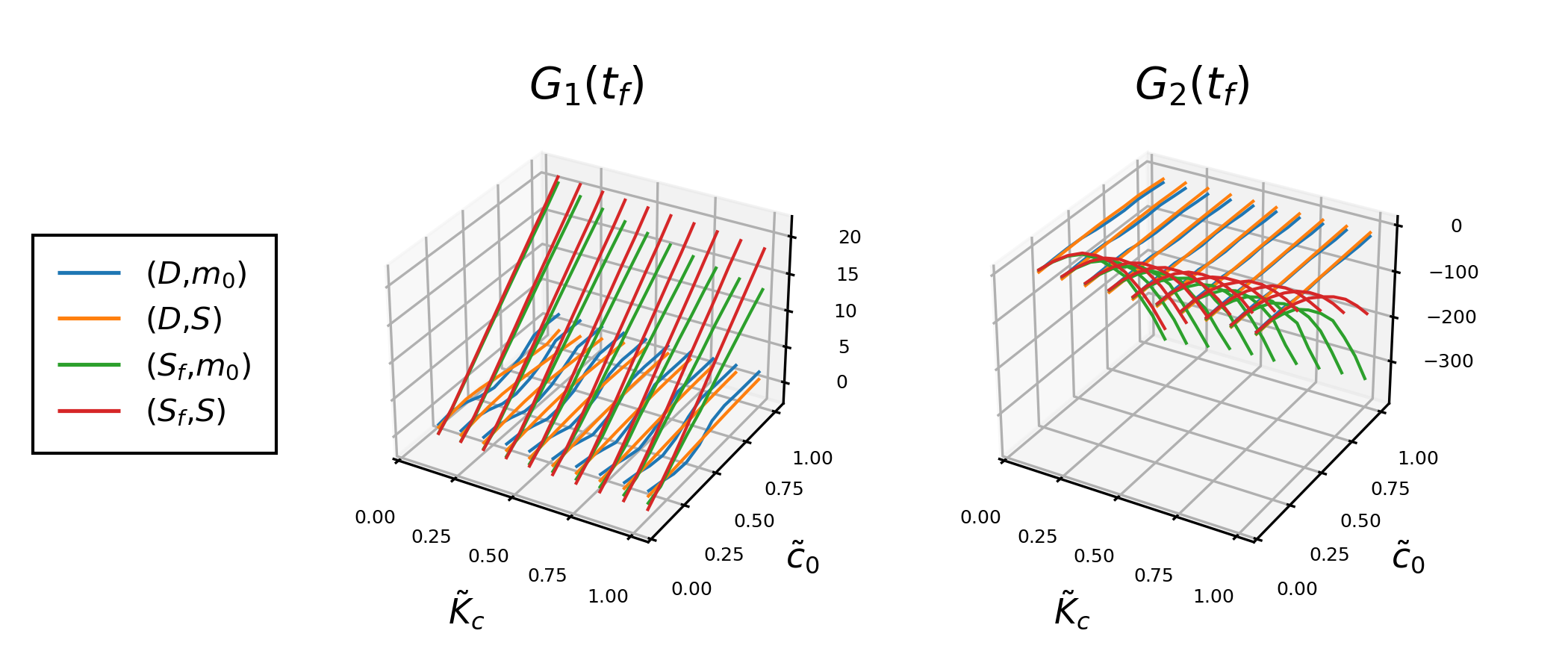}
    \caption{Using an unweighted combination of economic and control objectives for $t_f=50$ hr with unit sensitivities $\alpha=\beta = \gamma_1 = \gamma_2 = 1$, the best performing controller was the ($D$,$S$) SISO loop. $\tilde{K}_c$ and $\tilde{c}_0$ are the non-dimensional gain and bias for each controller, respectively, taken as the dimensional gain $K_c$ (or bias $u_0$) divided by the maximal gain $K_{c,\textrm{max}}$ (or bias $u_{0,\textrm{max}}$) simulated for each of the four SISO controllers.}
    \label{fig:yeast-closedloop-best-controller}
\end{figure}

No single controller performed best for both objective functions simultaneously (Fig.\ \ref{fig:yeast-closedloop-best-controller}). For most controller tuning parameters, the ($D$,$m_0$) and ($D$,$S$) SISO loops outperformed on the control objective \eqref{eq:obj-func-yeast-controller} due to rapid stabilization of oscillations. In these cases, the system was pushed into stabilizing dilution rates that quickly dispersed cell sub-populations. However, because the dilution rate $D$ directly affects both the yeast VCD $m_0$ outflow in \eqref{eq:yeast-PBM} and substrate feed concentration $S$ inflow in \eqref{eq:yeast-substrate-balance}, changes in $D$ produce conflicting conditions for increasing biomass yields. Ideally, a continuous cell culture will sustain larger VCDs with higher reactor residence times (e.g., lower $D$) and higher reactor substrate concentrations (e.g., higher $D$). By the same token, the ($S_\textrm{f}$,$m_0$) and ($S_\textrm{f}$,$S$) SISO loops outperformed on the economic objective \eqref{eq:obj-func-yeast-econ} by sustaining biomass increases. Manipulating the substrate feed concentration $S_\textrm{f}$ only directly affects the substrate dynamics in \eqref{eq:yeast-substrate-balance}, which in turn allows the culture to adjust to higher substrate concentrations at constant species residence times. Less desirable is that the $S_\textrm{f}$ loop options were not able to attenuate oscillations as quickly as the $D$ loop options in either of the reactor outputs considered (Fig.\ \ref{fig:yeast-closedloop-best-controller}). More details are provided in Figs.\ S15--S18 in the Supplementary Information.

\begin{figure}[htbp]
    \centering
    \includegraphics[width= 0.95\linewidth]{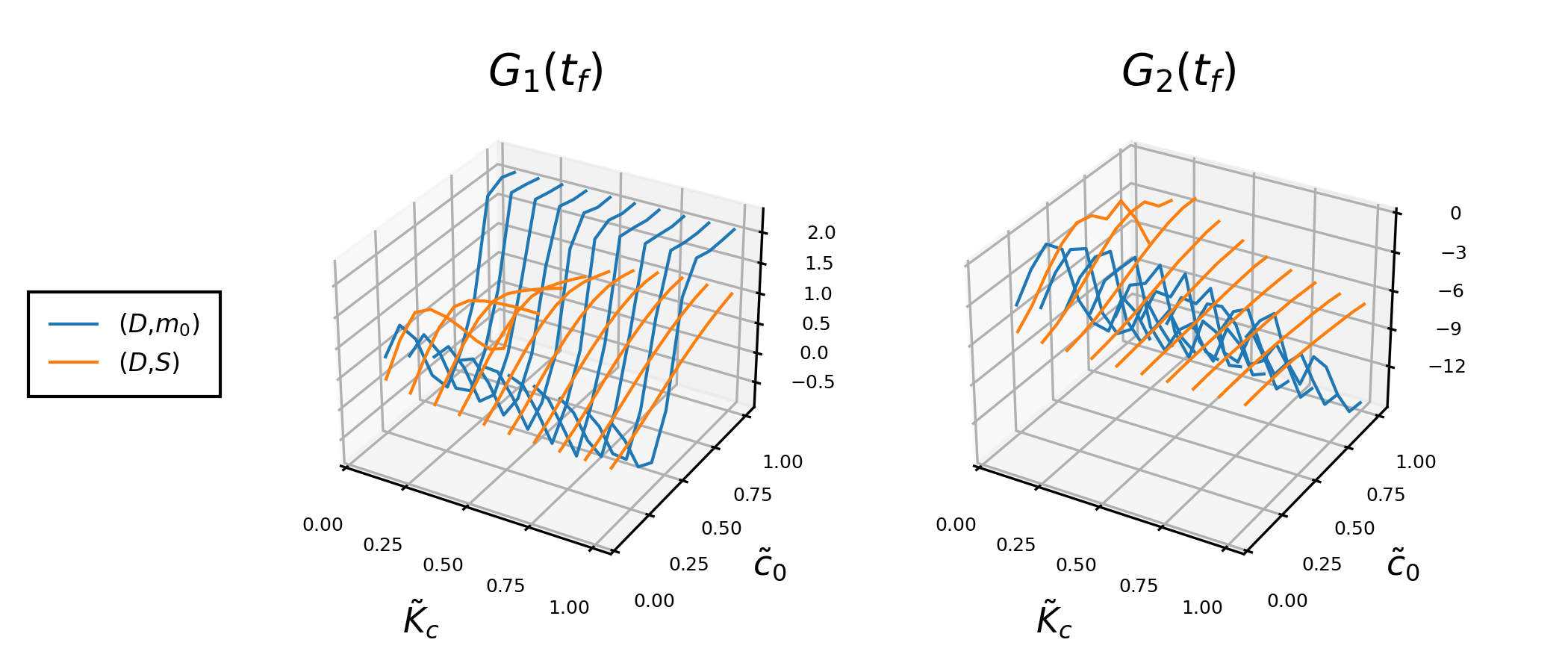}
    \caption{Comparison of the ($D$,$m_0$) and ($D$,$S$) SISO loops. See Fig.\ \ref{fig:yeast-closedloop-best-controller} for further description.}
    \label{fig:yeast-closedloop-best-controller-zoomed}
\end{figure}

With respect to the control objective $G_2(t_\textrm{f})$, which is the primary focus of this case study, the ($D$,$S$) SISO loop outperformed the ($D$,$m_0$) loop for most of the dilution rate controller biases $u_0$ (Fig.\ \ref{fig:yeast-closedloop-best-controller-zoomed}). Operating the feedback control at higher dilution rates (i.e., $0.4<D<0.85$ 1/hr) pushes the system into stable oscillations for both SISO control loops. Controlling the substrate concentration $S$ instead of the yeast VCD $m_0$ in most cases produces smaller amplitude and longer period oscillations for both outputs. In other words, less undesirable oscillatory character is observed with the ($D$,$S$) control loop pairing (see Figs.\ S15 and S16 in the Supplementary Information). One possible explanation is that controlling the substrate concentration $S$ directly impacts the yeast cell distribution through the direct dependency of the single cell growth rate on the substrate concentration \eqref{eq:yeast-growth-rate-function}. The substrate concentration $S$ is a more effective measurement because its dynamic response to changes in the bioreactor operation is faster than for the VCD $m_0$ (cf., \ref{eq:yeast-zeroth-moment-balance} and \ref{eq:yeast-substrate-balance}).

\section{Conclusions}

A simple single-loop output feedback control strategy has been demonstrated to attenuate (in some cases even eliminate) oscillatory behavior in multiple classes of bioreactor models. A well-designed feedback controller is more robust than designing and/or operating in open loop at known stable conditions, and feedback enables steady operation at higher productivity. In some cases (e.g., Model 3 in Section~\ref{sec:frensing}), there is a tradeoff between attenuating the oscillations and process productivity. More advanced control strategies such as model predictive control and extremum-seeking control can be developed to further optimize the system using the feedback-stabilized system as a basis. 

We have also considered alternative process configurations that are able to bypass the source of the oscillatory behavior in the original bioreactor system. Although not as prevalent as stirred tank bioreactors in industrial biomanufacturing sites, the tubular bioreactors remove the potential for oscillations altogether, without the need for developing sensors for one or species concentrations.

Collectively, we have shown that these control and design approaches for the suppression of oscillations have rather general utility, being demonstrated for four classes of oscillations reported in bioreactors. 
 
\section*{Availability of Code}
All the code used to perform the various simulations and generate the figures in the manuscript can be found at \url{https://github.com/pavaninguva/oscillatory_bioreactor}.

\section*{Acknowledgments}
Financial support is acknowledged from the Agency for Science, Technology and Research (A*STAR), Singapore. Furthermore, this material is based upon work supported by the U.S. Department of Energy, Office of Science, Office of Advanced Scientific Computing Research, Department of Energy Computational Science Graduate Fellowship under Award Number DE-SC0022158.

\section*{Disclaimer}\label{sec:disclaimer}

This report was prepared as an account of work sponsored by an agency of the United States Government. Neither the United States Government nor any agency thereof, nor any of their employees, makes any warranty, express or implied, or assumes any legal liability or responsibility for the accuracy, completeness, or usefulness of any information, apparatus, product, or process disclosed, or represents that its use would not infringe privately owned rights. Reference herein to any specific commercial product, process, or service by trade name, trademark, manufacturer, or otherwise does not necessarily constitute or imply its endorsement, recommendation, or favoring by the United States Government or any agency thereof. The views and opinions of authors expressed herein do not necessarily state or reflect those of the United States Government or any agency thereof.

\clearpage
\printbibliography

\end{document}


\maketitle
\footnotetext[3]{Authors contributed equally to this work.}
\footnotetext[1]{Corresponding author. Email: braatz@mit.edu}
\renewcommand{\thefootnote}{\arabic{footnote}} 

\section{Output feedback control of a continuous ethanol fermenter}

A series of simulations with different control parameters $K_{c}$ and $u_{0}$ for the different combinations of controlled and manipulated variables are presented below. 

\begin{figure}[htb]
    \centering
    \begin{subfigure}{0.8\textwidth}
        \centering
        \includegraphics[width=\textwidth]{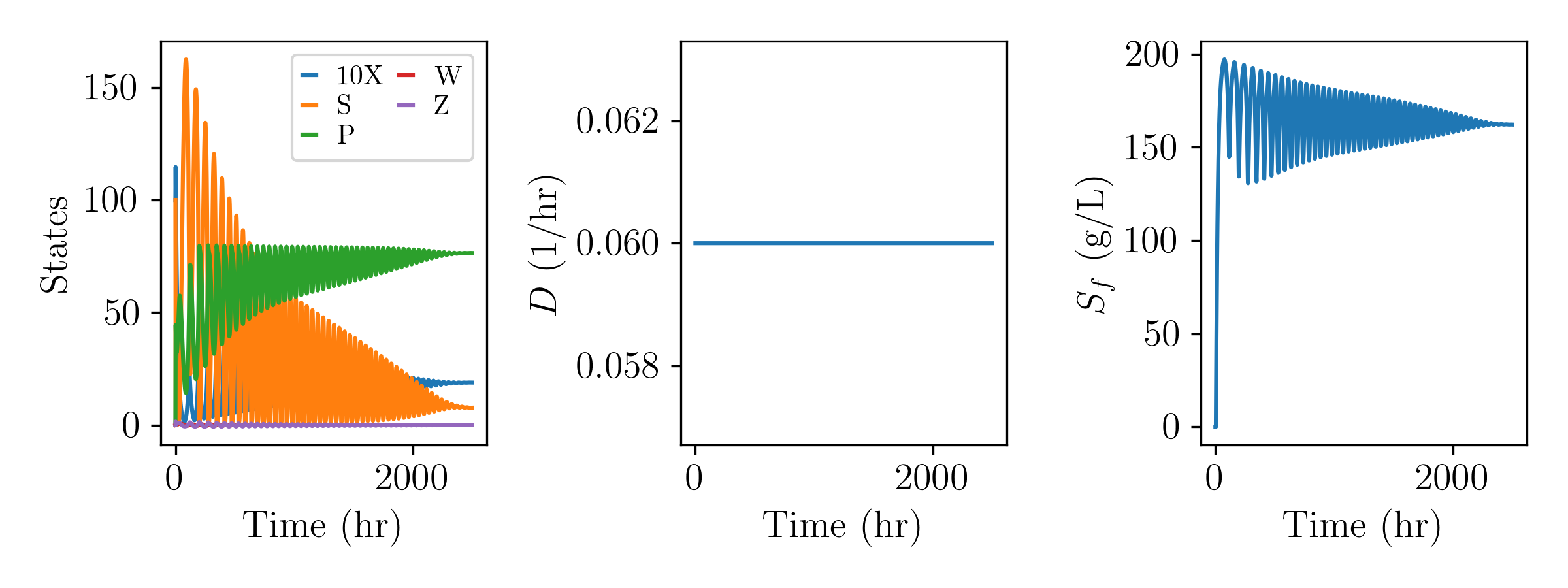} 
        \vspace{-0.2cm}
        \caption{$K_{c}= -20.0, u_{0} = 200.0$}
    \end{subfigure}
    \hfill
    \centering
    \begin{subfigure}{0.8\textwidth}
        \centering
        \includegraphics[width=\textwidth]{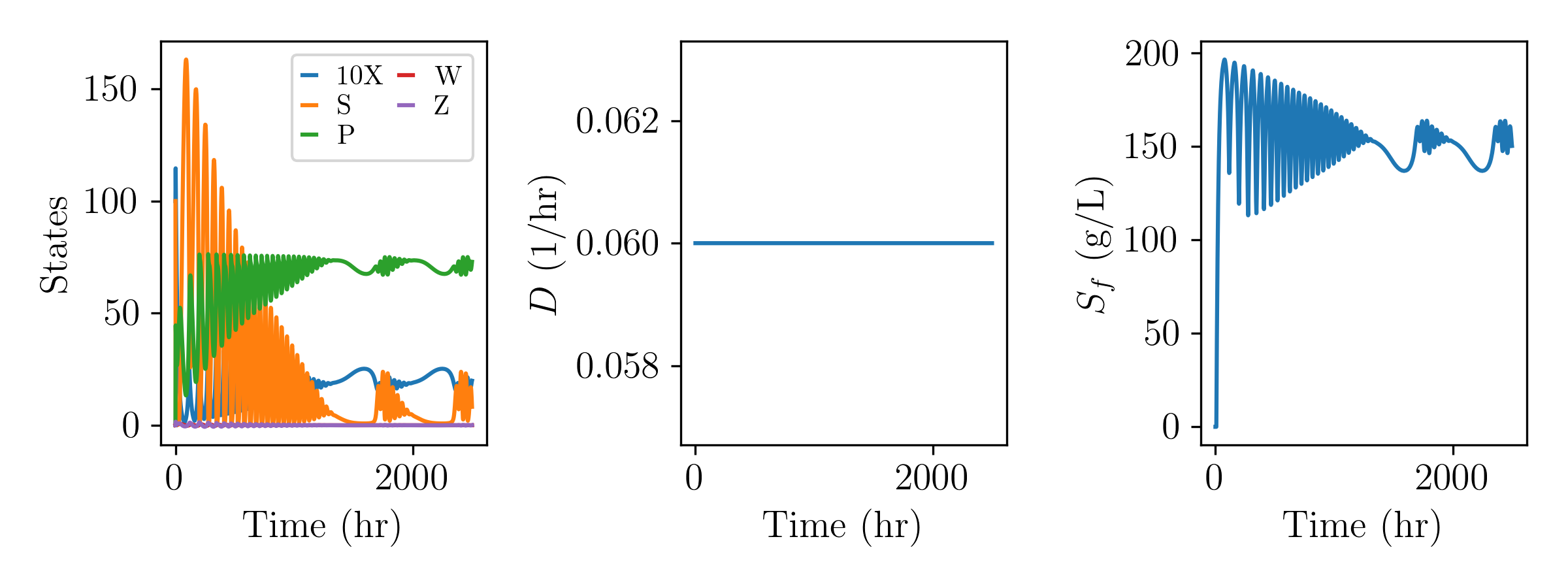} 
        \vspace{-0.2cm}
        \caption{$K_{c}= -25.0, u_{0} = 200.0$}
    \end{subfigure}
    \hfill
    \centering
    \begin{subfigure}{0.8\textwidth}
        \centering
        \includegraphics[width=\textwidth]{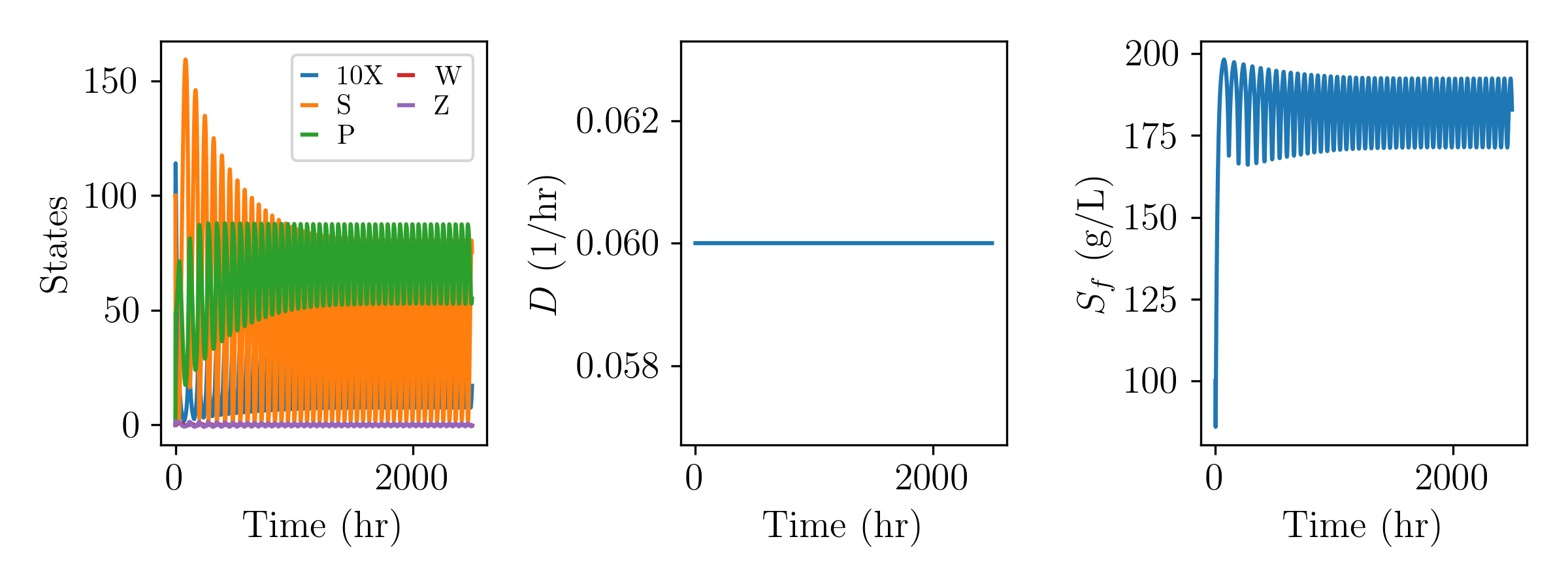} 
        \vspace{-0.2cm}
        \caption{$K_{c}= -10.0, u_{0} = 200.0$}
    \end{subfigure}
    \hfill
    \centering
    \begin{subfigure}{0.8\textwidth}
        \centering
        \includegraphics[width=\textwidth]{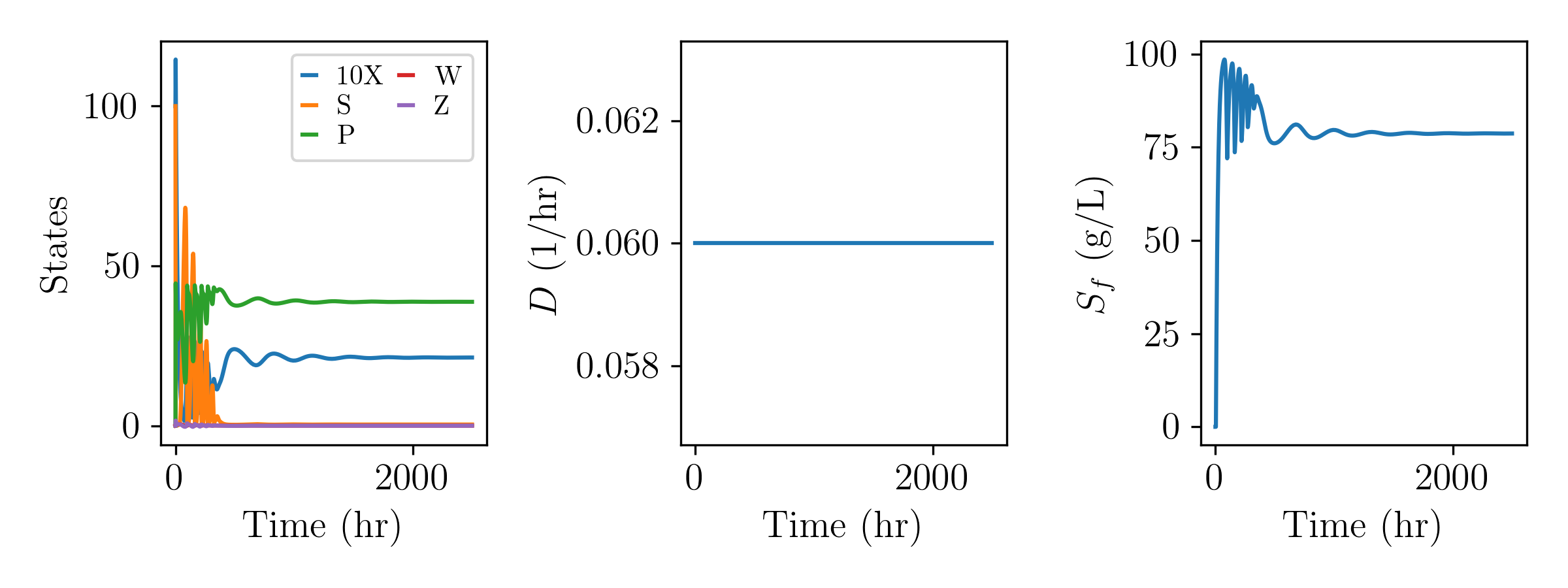} 
        \vspace{-0.2cm}
        \caption{$K_{c}= -10.0, u_{0} = 100.0$}
    \end{subfigure}
    \caption{Closed-loop feedback control simulations with a single loop feedback control law using $X$ as the controlled variable and $S_{f}$ as the manipulated variable.}
    \label{fig:ethanol_X_Sf}
\end{figure}

\begin{figure}[htb]
    \centering
    \begin{subfigure}{0.8\textwidth}
        \centering
        \includegraphics[width=\textwidth]{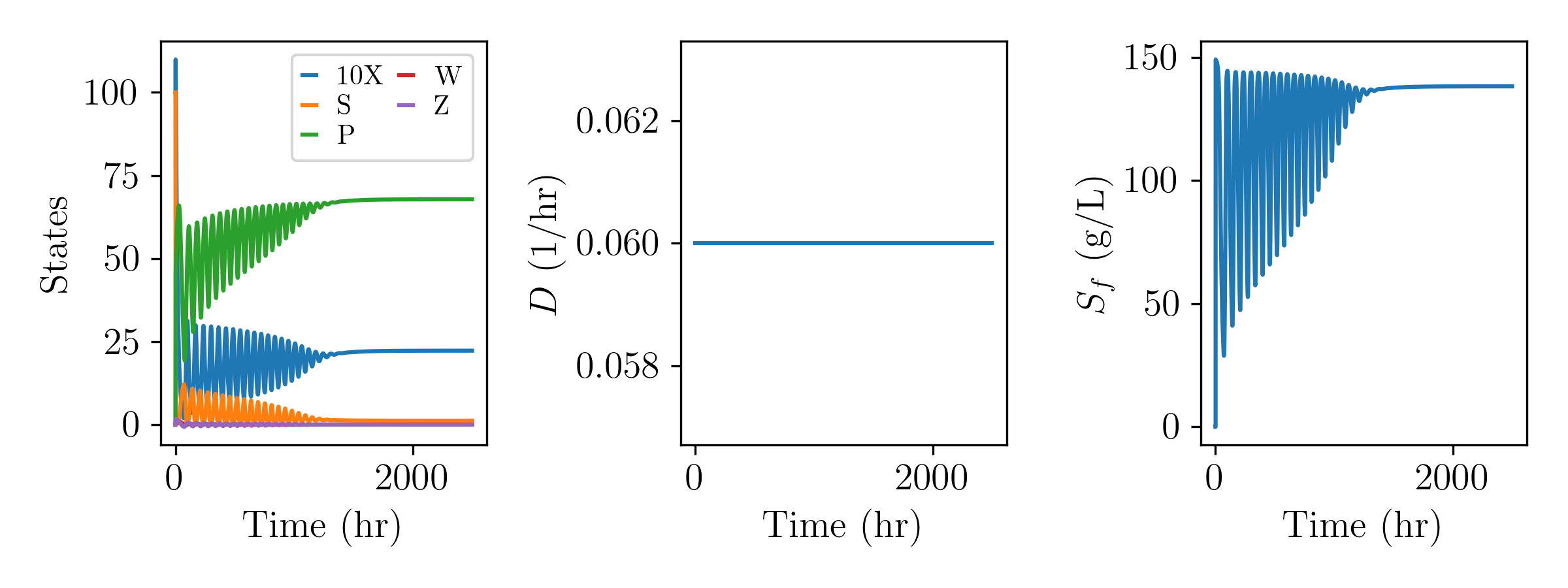} 
        \vspace{-0.2cm}
        \caption{$K_{c}= -10.0, u_{0} = 150.0$}
    \end{subfigure}
    \hfill
    \centering
    \begin{subfigure}{0.8\textwidth}
        \centering
        \includegraphics[width=\textwidth]{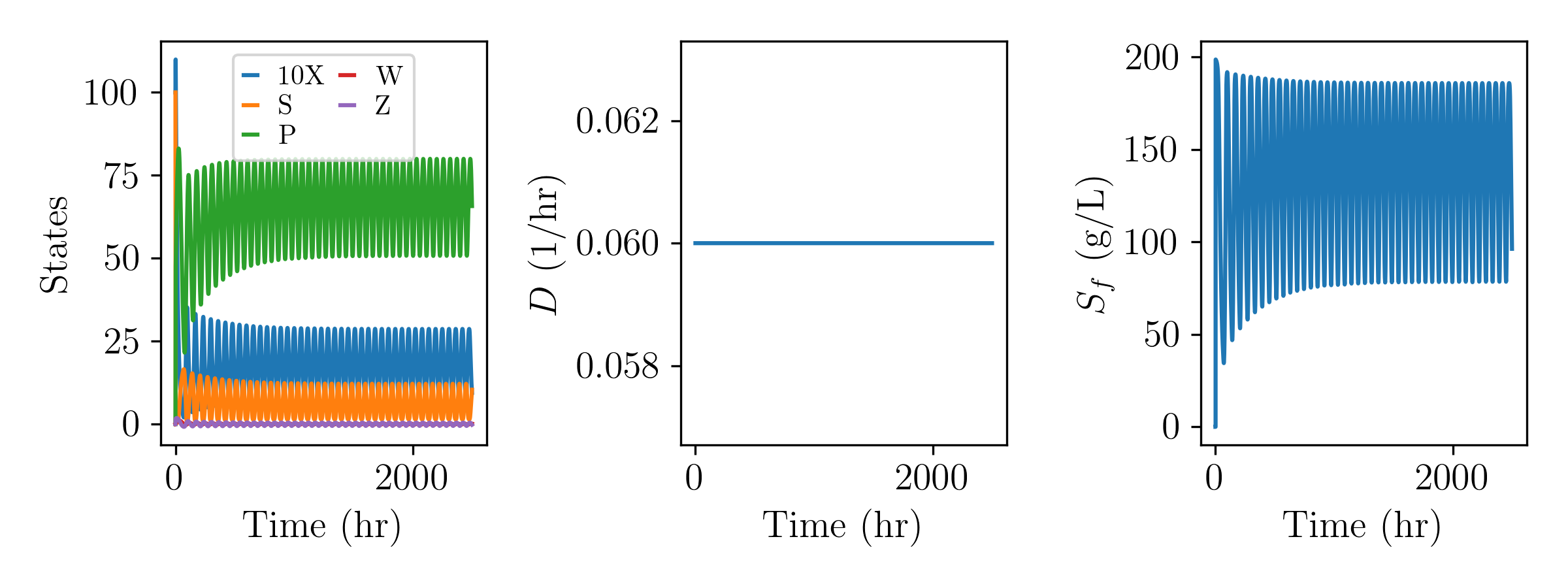} 
        \vspace{-0.2cm}
        \caption{$K_{c}= -10.0, u_{0} = 200.0$}
    \end{subfigure}
    \hfill
    \centering
    \begin{subfigure}{0.8\textwidth}
        \centering
        \includegraphics[width=\textwidth]{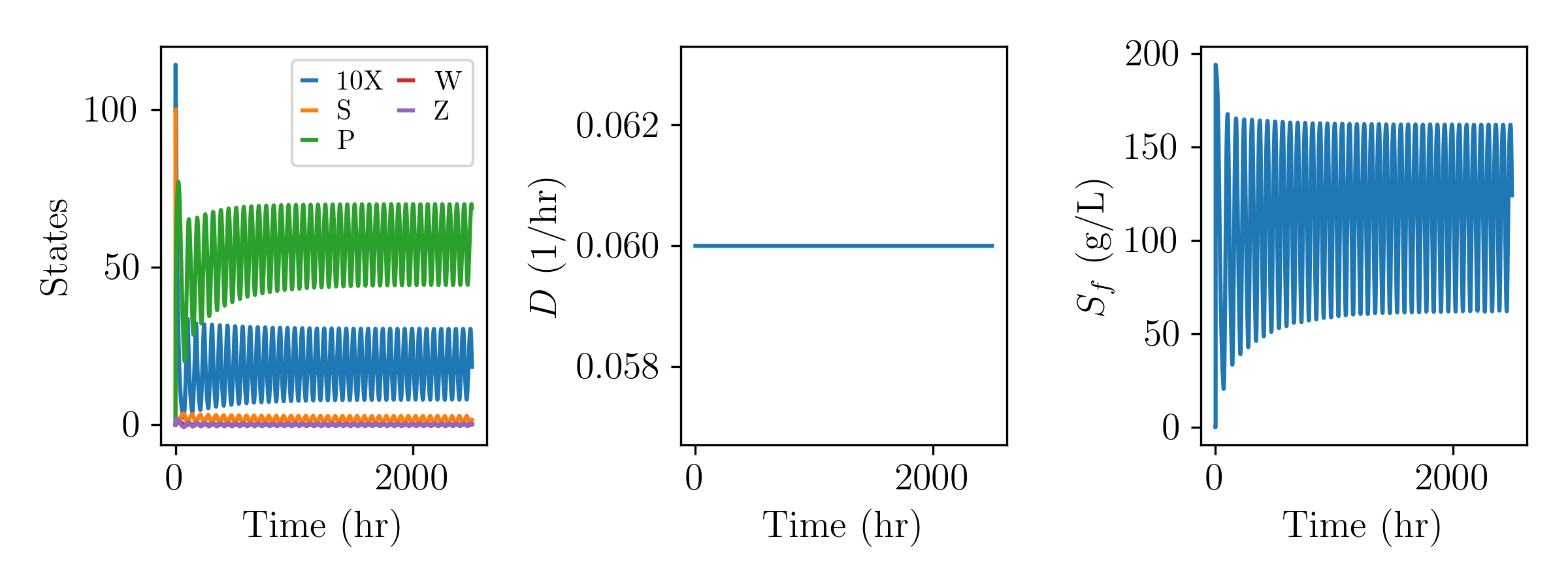} 
        \vspace{-0.2cm}
        \caption{$K_{c}= -50.0, u_{0} = 200.0$}
    \end{subfigure}
    \hfill
    \centering
    \begin{subfigure}{0.8\textwidth}
        \centering
        \includegraphics[width=\textwidth]{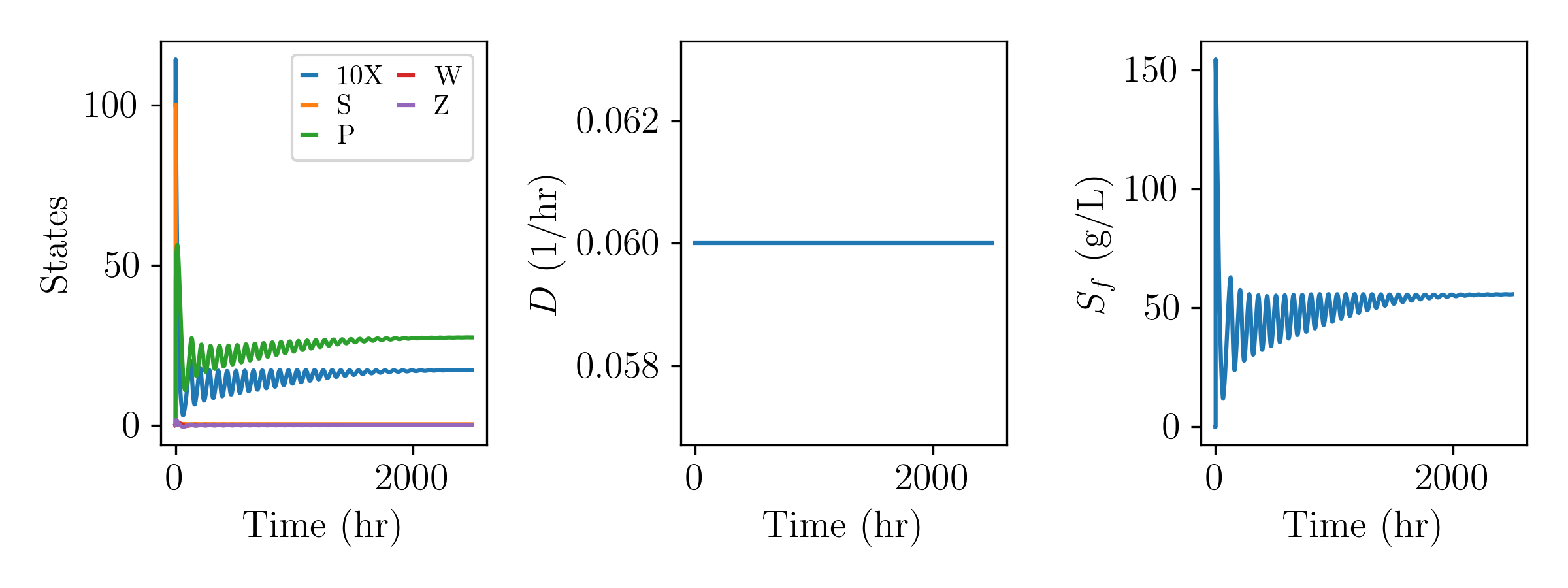} 
        \vspace{-0.2cm}
        \caption{$K_{c}= -500.0, u_{0} = 200.0$}
    \end{subfigure}
    \caption{Closed-loop feedback control simulations with a single loop feedback control law using $S$ as the controlled variable and $S_{f}$ as the manipulated variable.}
    \label{fig:ethanol_S_Sf}
\end{figure}

\begin{figure}[htb]
    \centering
    \begin{subfigure}{0.8\textwidth}
        \centering
        \includegraphics[width=\textwidth]{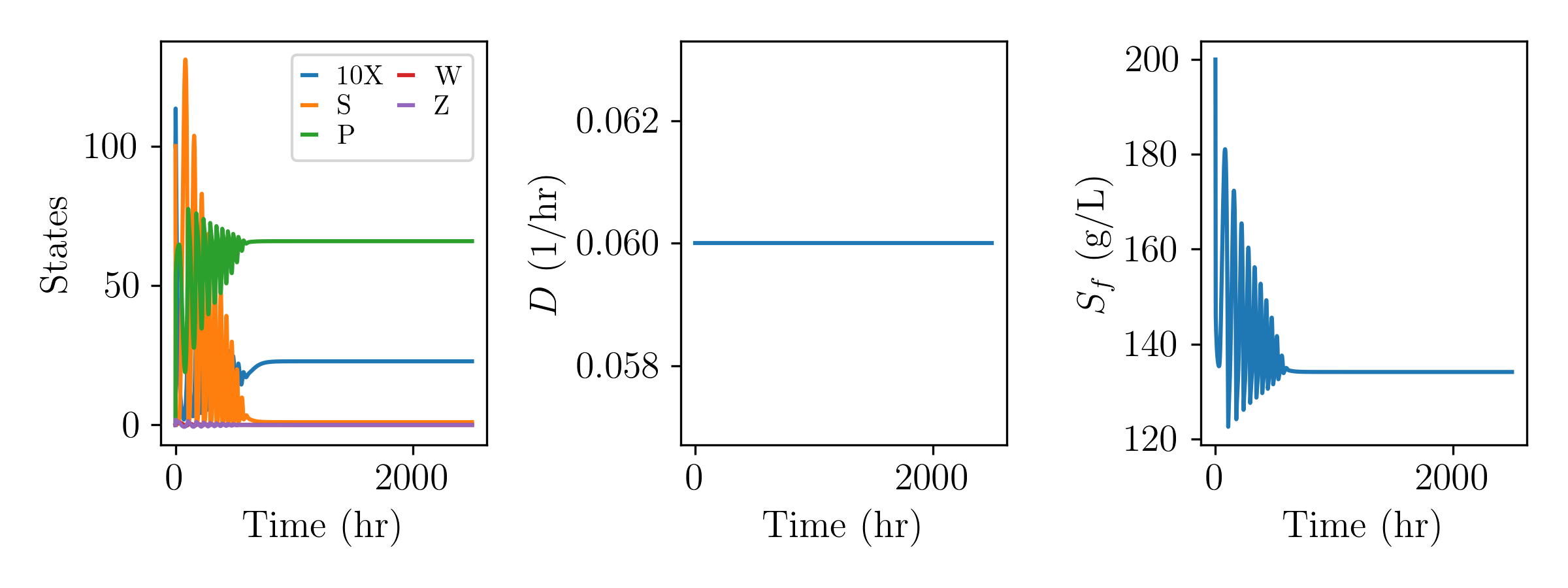} 
        \vspace{-0.2cm}
        \caption{$K_{c}= -1.0, u_{0} = 200.0$}
    \end{subfigure}
    \hfill
    \centering
    \begin{subfigure}{0.8\textwidth}
        \centering
        \includegraphics[width=\textwidth]{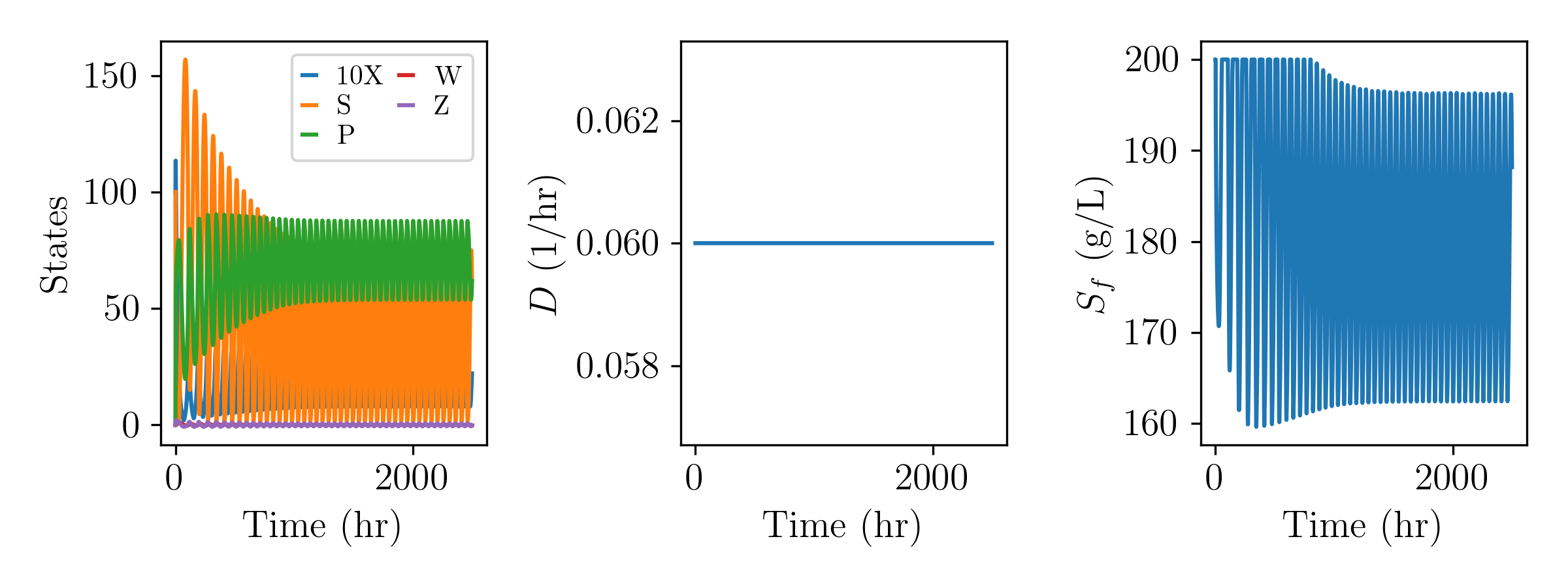} 
        \vspace{-0.2cm}
        \caption{$K_{c}= -1.0, u_{0} = 250.0$}
    \end{subfigure}
    \hfill
    \centering
    \begin{subfigure}{0.8\textwidth}
        \centering
        \includegraphics[width=\textwidth]{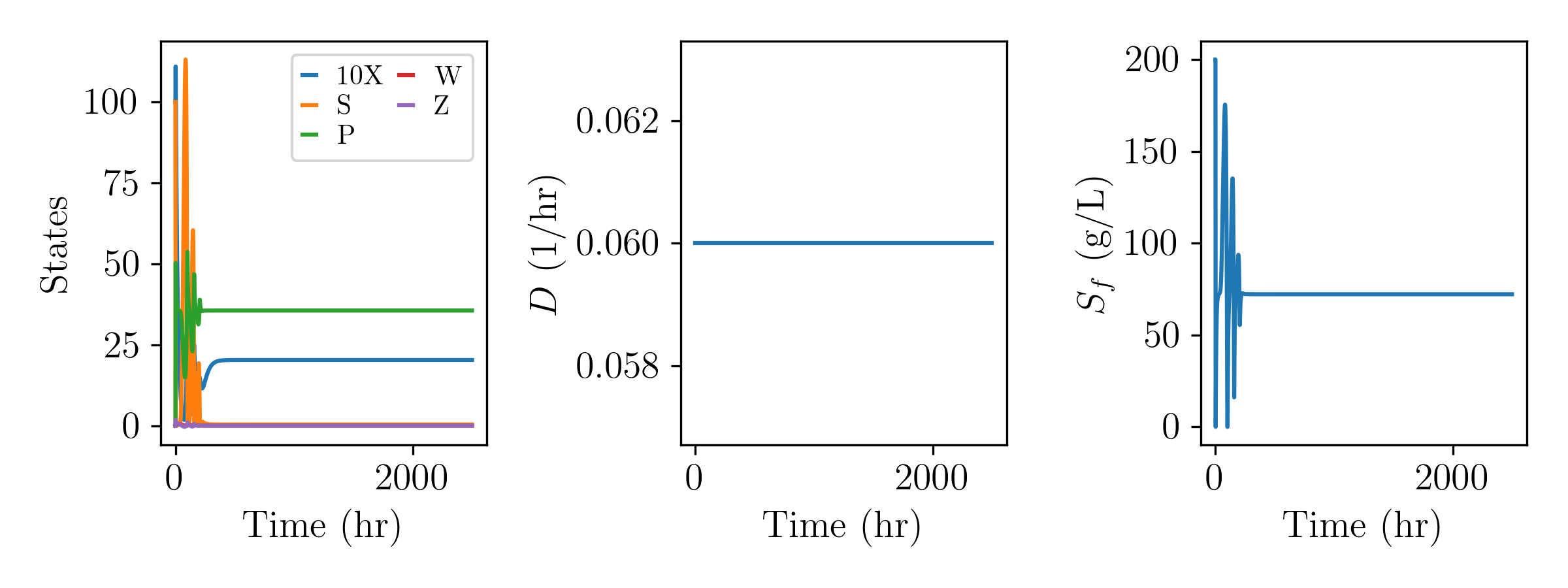} 
        \vspace{-0.2cm}
        \caption{$K_{c}= -5.0, u_{0} = 250.0$}
    \end{subfigure}
    \hfill
    \centering
    \begin{subfigure}{0.8\textwidth}
        \centering
        \includegraphics[width=\textwidth]{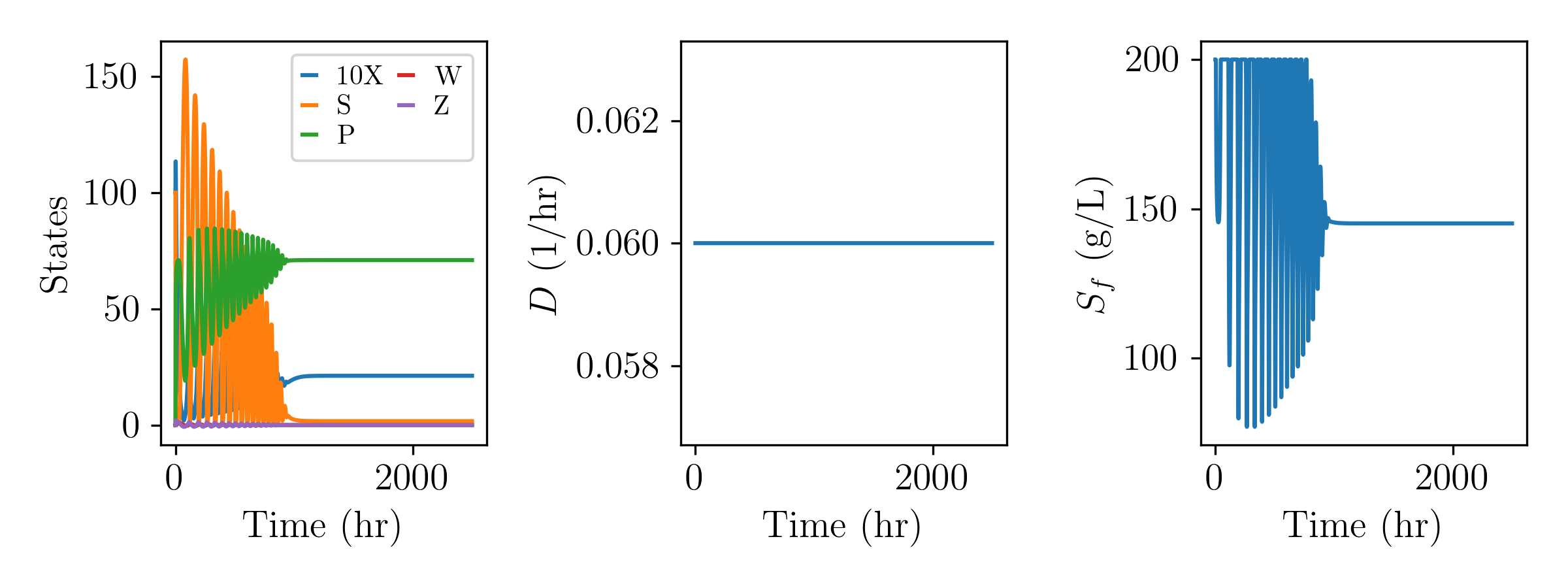} 
        \vspace{-0.2cm}
        \caption{$K_{c}= -5.0, u_{0} = 500.0$}
    \end{subfigure}
    \caption{Closed-loop feedback control simulations with a single loop feedback control law using $P$ as the controlled variable and $S_{f}$ as the manipulated variable.}
    \label{fig:ethanol_P_Sf}
\end{figure}

\begin{figure}[htb]
    \centering
    \begin{subfigure}{0.8\textwidth}
        \centering
        \includegraphics[width=\textwidth]{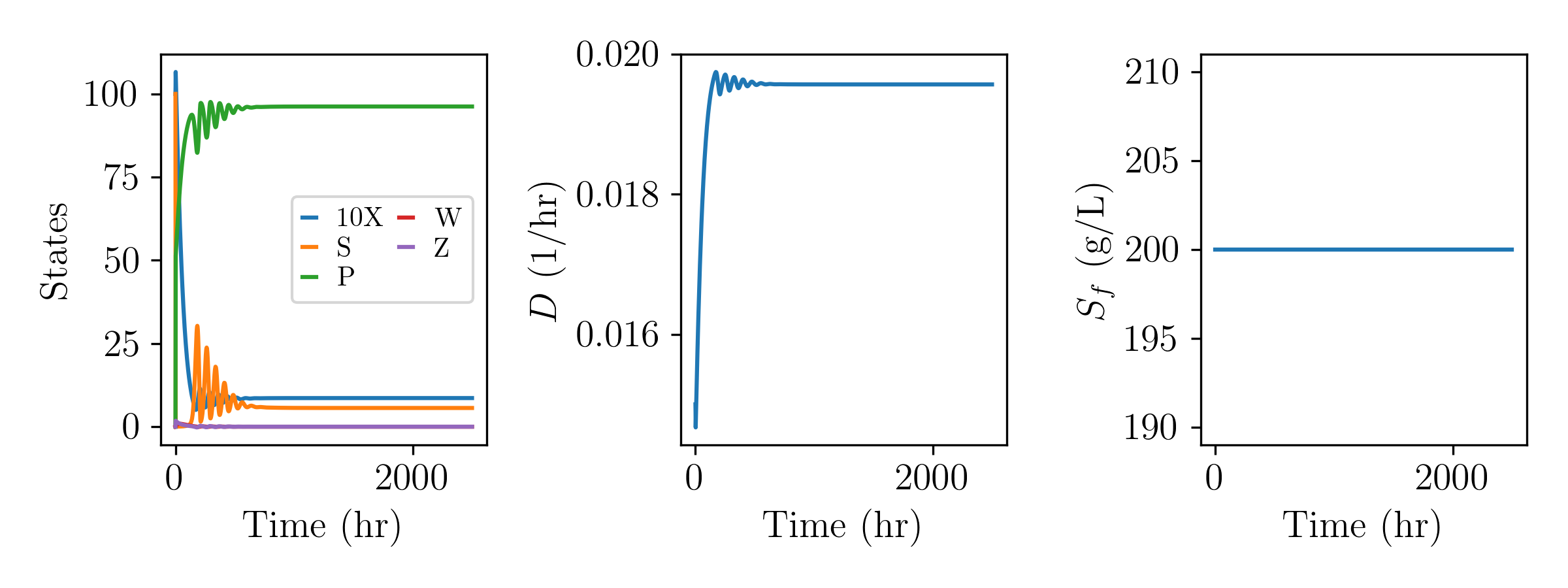} 
        \vspace{-0.2cm}
        \caption{$K_{c}= -0.0005, u_{0} = 0.02$}
    \end{subfigure}
    \hfill
    \centering
    \begin{subfigure}{0.8\textwidth}
        \centering
        \includegraphics[width=\textwidth]{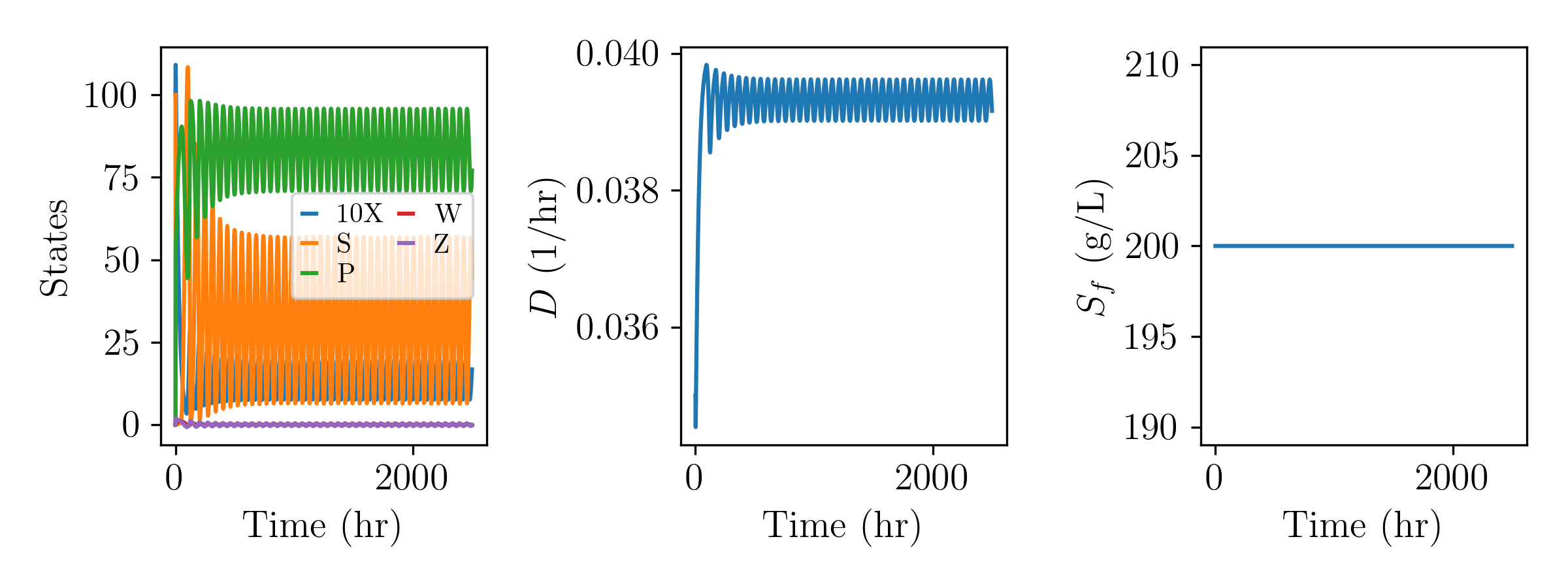} 
        \vspace{-0.2cm}
        \caption{$K_{c}= -0.0005, u_{0} = 0.04$}
    \end{subfigure}
    \hfill
    \centering
    \begin{subfigure}{0.8\textwidth}
        \centering
        \includegraphics[width=\textwidth]{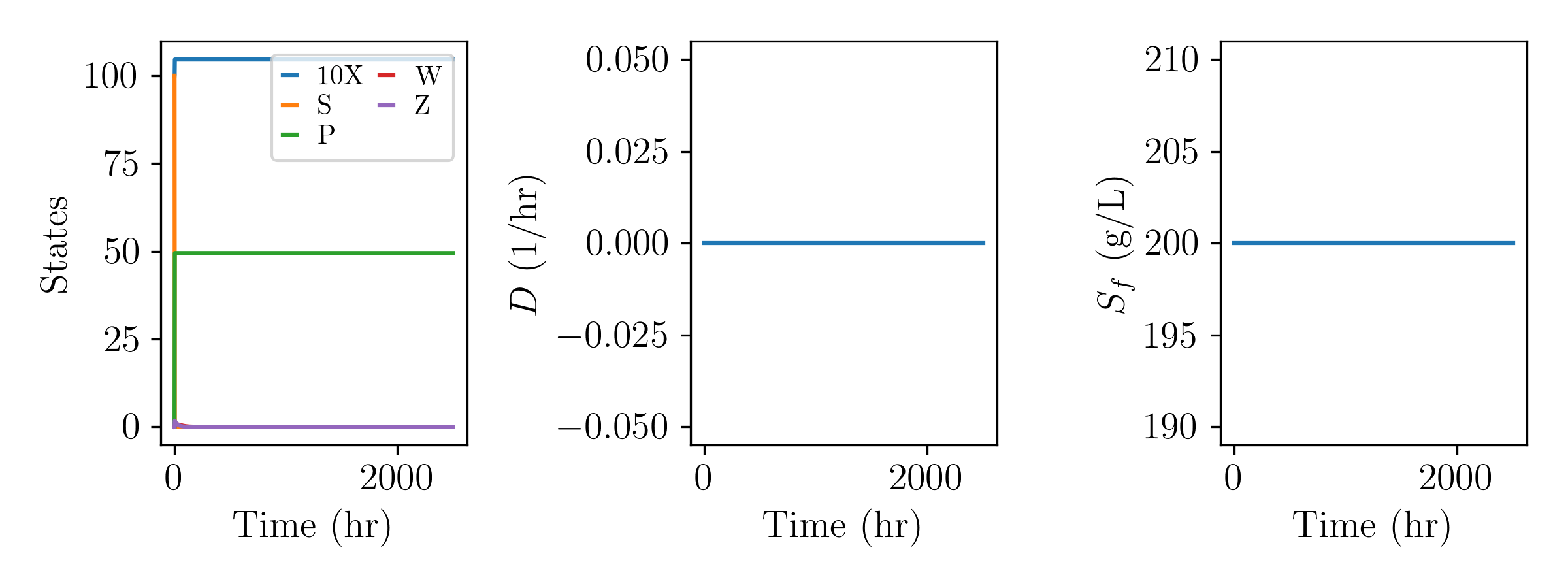} 
        \vspace{-0.2cm}
        \caption{$K_{c}= -0.005, u_{0} = 0.04$}
    \end{subfigure}
    \hfill
    \centering
    \begin{subfigure}{0.8\textwidth}
        \centering
        \includegraphics[width=\textwidth]{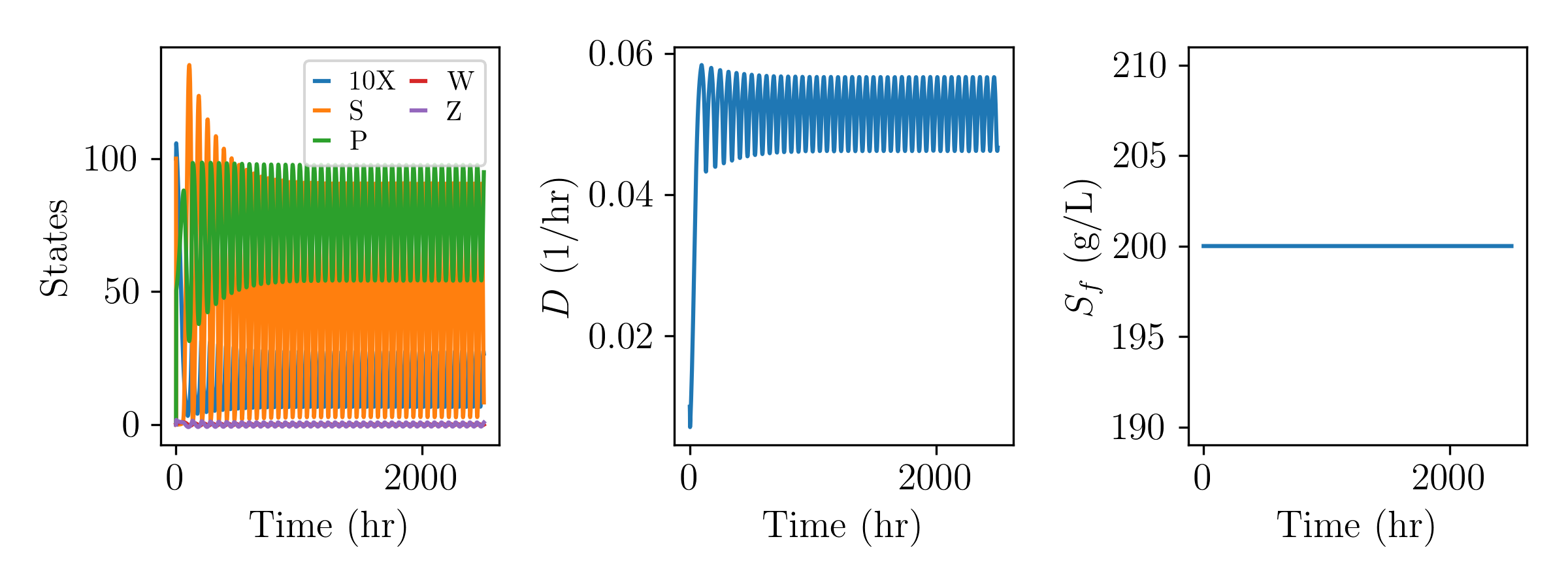} 
        \vspace{-0.2cm}
        \caption{$K_{c}= -0.005, u_{0} = 0.06$}
    \end{subfigure}
    \caption{Closed-loop feedback control simulations with a single loop feedback control law using $X$ as the controlled variable and $D$ as the manipulated variable.}
    \label{fig:ethanol_X_D}
\end{figure}

\begin{figure}[htb]
    \centering
    \begin{subfigure}{0.8\textwidth}
        \centering
        \includegraphics[width=\textwidth]{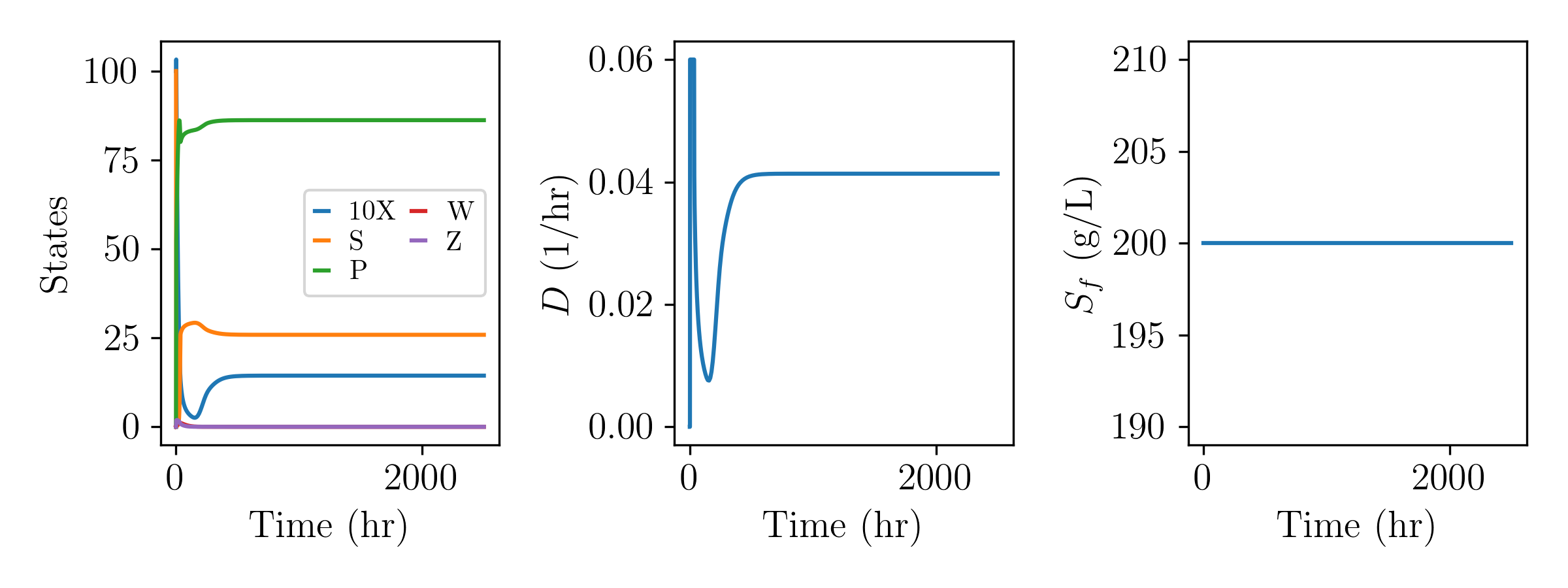} 
        \vspace{-0.2cm}
        \caption{$K_{c}= -0.01, u_{0} = 0.3$}
    \end{subfigure}
    \hfill
    \centering
    \begin{subfigure}{0.8\textwidth}
        \centering
        \includegraphics[width=\textwidth]{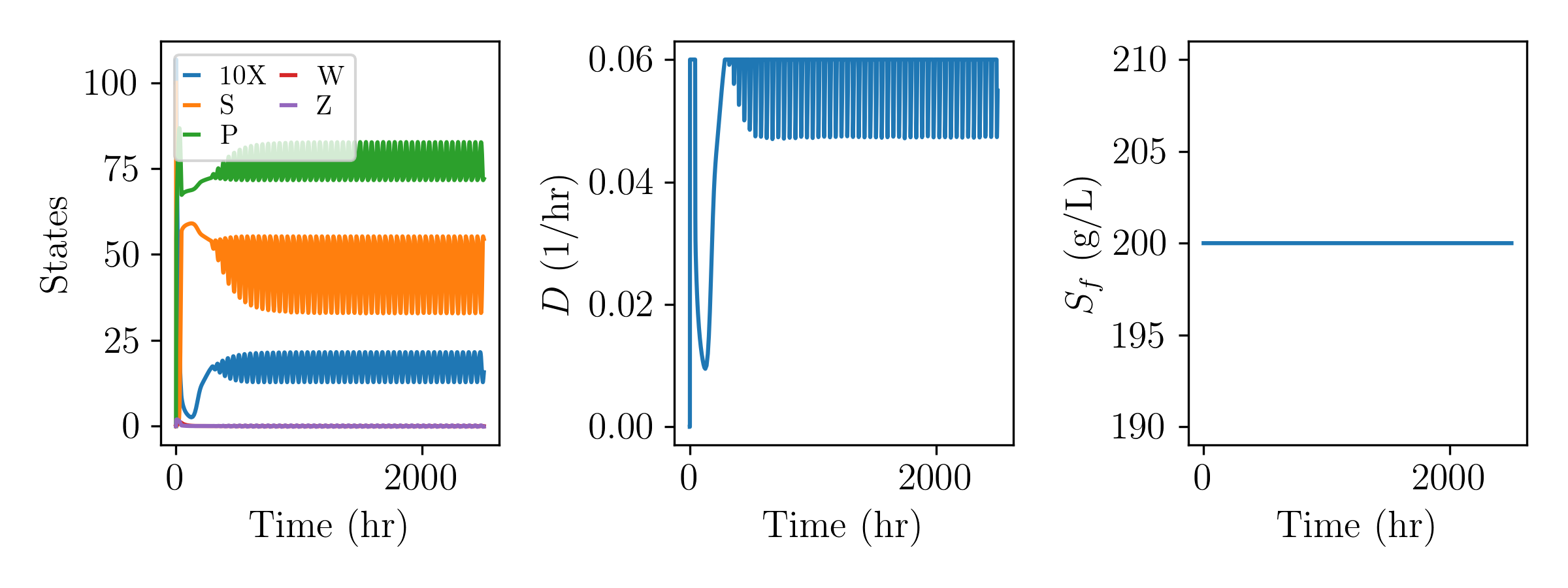} 
        \vspace{-0.2cm}
        \caption{$K_{c}= -0.01, u_{0} = 0.6$}
    \end{subfigure}
    \hfill
    \centering
    \begin{subfigure}{0.8\textwidth}
        \centering
        \includegraphics[width=\textwidth]{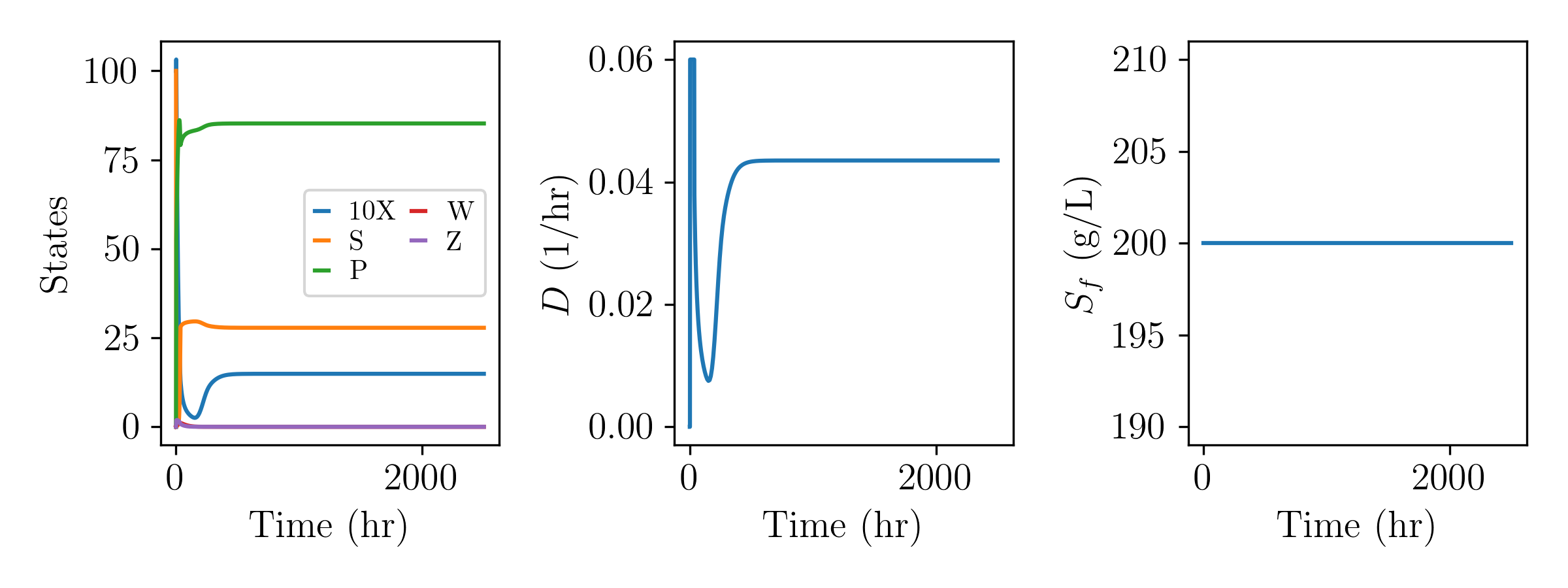} 
        \vspace{-0.2cm}
        \caption{$K_{c}= -0.02, u_{0} = 0.6$}
    \end{subfigure}
    \hfill
    \centering
    \begin{subfigure}{0.8\textwidth}
        \centering
        \includegraphics[width=\textwidth]{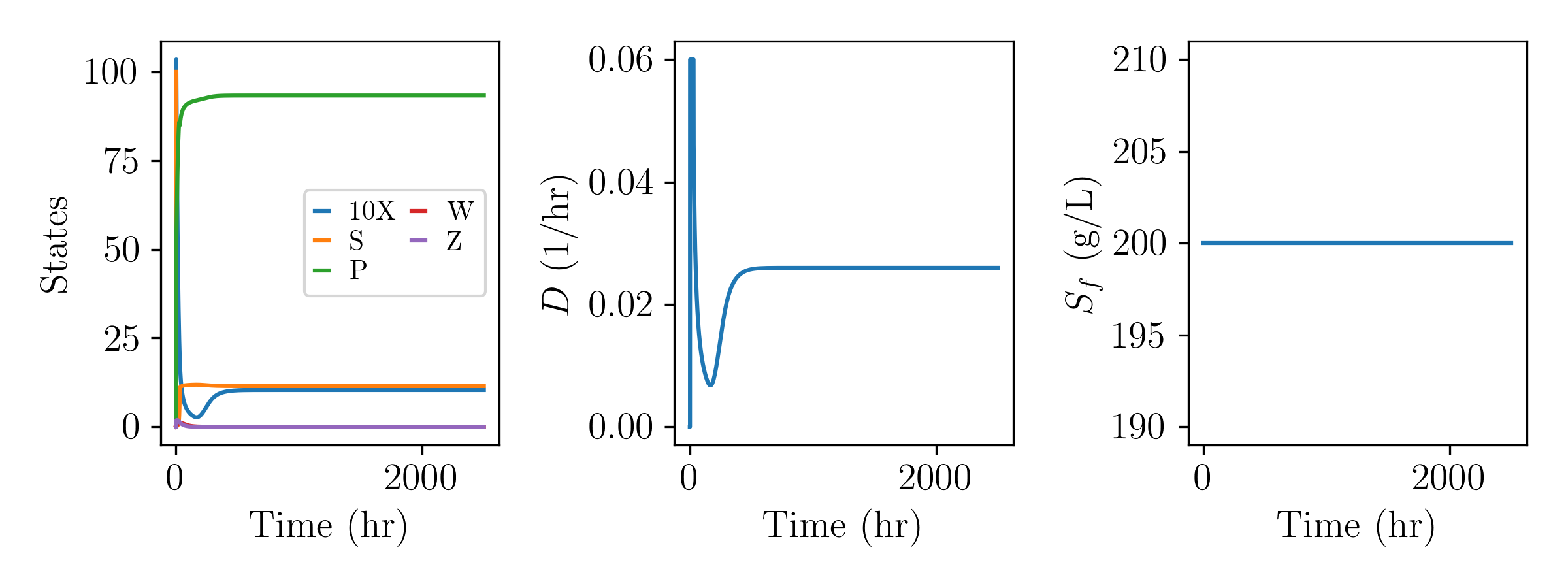} 
        \vspace{-0.2cm}
        \caption{$K_{c}= -0.05, u_{0} = 0.6$}
    \end{subfigure}
    \caption{Closed-loop feedback control simulations with a single loop feedback control law using $S$ as the controlled variable and $D$ as the manipulated variable.}
    \label{fig:ethanol_S_D}
\end{figure}

\begin{figure}[htb]
    \centering
    \begin{subfigure}{0.8\textwidth}
        \centering
        \includegraphics[width=\textwidth]{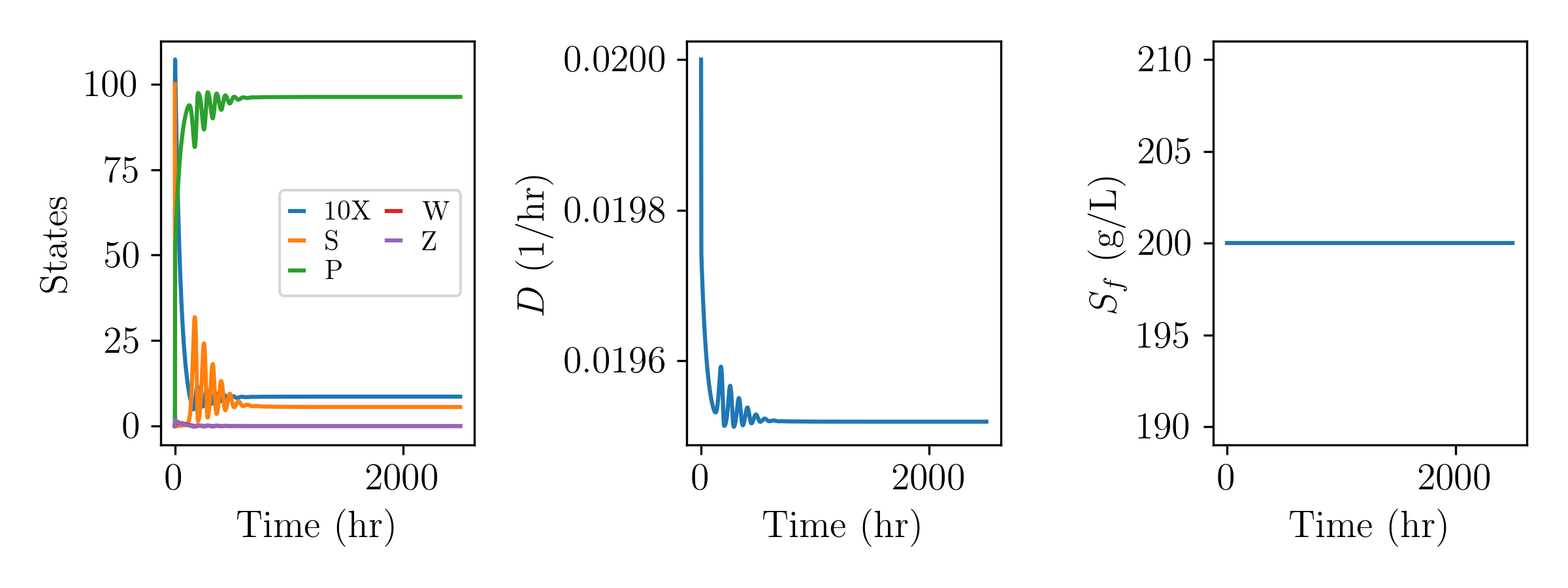} 
        \vspace{-0.2cm}
        \caption{$K_{c}= -5$ $\times$ $10^{-6}$, $ u_{0} = 0.02$}
    \end{subfigure}
    \hfill
    \centering
    \begin{subfigure}{0.8\textwidth}
        \centering
        \includegraphics[width=\textwidth]{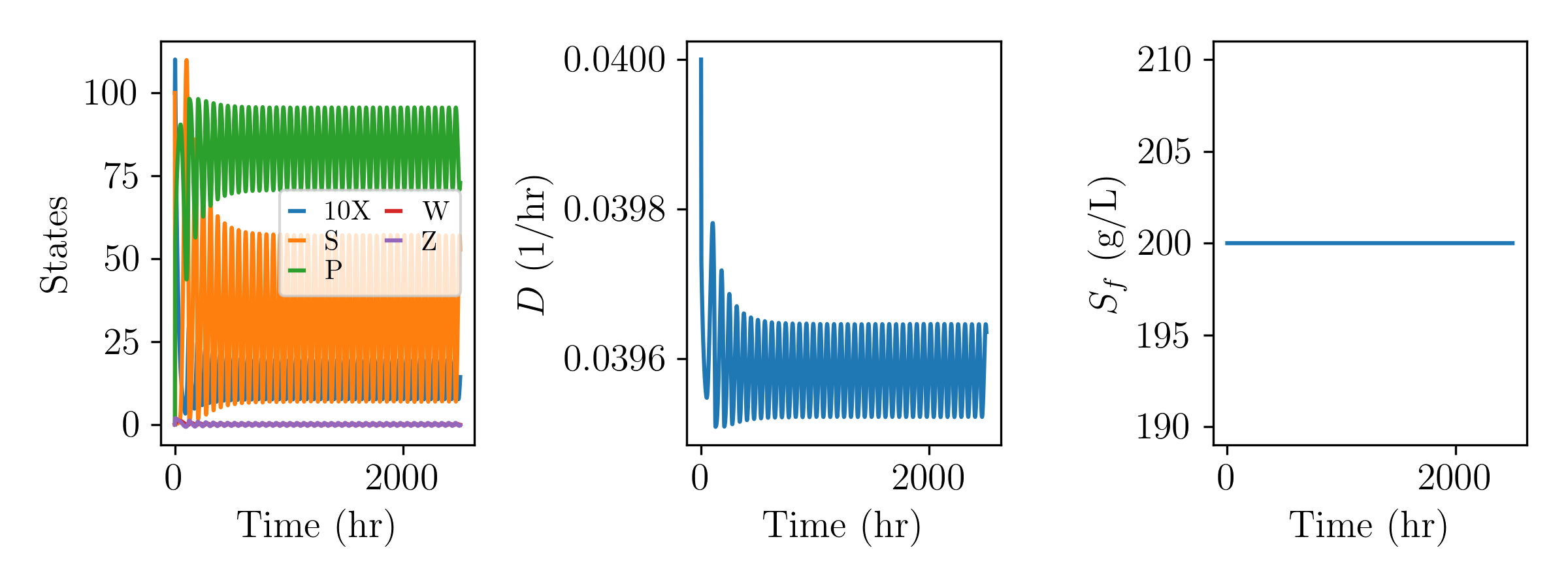} 
        \vspace{-0.2cm}
        \caption{$K_{c}= -5$ $\times$ $10^{-6}$, $ u_{0} = 0.04$}
    \end{subfigure}
    \hfill
    \centering
    \begin{subfigure}{0.8\textwidth}
        \centering
        \includegraphics[width=\textwidth]{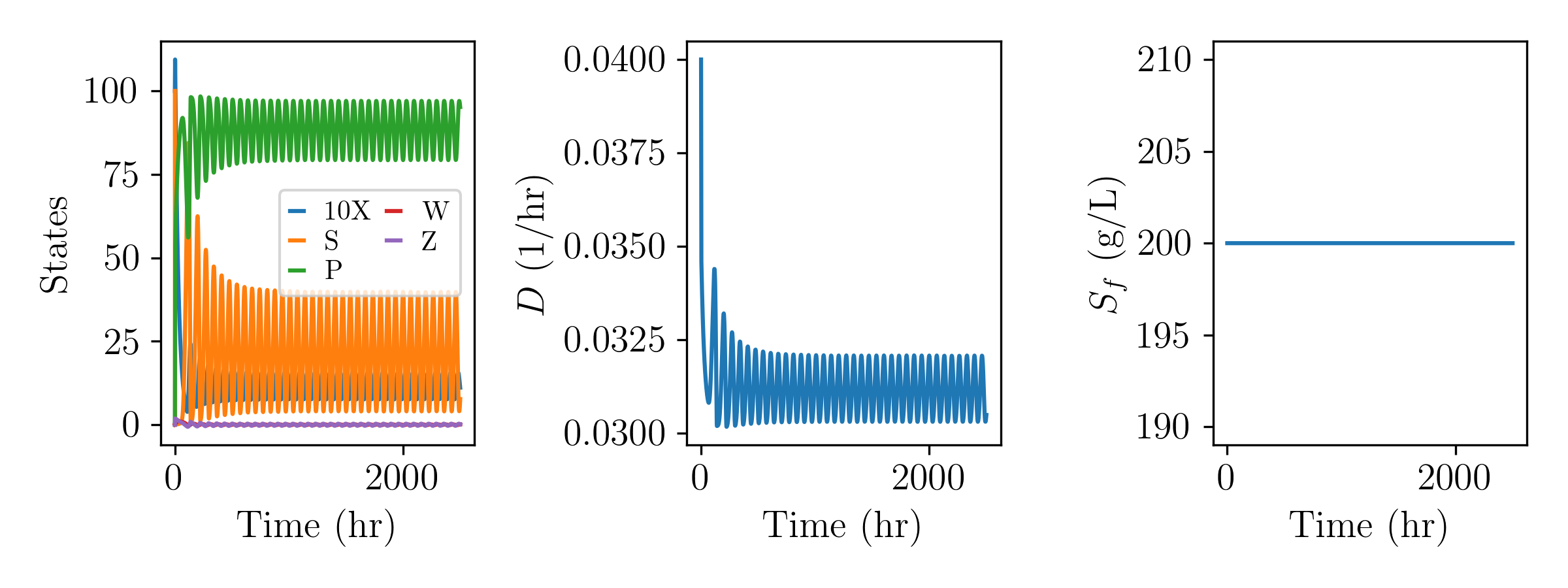} 
        \vspace{-0.2cm}
        \caption{$K_{c}= -1$ $\times$ $10^{-4}$, $ u_{0} = 0.04$}
    \end{subfigure}
    \hfill
    \centering
    \begin{subfigure}{0.8\textwidth}
        \centering
        \includegraphics[width=\textwidth]{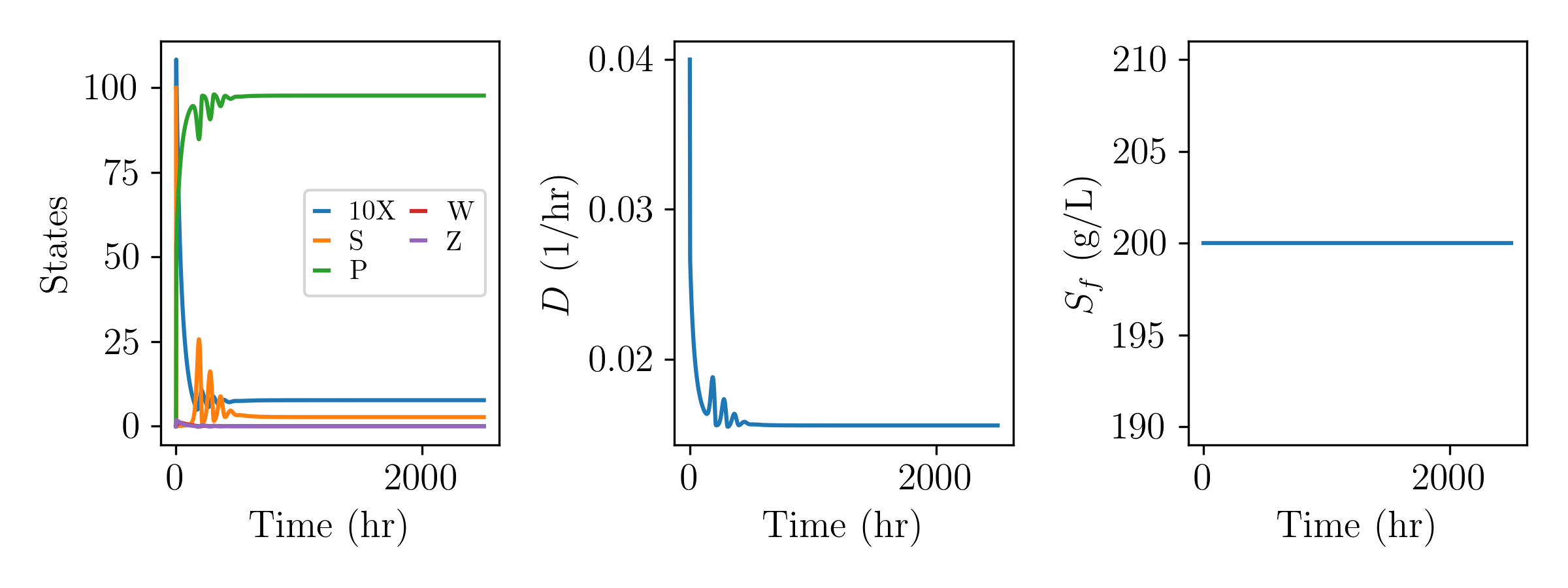} 
        \vspace{-0.2cm}
        \caption{$K_{c}= -2.5$ $\times$ $10^{-4}$, $ u_{0} = 0.04$}
    \end{subfigure}
    \caption{Closed-loop feedback control simulations with a single loop feedback control law using $P$ as the controlled variable and $D$ as the manipulated variable.}
    \label{fig:ethanol_P_D}
\end{figure}

\clearpage

\section{Output Feedback Control of a Continuous Viral Bioreactor}

A series of simulations with different values of the controller gain and bias $K_{c}$ and $u_{0}$ are presented below. 

\begin{figure}[htb]
    \centering
    \begin{subfigure}{0.8\textwidth}
        \centering
        \includegraphics[width=\textwidth]{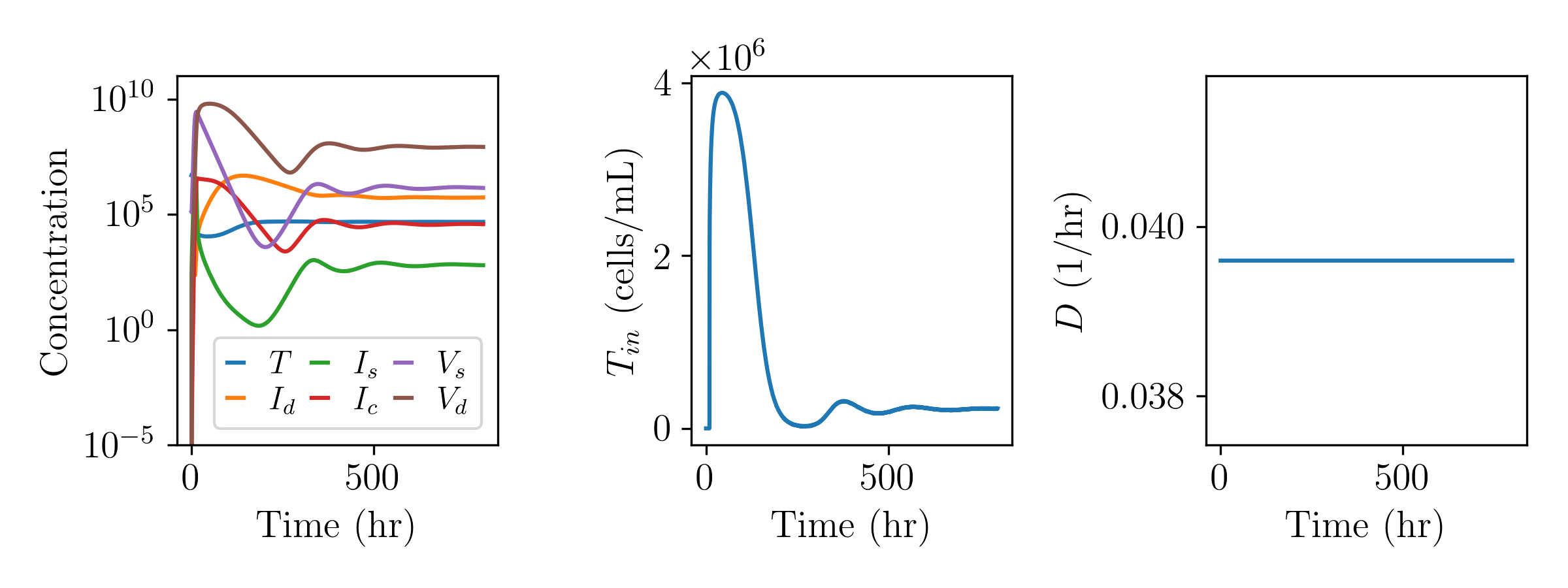} 
        \vspace{-0.2cm}
        \caption{$K_{c}= -100.0, u_{0} = 5$$\times$$10^{6}$}
    \end{subfigure}
    \hfill
    \centering
    \begin{subfigure}{0.8\textwidth}
        \centering
        \includegraphics[width=\textwidth]{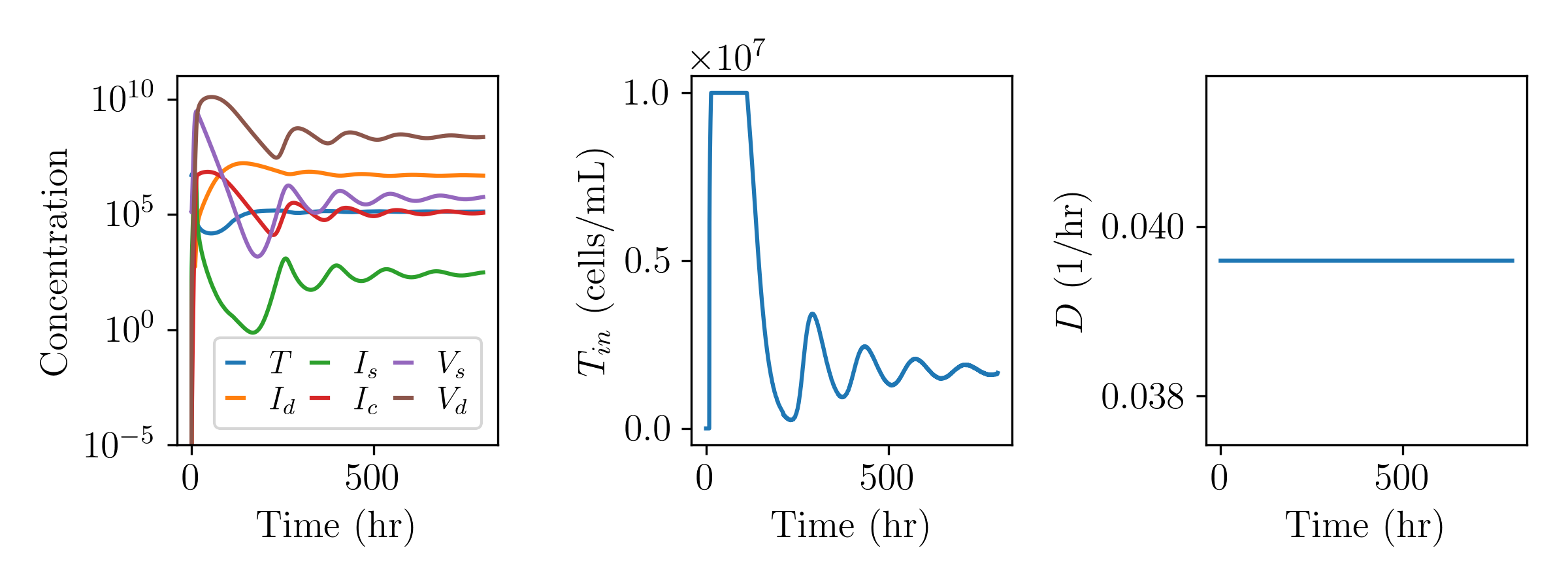} 
        \vspace{-0.2cm}
        \caption{$K_{c}= -100.0, u_{0} = 1.5$$\times$$10^{7}$}
    \end{subfigure}
    \hfill
    \centering
    \begin{subfigure}{0.8\textwidth}
        \centering
        \includegraphics[width=\textwidth]{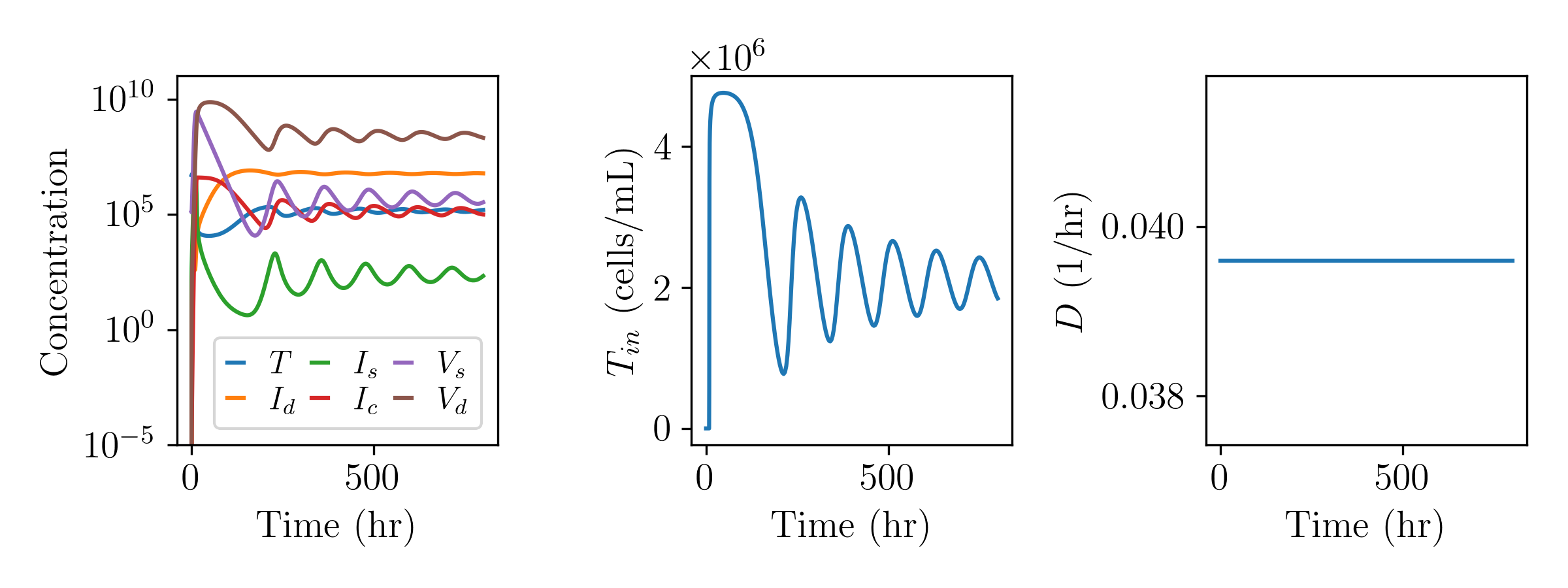} 
        \vspace{-0.2cm}
        \caption{$K_{c}= -20.0, u_{0} = 5$$\times$$10^{6}$}
    \end{subfigure}
    \hfill
    \centering
    \begin{subfigure}{0.8\textwidth}
        \centering
        \includegraphics[width=\textwidth]{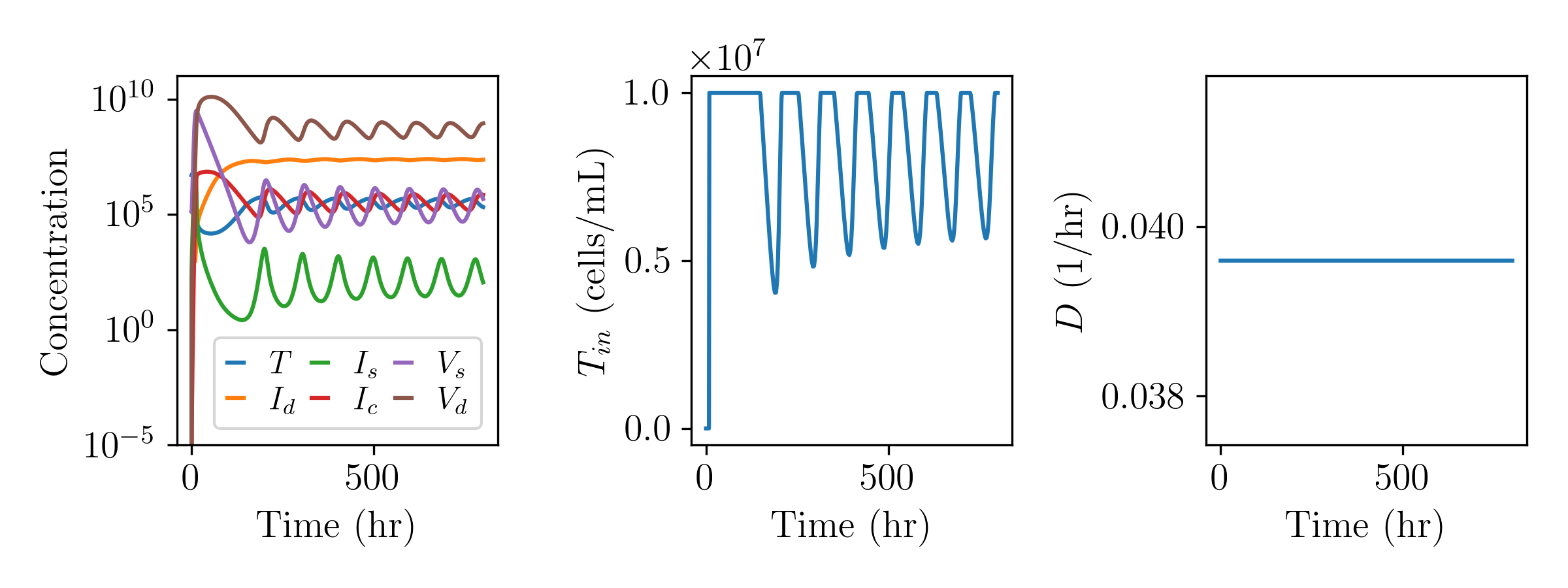} 
        \vspace{-0.2cm}
        \caption{$K_{c}= -20.0, u_{0} = 1.5$$\times$$10^{7}$}
    \end{subfigure}
    \caption{Closed-loop feedback control simulations with a single loop feedback control law using $T$ as the controlled variable and $T_{in}$ as the manipulated variable.}
    \label{fig:Frensing_T_Tin}
\end{figure}

\begin{figure}[htb]
    \centering
    \begin{subfigure}{0.8\textwidth}
        \centering
        \includegraphics[width=\textwidth]{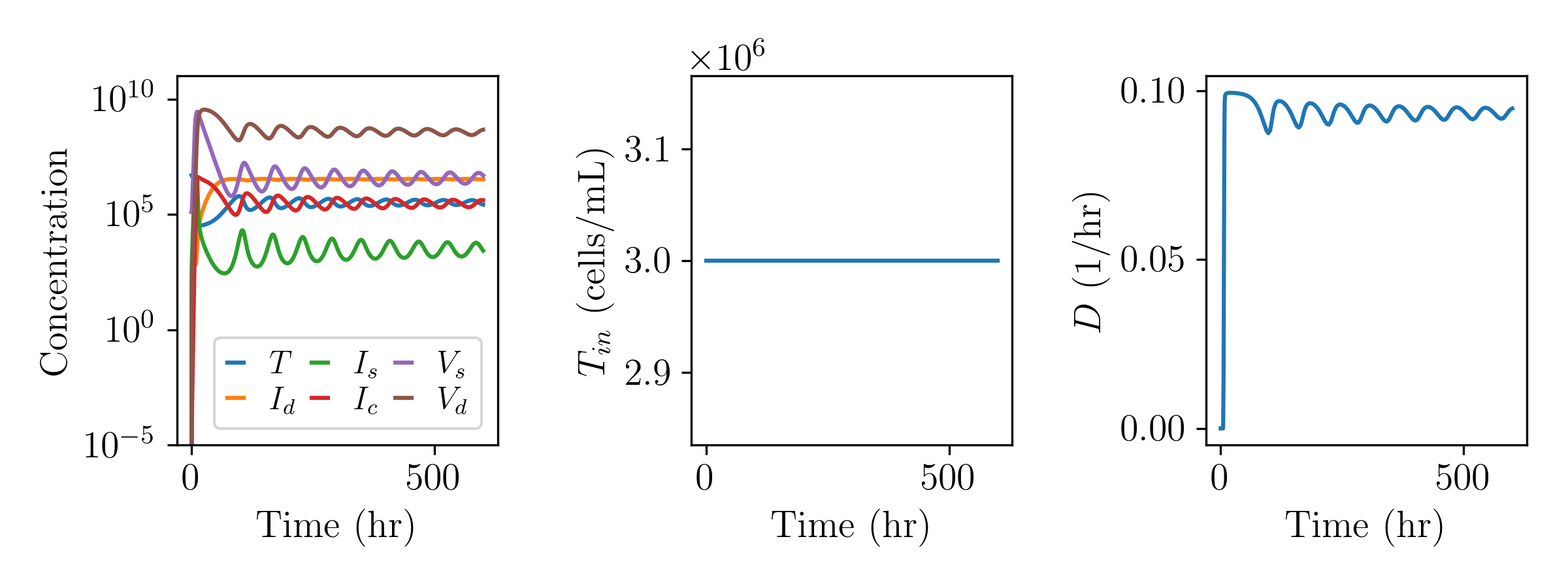} 
        \vspace{-0.2cm}
        \caption{$K_{c}= -2$$\times$$10$$^{-8}$, $u_{0} = 0.1$}
    \end{subfigure}
    \hfill
    \centering
    \begin{subfigure}{0.8\textwidth}
        \centering
        \includegraphics[width=\textwidth]{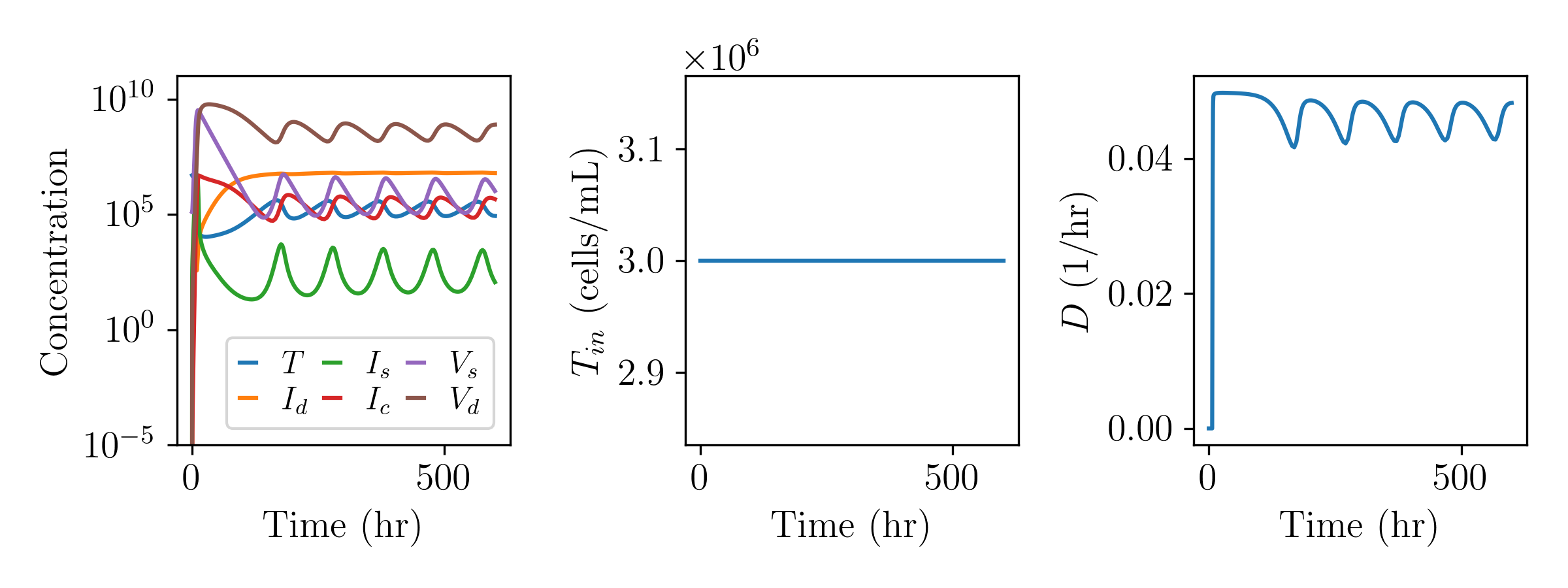} 
        \vspace{-0.2cm}
        \caption{$K_{c}= -2$$\times$$10$$^{-8}$, $u_{0} = 0.05$}
    \end{subfigure}
    \hfill
    \centering
    \begin{subfigure}{0.8\textwidth}
        \centering
        \includegraphics[width=\textwidth]{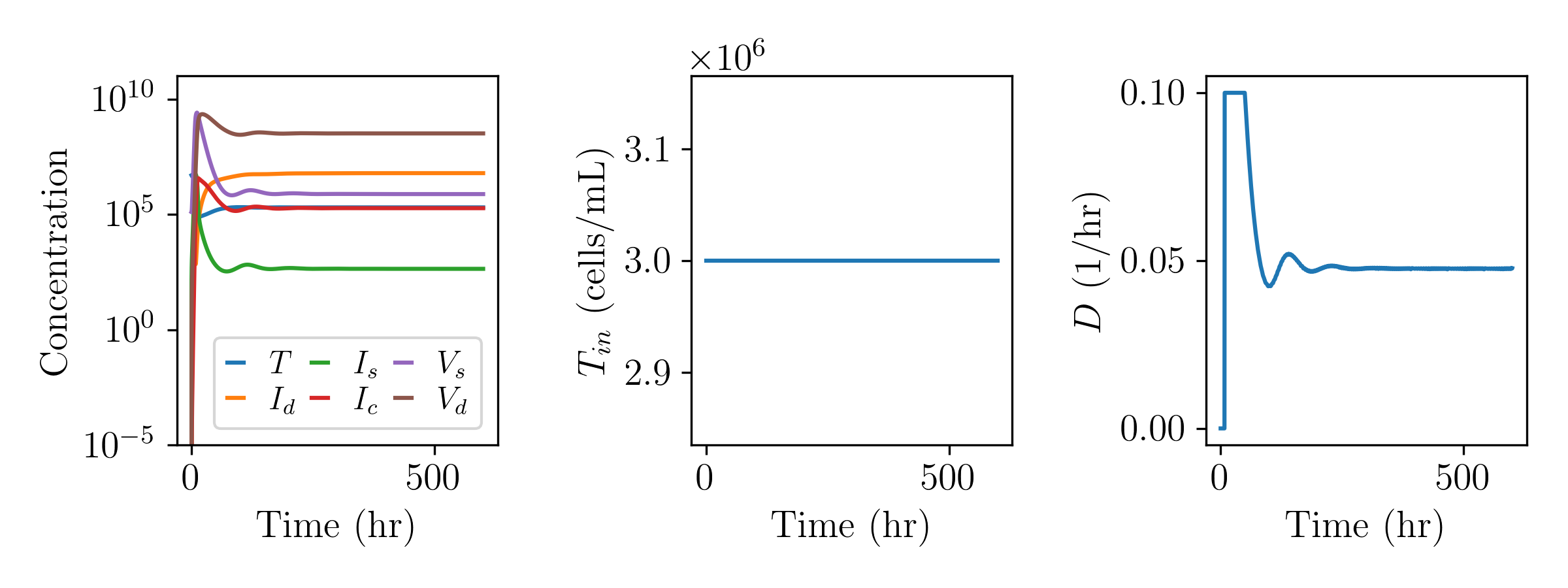} 
        \vspace{-0.2cm}
        \caption{$K_{c}= -1$$\times$$10$$^{-6}$, $u_{0} = 0.25$}
    \end{subfigure}
    \hfill
    \centering
    \begin{subfigure}{0.8\textwidth}
        \centering
        \includegraphics[width=\textwidth]{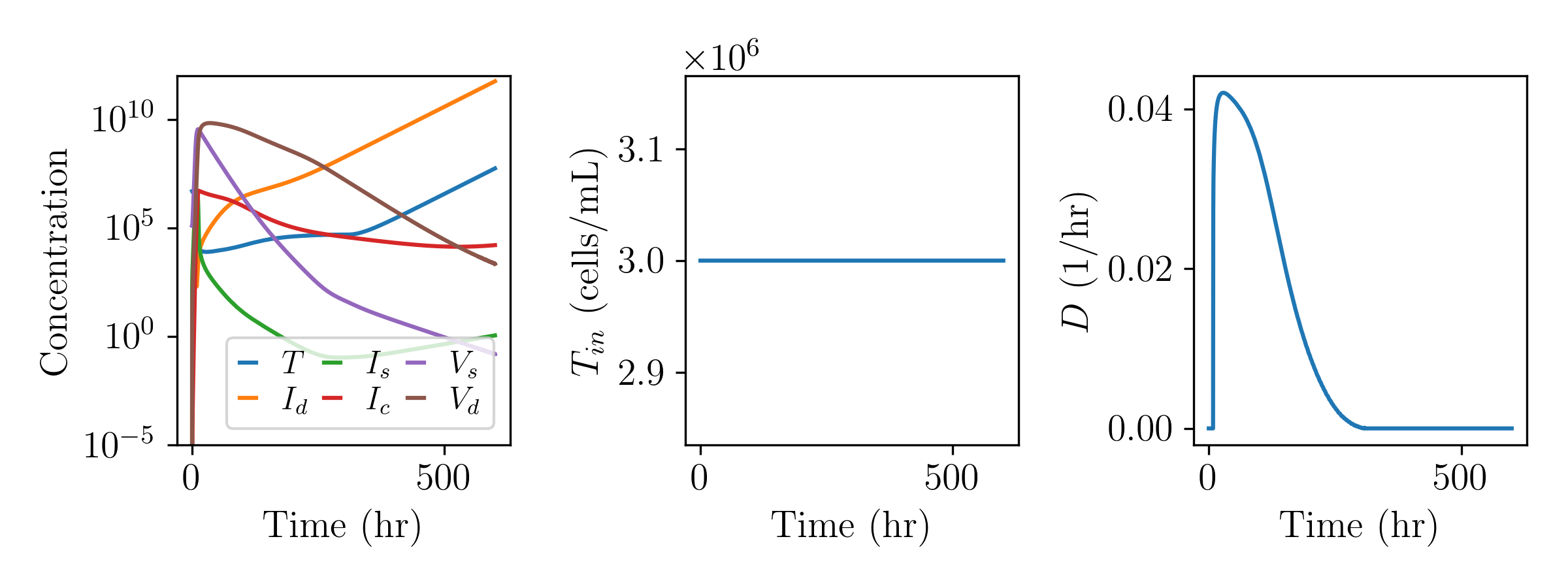} 
        \vspace{-0.2cm}
        \caption{$K_{c}= -1$$\times$$10$$^{-6}$, $u_{0} = 0.05$}
    \end{subfigure}
    \caption{Closed-loop feedback control simulations with a single loop feedback control law using the uninfected target cell concentration $T$ as the controlled variable and the dilution rate $D$ as the manipulated variable.}
    \label{fig:Frensing_T_D}
\end{figure}

\clearpage
\section{Continuous Yeast Cell Bioreactor Dynamics Analysis}

Single-parameter variation studies for probing system dynamics are provided below. Model (40) was simulated at all nominal conditions and parameters presented in Table 4, with the dilution rate $D$ varied in the controller actuator limits $D\in[0,1]$ 1/hr.

\begin{figure}[htbp]
    \centering
    \includegraphics[width= 0.95\linewidth]{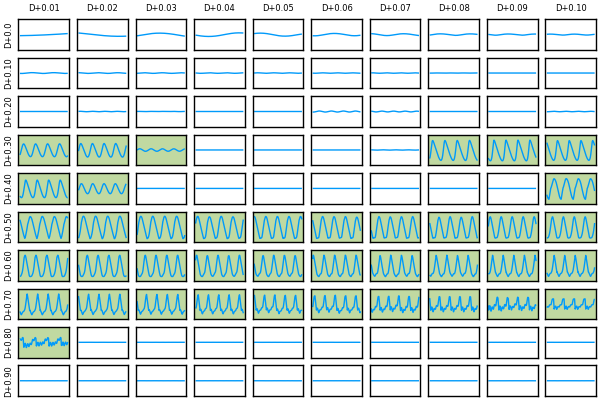}
    \caption{Time-series of yeast VCD $m_0$ state values for the last 5 hours of simulation with $t_f=100$ hr. Plot shading indicates detected stable oscillations.}
    \label{fig:yeast-m0-oscillations-one-peak}
\end{figure}

\begin{figure}[htbp]
    \centering
    \includegraphics[width= 0.95\linewidth]{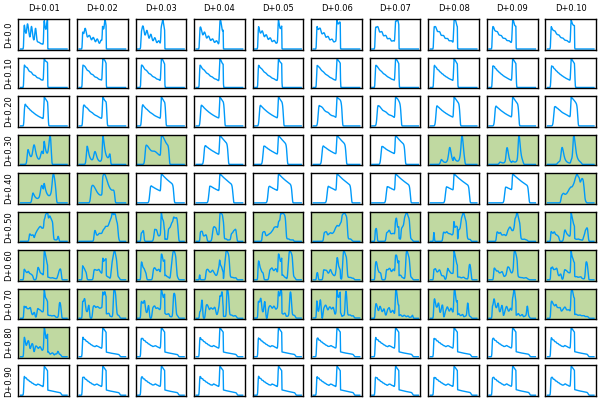}
    \caption{Corresponding final time $t_\textrm{f}$ yeast cell mass distributions $N(m,t_\textrm{f})$ for the last five hours of simulation with $t_f=100$ hr. Plot shading indicates detected stable oscillations in the corresponding yeast VCD $m_0$ state.}
    \label{fig:yeast-N-mt-oscillations-one-peak}
\end{figure}

\begin{figure}[htbp]
    \centering
    \includegraphics[width= 0.95\linewidth]{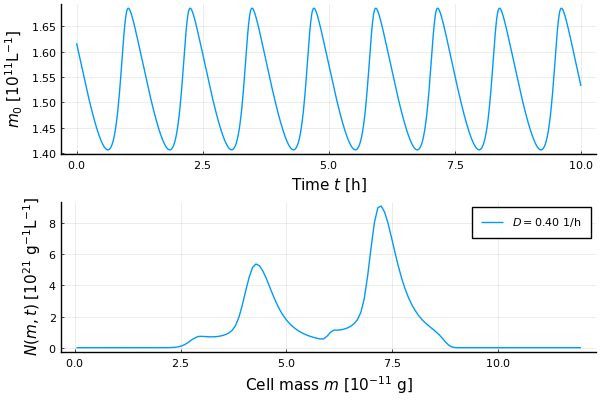}
    \caption{Yeast cell VCD $m_0$ open loop oscillations simulated forward 10 hours after initial simulation to $t_f=100$ hr at the nominal conditions in the case study. Oscillations persist, and the corresponding cell distribution after 10 hours demonstrates a high degree of synchronicity that persists.}
    \label{fig:yeast-two-peak-IC}
\end{figure}

\begin{figure}[htbp]
    \centering
    \includegraphics[width= 0.95\linewidth]{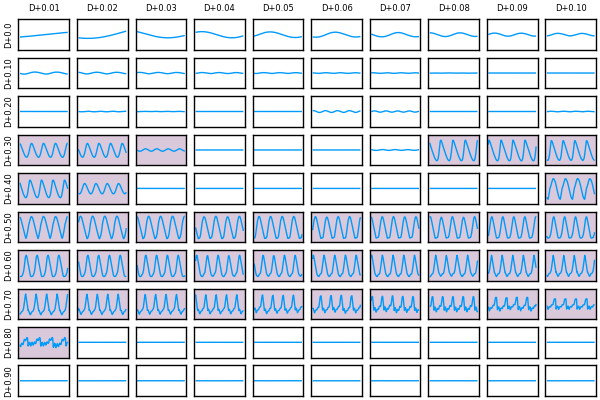}
    \caption{Using the initial conditions and cell distribution from the simulation in Fig. \ref{fig:yeast-two-peak-IC}, the system is simulated forward for 50 hours to investigate any possible changes in dynamics due to initialization. The time-series of yeast cell VCD $m_0$ state values shows oscillatory behavior for the same dilution rate $D$ parameter values as in Fig. \ref{fig:yeast-m0-oscillations-one-peak}, regardless of initial synchronization in the cell distribution.}
    \label{fig:yeast-m0-oscillations-two-peak}
\end{figure}

\begin{figure}[htbp]
    \centering
    \includegraphics[width= 0.95\linewidth]{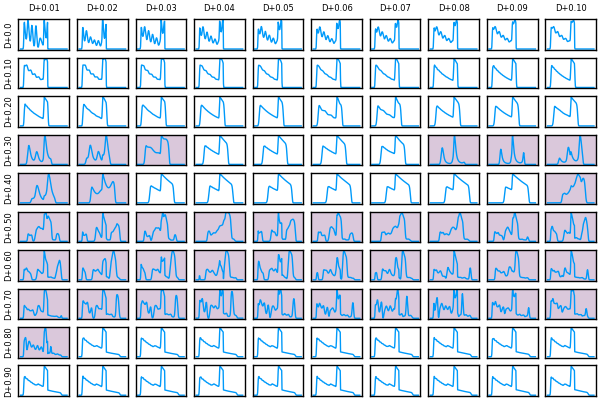}
    \caption{Corresponding final time $t_\textrm{f}$ yeast cell mass distributions $N(m,t_\textrm{f})$ for the last five hours of simulations given in Fig. \ref{fig:yeast-m0-oscillations-two-peak}.}
    \label{fig:yeast-N-mt-oscillations-two-peak}
\end{figure}

\clearpage
\section{Continuous Yeast Bioreactor Output Feedback Control Analysis}

The simulations below include open and closed loop responses for attenuating oscillations in the continuous yeast culture reactor for a variety of single-input, single-output (SISO) controller choices. Stable oscillations in the continuous yeast culture reactor substrate concentration $S$ and yeast VCD $m_0$ occur after initial transient modes for reactor start-up response relax. All of the following simulations are initialized using the final time ($t_f=100$ hr) cell distribution, substrate concentrations, and yeast VCD obtained from the open-loop simulation at nominal operating conditions given in Table 4 (Fig. \ref{fig:yeast-two-peak-IC}).

\begin{figure}[htbp]
    \centering
    \includegraphics[width= 0.95\linewidth]{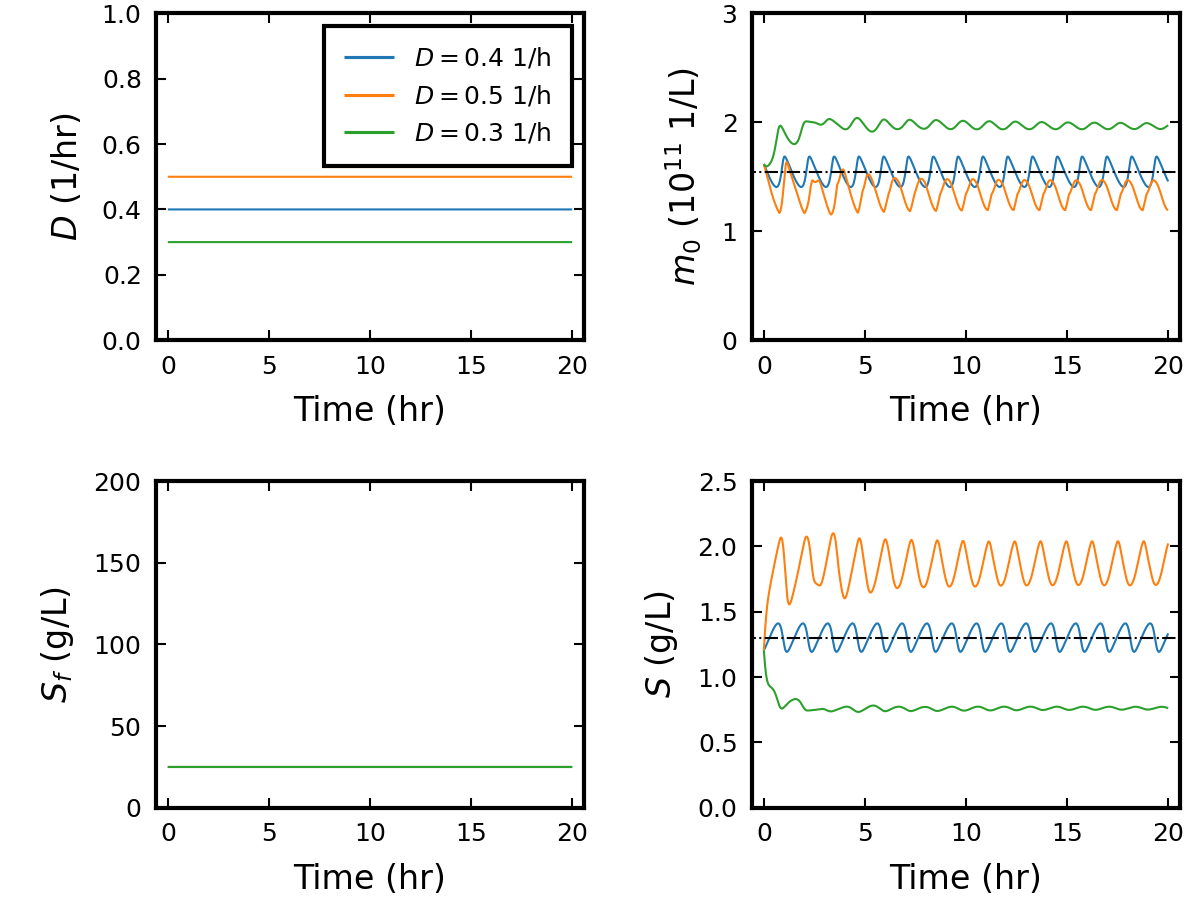}
    \caption{The open-loop step response for stepping the dilution rate $D$ by $\pm 0.1$ 1/hr about the nominal operating conditions causing oscillations ($D=0.4$ 1/hr and $S_f = 25$ g/L). These process gain values are conditional on the nominal operating conditions producing the oscillatory trajectories due to system nonlinearities.}
    \label{fig:yeast-openloop-D}
\end{figure}

\begin{figure}[htbp]
    \centering
    \includegraphics[width= 0.95\linewidth]{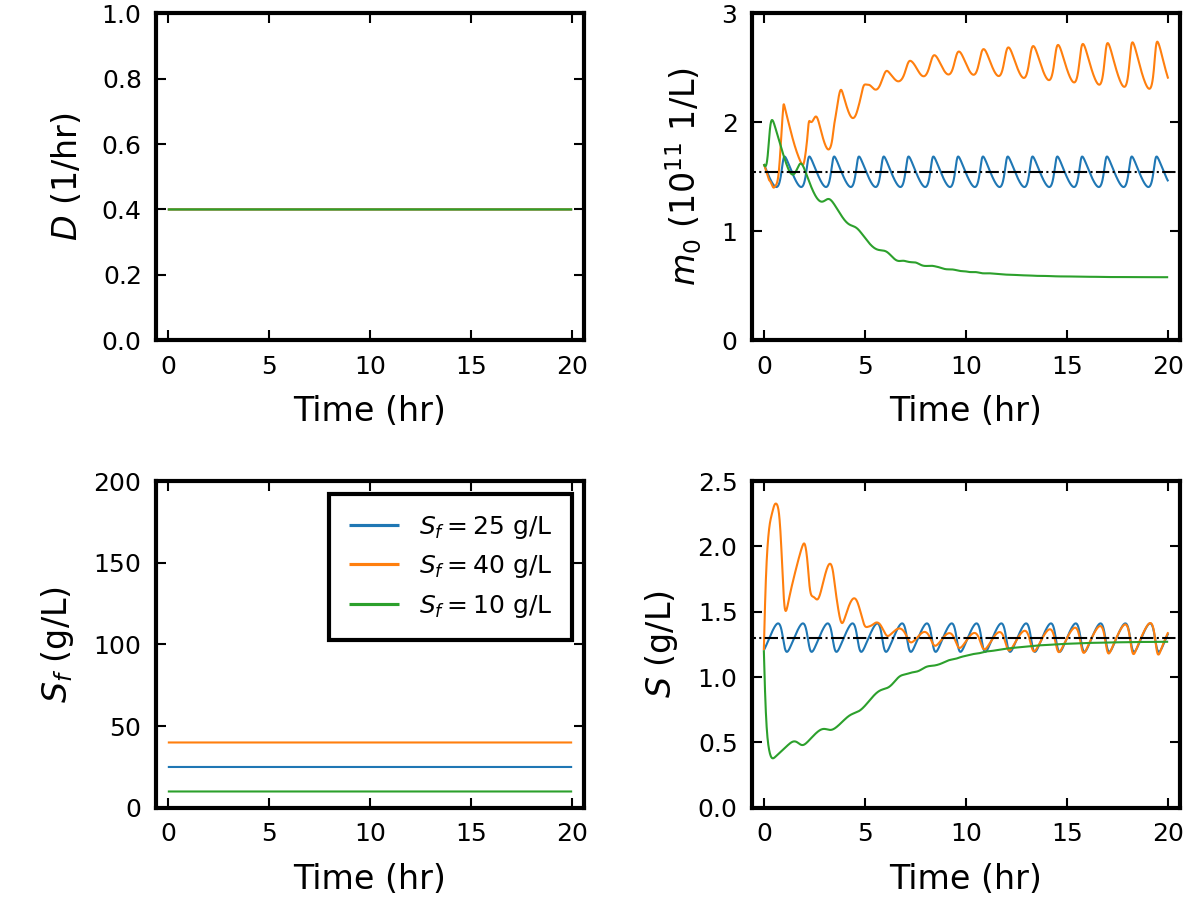}
    \caption{The open-loop step response for stepping the substrate feed concentration $S_\textrm{f}$ by $\pm 15$ g/Lr about the nominal operating conditions causing oscillations ($D=0.4$ 1/hr and $S_f = 25$ g/L). These process gain values are conditional on the nominal operating conditions producing the oscillatory trajectories due to system nonlinearities.}
    \label{fig:yeast-openloop-Sf}
\end{figure}

Step-response studies in Figs. \ref{fig:yeast-openloop-D} and \ref{fig:yeast-openloop-Sf} provide graphical means for approximating the four different process gains $K_\textrm{p}$ possible for this two-input, two-output system. In simulations where stable oscillations occur, the half-width of oscillations in the oscillating state variables is taken as the response value for process gain estimation.

\begin{figure}[htbp]
    \centering
    \includegraphics[width= 0.95\linewidth]{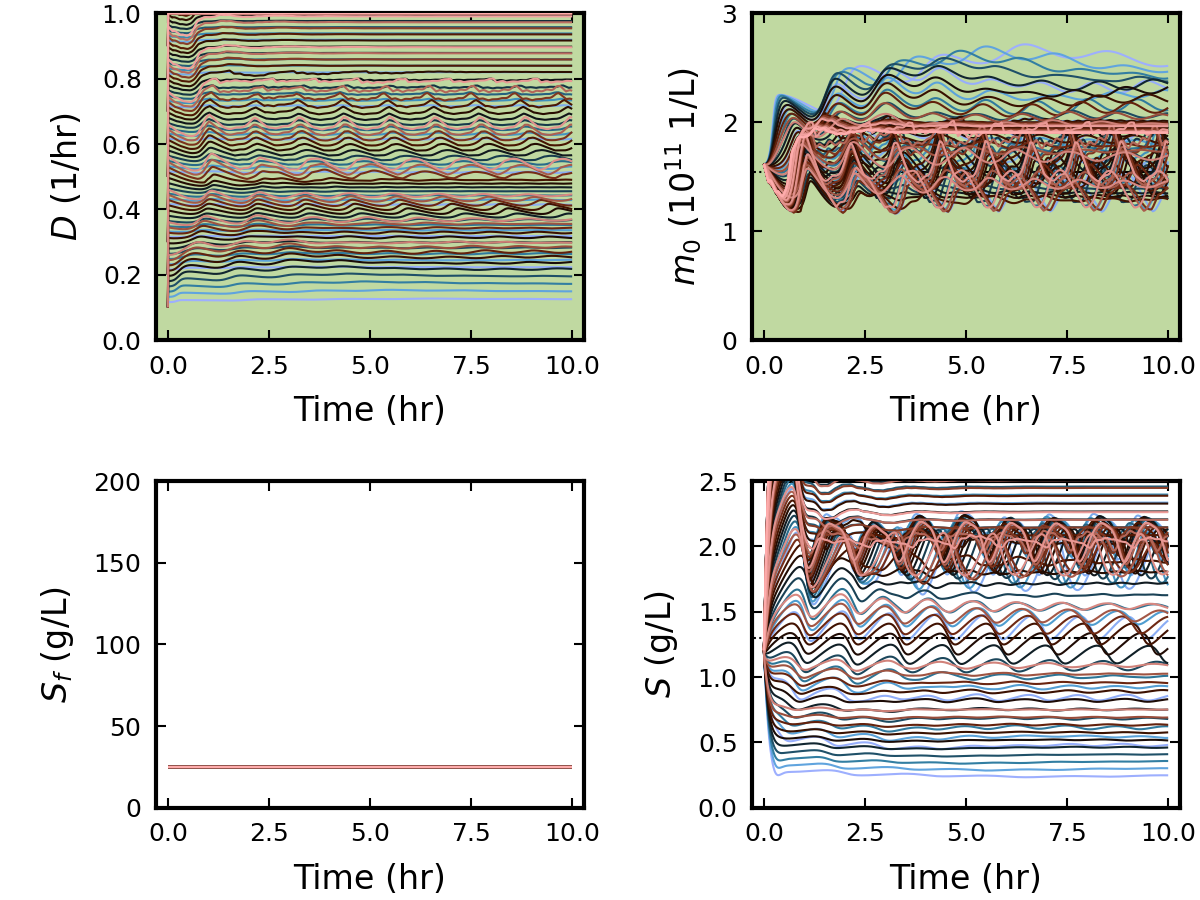}
    \caption{Controller tuning parameter ranges tested: 10 equally-spaced values for $K_c\in[1.0e-13,1.0e-12]$ with 10 equally-spaced values for $u_0\in[0.1,1]$ for 100 total combinations.}
    \label{fig:yeast-closedloop-Dm0-SISO-u0-varied}
\end{figure}

\begin{figure}[htbp]
    \centering
    \includegraphics[width= 0.95\linewidth]{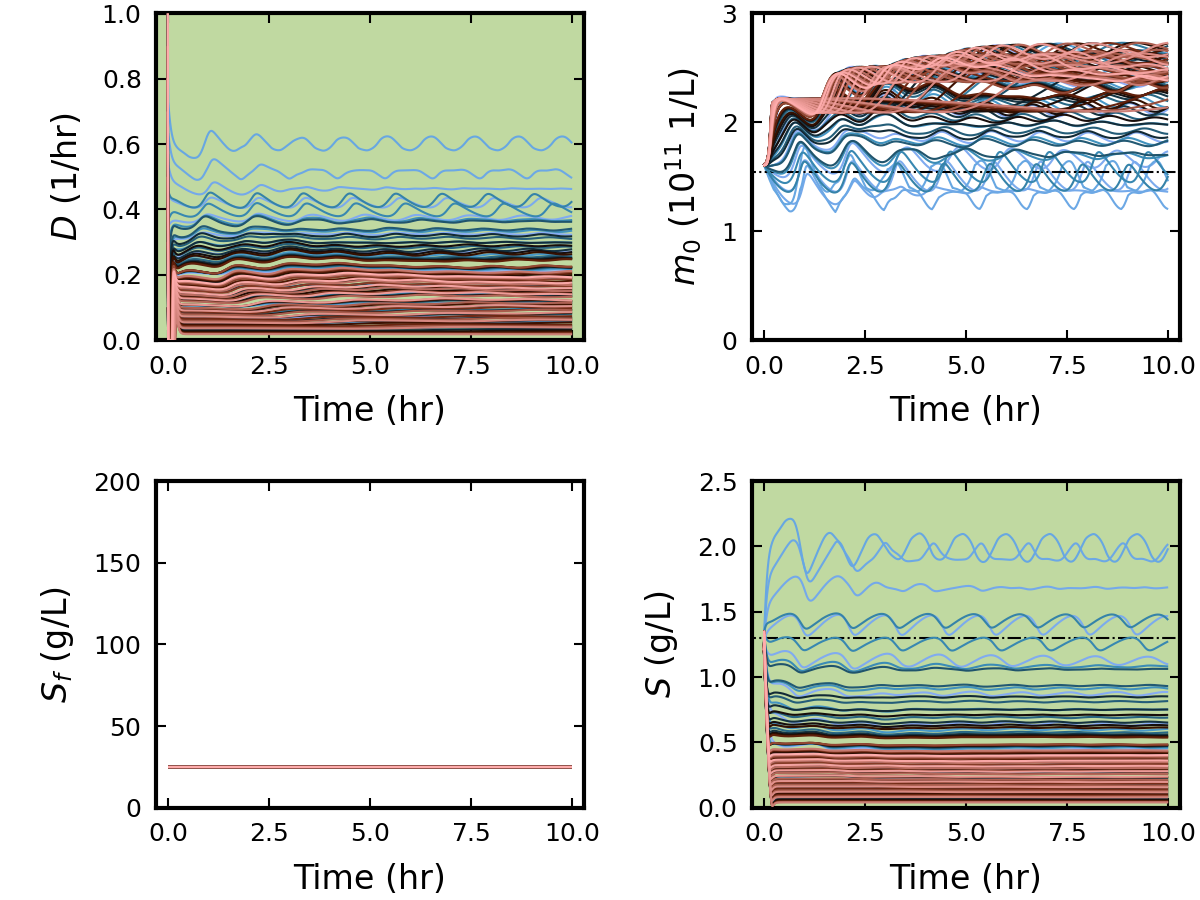}
    \caption{Controller tuning parameter ranges tested: 10 equally-spaced values for $K_c\in[-2,-0.2]$ with 10 equally-spaced values for $u_0\in[0.1,1]$ for 100 total combinations.}
    \label{fig:yeast-closedloop-DS-SISO-u0-varied}
\end{figure}

\begin{figure}[htbp]
    \centering
    \includegraphics[width= 0.95\linewidth]{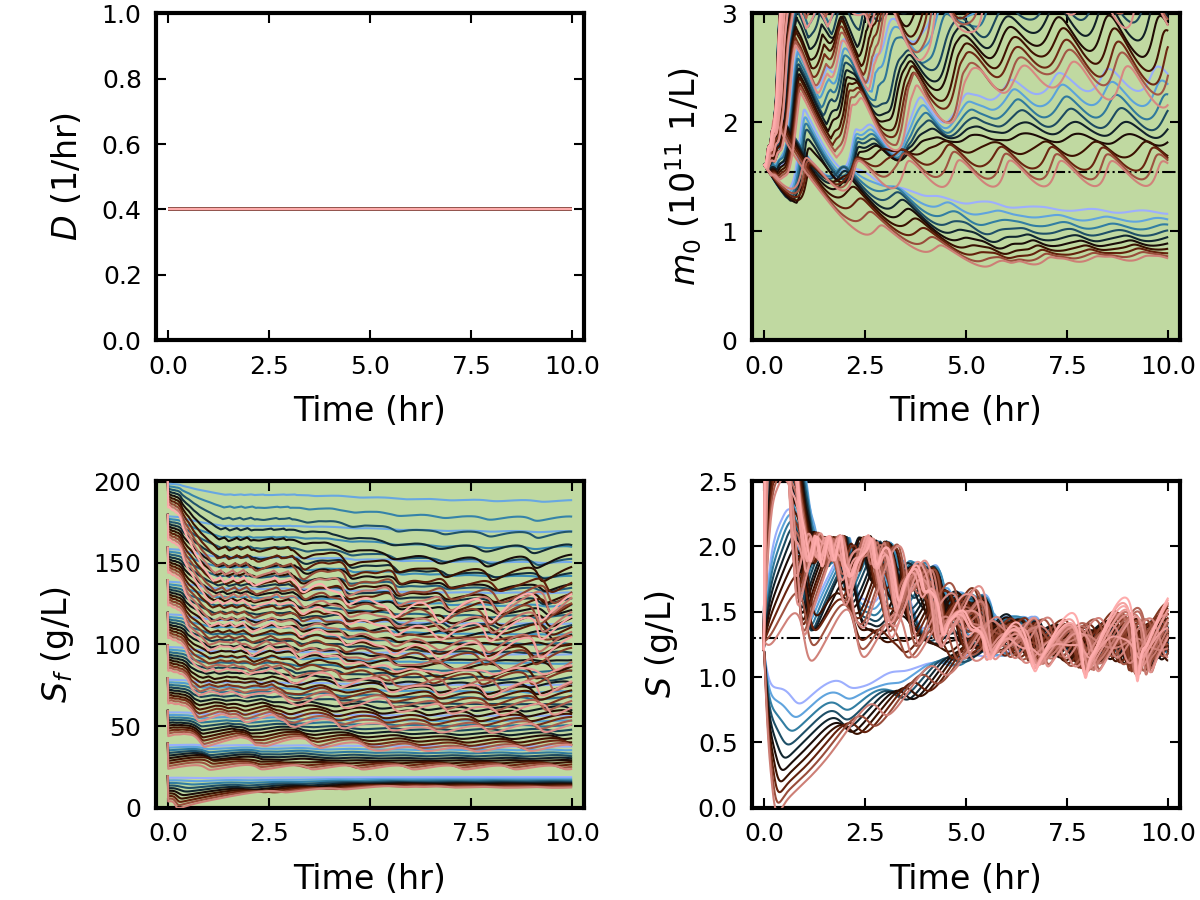}
    \caption{Controller tuning parameter ranges tested: 10 equally-spaced values for $K_c\in[-1.0e-11,-1.0e-10]$ with 10 equally-spaced values for $u_0\in[20,200]$ for 100 total combinations.}
    \label{fig:yeast-closedloop-Sfm0-SISO-u0-varied}
\end{figure}

\begin{figure}[htbp]
    \centering
    \includegraphics[width= 0.95\linewidth]{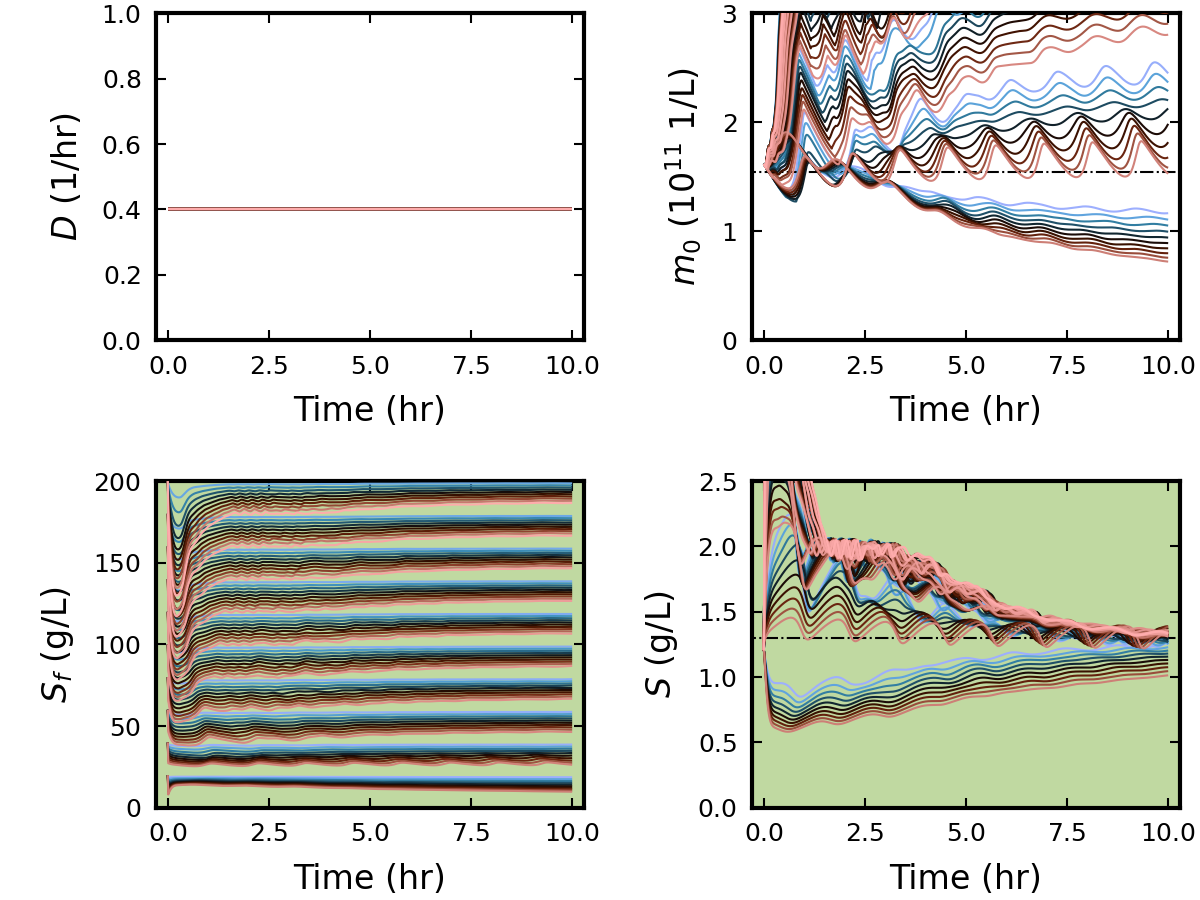}
    \caption{Controller tuning parameter ranges tested: 10 equally-spaced values for $K_c\in[-10,-1]$ with 10 equally-spaced values for $u_0\in[20,200]$ for 100 total combinations.}
    \label{fig:yeast-closedloop-SfS-SISO-u0-varied}
\end{figure}

Figs. \ref{fig:yeast-closedloop-Dm0-SISO-u0-varied}--\ref{fig:yeast-closedloop-SfS-SISO-u0-varied} provide closed-loop proportional feedback controller responses for the four different SISO loops. As described in Subsection 5.3, a range of proportional controller gains $K_c$ and controller biases $u_0$ is chosen, each containing the approximate $K_c$ obtained from Figs. \ref{fig:yeast-openloop-D} and \ref{fig:yeast-openloop-Sf}. Simulations are run for enough time, i.e., 10 hr, to discern between steady state and stable oscillation dynamics. Plot shading indicates the control loop input-output pairing.

